\documentclass[useAMS,usenatbib]{mn2e}
\usepackage{times}
\usepackage{amssymb}
\usepackage{graphicx}
\usepackage{rotating}
\usepackage{lscape}
\usepackage{multirow}
\usepackage{fixltx2e}

\begin{document}

%
%
%
%



\title[HiZELS merger rates]{The merger rates and sizes of galaxies across the peak epoch of star formation from the HiZELS survey.} 
\author[J.P. Stott et al.]{John P. Stott$^{1}$\thanks{E-mail: j.p.stott@durham.ac.uk}, David Sobral$^{2}$, Ian Smail$^{1}$, Richard Bower$^{1}$, Philip N. Best$^{3}$, \newauthor James E. Geach$^{4}$
\\
$^{1}$ Institute for Computational Cosmology, Durham University, South Road, Durham, DH1 3LE, UK\\
$^{2}$ Leiden Observatory, Leiden University, P.O. Box 9513, NL-2300 RA Leiden, The Netherlands\\
$^{3}$ SUPA, Institute for Astronomy, Royal Observatory of Edinburgh, Blackford Hill, Edinburgh, EH9 3HJ, UK\\
$^{4}$ Department of Physics, McGill University, Ernest Rutherford Building, 3600 Rue University, Montr«eal, Quebec, H3A 2T8, Canada\\
}

\date{}

\pagerange{\pageref{firstpage}--\pageref{lastpage}} \pubyear{2012}

\maketitle

\label{firstpage}

\begin{abstract}

We use the HiZELS narrow-band $\rm H\alpha$ survey in combination with CANDELS, UKIDSS and WIRDS near-infrared imaging, to investigate the morphologies, merger rates and sizes of a sample of $\rm H\alpha$ emitting galaxies in the redshift range $z=0.40- 2.23$, an epoch encompassing the rise to the peak of the star formation rate density. Merger rates are estimated from space- and ground-based imaging using the $\rm M_{20}$ coefficient. To account for the increase in the specific star-formation rate (sSFR) of the star forming `main-sequence' with redshift, we normalise the star-formation rate of galaxies at each epoch to the typical value derived from the $\rm H\alpha$ luminosity function. Once this trend in sSFR is removed we see no evidence for an increase in the number density of star-forming galaxies or the merger rate with redshift. We thus conclude that neither is the main driver of the enhanced star-formation rate density at $z\sim1-2$, with secular processes such as instabilities within efficiently fuelled, gas-rich discs or multiple minor mergers the most likely alternatives. However, we find that $\sim40-50\%$ of starburst galaxies, those with enhanced specific star formation at their epoch, are major mergers and this fraction is redshift independent. Finally, we find the surprising result that the typical size of a star-forming galaxy of a given mass does not evolve across the redshift range considered, suggesting a universal size-mass relation. Taken in combination, these results indicate a star-forming galaxy population that is statistically similar in physical size, merger rate and mass over the $\sim6$ Gyr covered in this study, despite the increase in typical sSFR.
 
\end{abstract}


\section{Introduction}
The peak in the volume averaged star formation rate for galaxies occurs in the redshift range $z=1-3$ \citep{lilly1996,madau1996,sobral2012b}. At this epoch, the star formation rate (SFR) in typical galaxies is an order of magnitude higher than in the local Universe \citep{reddy2009}. This is  the era when most of the stars in the Universe were formed and represents the peak in black hole activity. The task is now to address `how' and `why' the Universe was so different then.

A picture is emerging in which the dominant mode of star formation at this earlier epoch is very different to that in the local Universe.  Rather than the quiescent formation of stars that is the norm in today's Universe, violent  episodes of star formation are dominated by the formation of super-star clusters (e.g. \citealt{swinbank2010a}). However, the origin of these differences is somewhat controversial: one picture, which has some observational support, is that they are driven by an increase in the galaxy merger rate (e.g. \citealt{somerville2001,hopkins2006,conselice2003,conselice2008}), but other theories have suggested that it is the result of the higher rate of gas accretion expected in the high-redshift Universe \citep{keres2005,dekel2009}. It is therefore important to study the SFR, merger fractions and gas content of these galaxies in order to identify the processes responsible for driving this epoch of enhanced activity.

In recent years the presence of a star forming `main-sequence' seen in the local Universe (e.g. \citealt{brinchmann2004}) has been confirmed at increasingly high redshift \citep{elbaz2007,elbaz2011,daddi2007,rodig2011,sargent2012}. This is a relation between SFR and stellar mass for star forming galaxies, with a typical specific star formation rate (sSFR, the ratio of the star formation rate to the stellar mass of the galaxy) found to increase with redshift \citep{elbaz2011}. Galaxies that lie off this relation with sSFRs too high to be in the typical star-forming population are often described as `starbursts' and are thought to be triggered by violent events such as major mergers \citep{hopkins2006,elbaz2007,rodig2011}.

From a theoretical perspective, in the $\Lambda$ Cold Dark Matter ($\Lambda$CDM) paradigm dark matter halos merge hierarchically from the bottom up, with the largest halos created at later times (e.g. \citealt{lacey1993, cole2000, springel2005}). As the galaxies trace the underlying dark matter we therefore expect those to merge hierarchically also. However, it has been known for sometime that the most massive galaxies appear to have older stellar populations than their less massive counterparts \citep{cowie1996,bower2006,gilbank2010}. Environment also plays a key role with massive quiescent galaxies typically living in denser environments than lower mass star-forming galaxies \citep{dressler1980}. There are several ways to reconcile these observations with hierarchical merging which are implemented in phenomenological, semi-analytic models that seek to reproduce observations of galaxy evolution by populating dark matter halos from N-body simulations with mock galaxies (e.g. \citealt{bower2006,croton2006}). A reasonable match is achieved through interactions and feedback mechanisms that cease star formation in massive galaxies within massive dark matter halos, requiring that these galaxies build up their stellar mass at late times by so called `dry' mergers which trigger no significant new star formation due to the lack of available cold gas \citep{delucia2007}. 

In the high-redshift Universe the cold gas fraction in galaxies is higher than at low-redshift and thus there is more fuel for star formation \citep[e.g.][]{tacconi2010,geach2011}. It is therefore possible to more easily trigger significant star-forming events during mergers \citep{somerville2001} or through high gas accretion rates and disk instabilities in isolated galaxies \citep{keres2005,bower2006,dekel2009,forster2011,Cacciato2012}. The latter process leads to the intriguing possibility of the enhanced star-formation rates at high redshift being dominated by secular evolution rather than mergers. In fact while some observations suggest an increase in the merger fraction with redshift \citep{conselice2003} others seem to prefer in-situ galactic processes over galaxy-galaxy merging, or at least a mixture of these processes \citep{lotz2008,elbaz2007}

To test whether it is galaxy mergers or secular processes that dominate and drive galaxy evolution at the peak era for star formation, a method to distinguish between galaxy mergers and non-mergers needs to be implemented. The two main methods of estimating the merger fraction are counting close pairs of galaxies, under the assumption that they will subsequently merge (e.g. \citealt{lefevre2000,lin2008,bluck2009}), and using a method of identifying galaxies with a merging morphology (e.g. \citealt{conselice2003,lotz2004,conselice2008,lotz2008,conselice2009}). The results of both of these methods often suggest that the merger fraction increases with redshift and, depending on the mass range considered, the merger fraction at $z\gtrsim1$, where the star formation rate density peaks, is roughly $\sim0.1-0.3$ on average (but with some systematic offsets between studies) compared to a fraction $\lesssim0.1$ in the local Universe. A third, potentially more reliable, method is to employ detailed integrated field unit observations of $z=1-2$ galaxies to look for merger signatures in the dynamics of the galaxies. Such studies, although generally smaller in sample size, also find a merger fraction of $\sim0.3$ (e.g. \citealt{forster2009,shapiro2008}). 

In order to study the star-forming population, an excellent tracer of ongoing star formation is the $\rm H{\alpha}$ emission line which is less affected by dust obscuration than shorter wavelength star-formation tracers (e.g. UV continuum light or [OII]3727). Unfortunately beyond $z=0.4$, $\rm H{\alpha}$ is redshifted out of the optical window, thus high redshift studies of star formation have been limited to either using the obscuration-effected short wavelength tracers or studying small samples of $\rm H{\alpha}$ emitters using conventional near-infrared spectrographs. However, in the last few years panoramic narrow-band surveys have started to provide large samples of $\rm H{\alpha}$-selected galaxies (e.g. the High-redshift (Z) Emission Line Survey, HiZELS, \citealt{geach2008,geach2012,garn2010a,sobral2009,sobral2010a,sobral2012a,sobral2012b} and the studies of \citealt{villar2008} and \citealt{ly2011}). Narrow-band surveys provide a well understood, volume-selected sample of star-forming galaxies allowing for straight-forward analysis of trends with SFR, mass and size etc. They provide emission line information over large areas of the sky and are thus able to probe a significant range of the $\rm H{\alpha}$ luminosity and stellar mass functions for star-forming galaxies, required for an unbiased analysis of the star formation rate density (SFRD, e.g. \citealt{geach2008,sobral2009,sobral2012a,sobral2012b}). This selection method has also been shown to be extremely effective at detecting intrinsically faint galaxies, helping to overcome the bias towards massive galaxies associated with photometric redshift selection.

In this study we use the $z=0.4-2.23$ HiZELS sample presented in \cite{sobral2012b}, to not only analyse the merger rate as a function of redshift and stellar mass but also as a function of the well-determined SFR. We can therefore test whether it is major mergers that drive the rise to enhanced activity seen at these epochs. In contrast to earlier studies, which analyse {\it Hubble Space Telescope} ({\it HST}) rest frame UV morphologies, with the advent of the WFC3 camera we can also study the rest-frame optical bands for a subsample of our galaxies that lie within the CANDELS region of our survey and use this to calibrate morphologies derived from deep, wide-field, ground based near-infrared imaging, better matched to the extent of the full HiZELS fields. We also analyse the size--mass relation for star-forming galaxies over this epoch in order to study the size evolution which may also indicate the merger history of these systems.

The structure of this paper is as follows. In \S\ref{sec:sample} we describe the HiZELS narrow band sample and the imaging data. We then derive SFR for the sample and analyse the evolution of the number density of galaxies above a given SFR. The size--mass relation is then studied in order to look for an evolution.  A method for automating morphological classification is defined and this is used to study the merger rates of the galaxies in our sample and how they evolve and depend on SFR and mass. Finally, we discuss our findings in the context of understanding the physical processes that occur within galaxies, that lead to the rapid downturn in the global volume averaged SFR below $z\sim1$. 
 
A $\Lambda$CDM cosmology ($\Omega_{\rm m}=0.27$, $\Omega_{\Lambda}=0.73$, $H_{0}=70$ km\,s$^{-1}$ Mpc$^{-1}$) is used throughout this work and all magnitudes are AB.   
 
\section{The sample and data}
\label{sec:sample}
\subsection{The HiZELS survey}
HiZELS \citep{geach2008,sobral2012b} is a Campaign Project using the Wide Field CAMera (WFCAM, \citealt{casali2007}) on the United Kingdom Infra-Red Telescope (UKIRT) and exploits specially designed narrow-band filters in the $J$ and $H$ bands (NBJ and NBH), along with the H$_{2}$S1 filter in the $K$ band, to undertake panoramic, moderate depth surveys for line emitters. HiZELS targets the $\rm H{\alpha}$ emission line redshifted into the near-infrared at $z = 0.84, 1.47 \rm \,and \,2.23$ using these filters. In addition, the UKIRT data are complemented by deeper narrow band observations with Subaru Suprime-Cam NB921 imaging \citep{sobral2012a,sobral2012b} to obtain $\rm H{\alpha}$ emitting galaxies at $z=0.4$ and the [OII] emission from the $z=1.47$ $\rm H{\alpha}$ sample, as well as deeper WFCAM and Very Large Telescope near-infrared imaging through the H$_{2}$S1 filter in selected fields. The survey is designed to trace star-formation activity across the likely peak of SFR density and provide detailed information about a well-defined statistical sample of star-forming galaxies at each epoch (see \citealt{best2010}).  

In this study we concentrate on the main HiZELS sample of $z = 0.4, 0.84, 1.47 \rm \,and \,2.23$ $\rm H\alpha$ emitters in both the UKIRT Infrared Deep Sky Survey, Ultra Deep Survey (UKIDSS UDS, \citealt{lawrence2007}, Almaini et al. in prep.) and The Cosmic Evolution Survey (COSMOS, \citealt{cosmos2007}) fields as described in \cite{sobral2012b} and we refer the reader to that paper for full details of the catalogues used. These data cover areas of $0.6-1.6$ square degrees depending on the field and waveband. The narrow band excess sources are visually inspected to remove image artefacts and, to ensure the galaxies are at the desired redshift, spectral energy distribution (SED) fitting and optimised colour-colour selections are used to provide clean samples of $\rm H\alpha$ emitters in the four redshift slices \citep{sobral2012b}. The excess narrow-band flux is then converted into an emission line luminosity. For the analyses in this paper we take these cleaned catalogues and introduce cuts to ensure that the data in each narrow-band filter are complete to the same flux limit across the entire area observed. These final catalogues contain: 428 $\rm H\alpha$ emitters at $z=0.40$, 595 at $z=0.84$, 420 at $z=1.47$ and 372 at $z=2.23$ down to the SFR limits $\sim 0.2, 3.0, 12.0 \rm  \,and \,25.0 \,M_{\odot} yr^{-1}$ respectively (assuming $A_{\rm H\alpha}=1.0$), to an $\rm H\alpha$ equivalent width lower limit of 25\AA.

The star formation rates for the HiZELS sample are calculated from the $\rm H{\alpha}$ luminosity and the relation of \cite{kennicutt1998} (${\rm SFR ( M_{\odot}yr^{-1})}=7.9\times10^{-42}L(H\alpha) \rm(erg \,s^{-1})$), assuming a dust extinction $A_{H{\alpha}}=1$\,mag \citep[see][]{sobral2012b}. Stellar masses are computed by fitting SEDs to the rest-frame UV, optical and near-infrared data available ($FUV, NUV, U, B, g, V, R, i, I, z, Y, J, H, K$, $3.6\mu \rm m, 4.5\mu \rm m, 5.8\mu \rm m, 8.0\mu \rm m$ collated in \citealt{sobral2012b}, see references therein), following \cite{sobral2011} and the reader is referred to that paper for more details. The SED templates are generated with the \cite{bc2003} package using Charlot \& Bruzual (2007, unpublished) models, a \cite{chabrier2003} IMF, and an exponentially declining star formation history with the form $e^{-t/\tau}$, with $\tau$ in the range 0.1 Gyrs to 10 Gyrs. The SEDs were generated for a logarithmic grid of 200 ages (from 0.1 Myr to the maximum age at each redshift being studied). Dust extinction was applied to the templates using the \cite{calzetti2000} law with $E(B-V)$ in the range 0 to 0.5 (in steps of 0.05), roughly corresponding to A$_{\rm H\alpha}\sim0-2$. The models are generated with different metallicities, including solar; the reader is referred to \cite{sobral2011} for further details. For each source, the stellar mass is computed as the median of stellar masses of the $1\,\sigma$ best-fits over the range of parameters.

\begin{figure*}
   \centering
\includegraphics[scale=0.5]{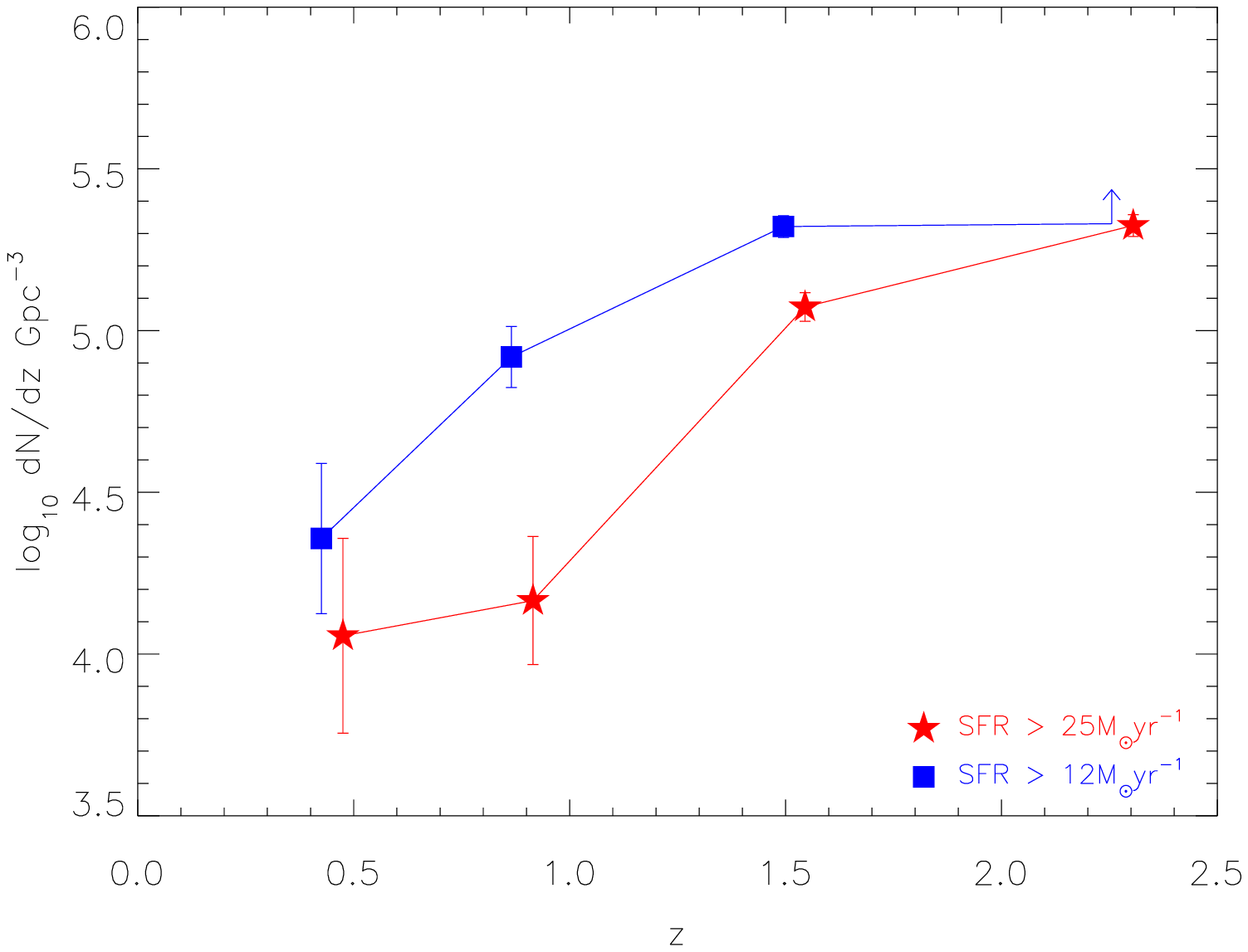}\includegraphics[scale=0.5]{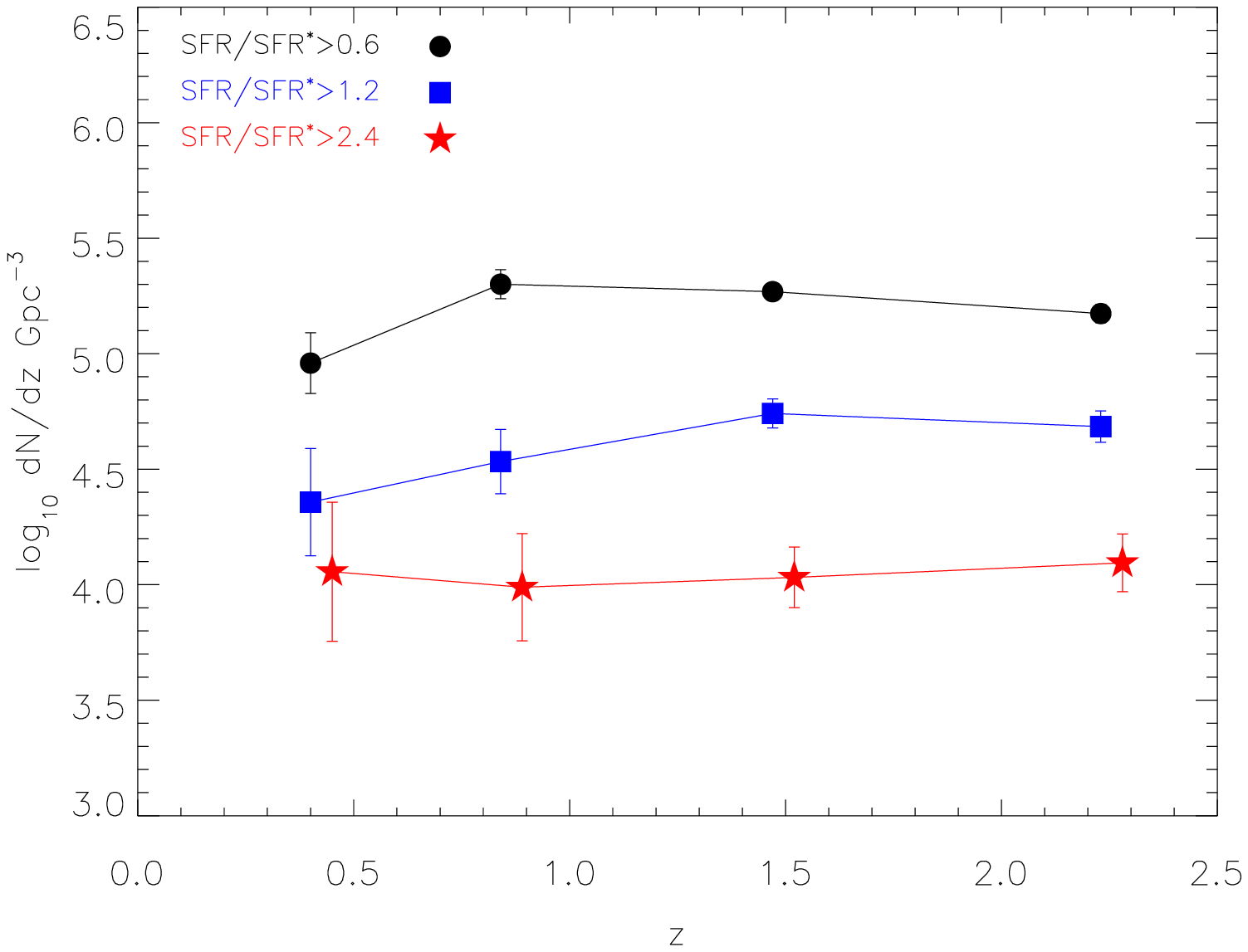}  
\caption[]{{\it Left}: The number density of HiZELS galaxies above a stellar mass of $\rm 10^{10}M_{\odot}$ and a given SFR plotted against redshift. The SFR$\,>25\, \rm M_{\odot} yr^{-1}$ lines are offset slightly in $z$ for clarity. {\it Right}: The number density of $\rm >10^{10}M_{\odot}$ galaxies above an epoch normalised star formation (ENSFR) threshold. We define ENSFR as the ratio of SFR to SFR$^{\star}(z)$ (with SFR$^{\star}(z)$derived from the $L_{\rm H\alpha}^{\star}$ ie the typical SFR from the $\rm H\alpha$ luminosity function at that redshift, \citealt{sobral2012b}). In this way we remove the trend that the average sSFR of galaxies increases with redshift. As there is no evidence of a significant trend this demonstrates that the number density of typical star-forming galaxies does not evolve significantly with redshift and thus the increase in the SFRD is purely an effect of increased typical sSFR.  }
   \label{fig:sfrdown}
\end{figure*}

In Figure \ref{fig:sfrdown} ({\it left}) we plot the number density of galaxies, from the combined UDS and COSMOS fields, above a stellar mass of $\rm 10^{10}M_{\odot}$ and a given SFR, against redshift. From this we can see that for a given SFR limit the number density increases rapidly with redshift. This is a manifestation of the fact that a typical star-forming galaxy has a greater sSFR at higher redshift, forming stars more rapidly for a given mass. In order to look for trends with redshift we therefore define a quantity which we term the epoch-normalised star formation rate ENSFR which is the SFR of a galaxy divided by the SFR$^{\star}(z)$. SFR$^{\star}(z)$ is the star-formation rate derived from the quantity $L_{\rm H\alpha}^{\star}$ found by fitting a Schechter function to the $\rm H\alpha$ luminosity function at a given redshift, which we take from \cite{sobral2012b}. We note that normalising the SFR to SFR$^{\star}(z)$ accounts, to first order, for the increase in sSFR with redshift. However, significant evolution in either the slope of the SFR -- stellar mass relation or the dust obscuration would invalidate this.

The values of SFR$^{\star}$ essentially double for each HiZELS redshift interval considered with SFR$^{\star}\sim7.0, 14.0, 29.0, \rm \,and \,57.0 \,M_{\odot}\,yr^{-1}$ for $z = 0.4, 0.84, 1.47 \rm \,and \,2.23$ respectively. Interestingly, this same behaviour is seen in the evolution of the typical sSFR from \cite{elbaz2011} with $\rm sSFR\sim0.2, 0.4, 0.8, \rm \,and\, 2.0 \,yr^{-1}$,  again at these redshifts. We suggest that this is because the $\rm H\alpha$ luminosity (and thus SFR) function evolves significantly more than the stellar mass function.

In Figure \ref{fig:sfrdown} ({\it right}) we plot the number density of galaxies of a given mass above the thresholds SFR/SFR$^{\star}(z)=0.6, 1.2, 2.4$. From this plot one can clearly see that the number of star-forming galaxies with their SFR normalised to the typical SFR at that epoch is broadly constant. This means that the number density of star-forming galaxies of a given mass and ENSFR does not evolve significantly over the period studied here. This demonstrates that the star-forming population is constant with redshift but simply evolves in sSFR. This is similar to the result found in \cite{sobral2012b} in which there is no strong evolution in the Schechter parameterisation of the normalisation of the H$\alpha$ luminosity function, $\phi^{\star}_{\rm H{\alpha}}$. We discuss the implications of this in \S\ref{sec:disc}.

\subsection{Imaging data}
In this study we analyse near-infrared imaging from the space-based {\it HST}/WFC3 Cosmic Assembly Near-infrared Deep Extragalactic Legacy Survey (CANDELS, \citealt{candels2011,candels2}) and the ground-based UKIDSS UDS and the WIRCam Deep Survey (WIRDS, \citealt{bielby2011}). 

The CANDELS imaging we use is from WFC3 F160W covering a 2-orbit depth over 720 sq. arcmin of the UDS. The CANDELS imaging has a pixel scale of 0.06 arcsec and a point spread function (PSF) with a FWHM of $0.18''$. The CANDELS data are well suited to this project for which we require high resolution imaging in the rest-frame optical, however to obtain the wider area coverage needed to build up a statistical sample of rarer high-mass systems from HiZELS we also need to use ground-based near-infrared imaging.

The UKIDSS UDS $K$-band imaging covers an area of 0.8 square degrees, to a depth of $K=24.6$ ($5\sigma$, AB) with a pixel scale of 0.13 arcsec and a PSF FWHM of $0.7''$. The WIRDS $K$-band imaging covers a total effective area of 2.1 square degrees and reaches an AB 50\% completeness limits of $\sim24.5$ across the COSMOS field, it has a pixel scale of 0.15 arcsec and a PSF FWHM of $0.7''$ and is thus comparable to the UKIDSS UDS. 

The combination of these three near-infrared imaging datasets allows us to probe the rest-frame optical morphologies and sizes of the HiZELS galaxies over a wide range in luminosity while at the same time providing a rest-frame optical view of the galaxies' stellar distribution.

\section{Analysis}

\subsection{Sizes}
\label{sec:size}
Before studying the morphologies and the merger rates of the galaxies in the HiZELS sample, we first assess their typical sizes. This is interesting from a galaxy evolution perspective, as an increase in size with cosmic time may imply that mass is being built up either through mergers or accretion or that the mass is being redistributed somehow. If there is no direct evolutionary connection between the galaxy populations at each epoch then changes in typical size may suggest differing formation scenarios. Importantly, it will also help us to understand the reliability of the morphological classification as the smallest galaxies will be most  affected by the resolution of our ground-based imaging.  

The surface photometry of galaxies is often described by a S\'{e}rsic profile \citep{sersic1968}.

\begin{equation}
I(r)=I_{e}\,{\rm exp}\displaystyle \left\{-b_{n}\left[\left(\frac{r}{r_{e}}\right)^{1/n} -1\right]\right \},
\end{equation}

where $I(r)$ is the intensity, $r$ is the radius from the centre of the galaxy, $r_{e}$ is the scale radius, $I_{e}$ is the intensity at $r_{e}$, $n$ in the exponent is a free parameter widely known as the S\'{e}rsic index and $b_{n}=2n - 0.327$; a coefficient chosen so that $r_{e}$ is the half-light radius defined as the radius which encircles half the light from the galaxy (e.g. \citealt{graham1996}).

\begin{figure}
   \centering
\includegraphics[scale=0.5, trim=0 10 0 10, clip=true]{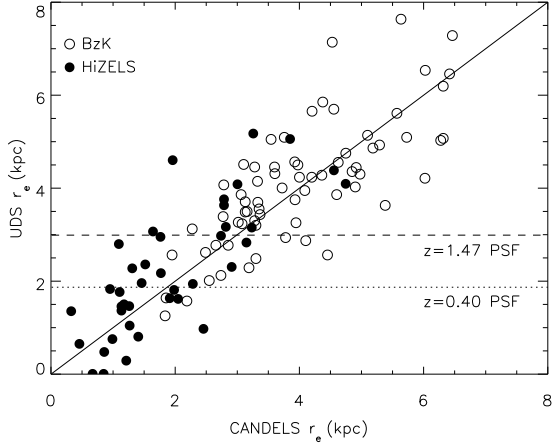} 
\caption[]{The half-light radius measured from the UKIDSS UDS ground based imaging plotted against that from the {\it HST}/WFC3 CANDELS data at all redshifts. Solid line is the 1-to-1 line. The open and filled circles represent $BzK$ and HiZELS galaxies respectively. The dashed and dotted lines represent the UKIDSS UDS PSF HWHM at $z=1.47$ and $z=0.4$ respectively, which bracket the other two epochs. This demonstrates that we can recover the sizes of galaxies by accounting for the ground-based PSF using {\sc galfit} \citep{peng2002}.}
   \label{fig:sizetest}
\end{figure}

To measure the sizes of the galaxies we fit a 2-dimensional S\'{e}rsic profile to the galaxy images using the {\sc galfit} (version 3) software package \citep{peng2002}. This software requires reasonable initial input parameters such as position, apparent magnitude and ellipticity, all of which are estimated by first running the {\sc sextractor} package \citep{bertin1996} so that the iterative fitting process converges to the correct solution in the shortest possible time. {\sc galfit} deconvolves the point spread function which is dominated either by the telescope itself, in the case of {\it HST}, or by the atmospheric seeing for the ground-based imaging. To this end we check that the effect of seeing has been correctly accounted for in the analysis of the ground-based imaging by comparing the CANDELS derived sizes to those from the UKIDSS UDS imaging for the same galaxies. Figure \ref{fig:sizetest} shows this comparison of galaxy sizes for a sample drawn from a combination of all four HiZELS redshift slices and a sample of $BzK$ \citep{daddi2004} galaxies in the UDS field (the photometry to select $BzK$ galaxies is taken from the UDS catalogues, Almaini et al., in prep). These two independent size measurements are correlated and scattered around the 1-to-1 line with $\Delta r_{e}/r_{e}\sim0.4$, which confirms that the sizes recovered are comparable, demonstrating that {\sc galfit} is able to successfully account for the seeing. 

We note that there may be some selection effects and biases in size measurements, in that galaxies with large-sizes can be missed due to low surface brightnesses and compact galaxies may have sizes overestimated \citep{barden2005}. The former is less likely as the HiZELS galaxies are selected on their $\rm H\alpha$ emission. Also, Figure \ref{fig:sizetest} demonstrates that there is no significant bias in size estimates between the ground and space-based analysis of the smallest galaxies so we take this as evidence that their sizes are not overestimated.

\begin{figure}
   \centering
\includegraphics[scale=0.43, trim=0 62 0 0, clip=true]{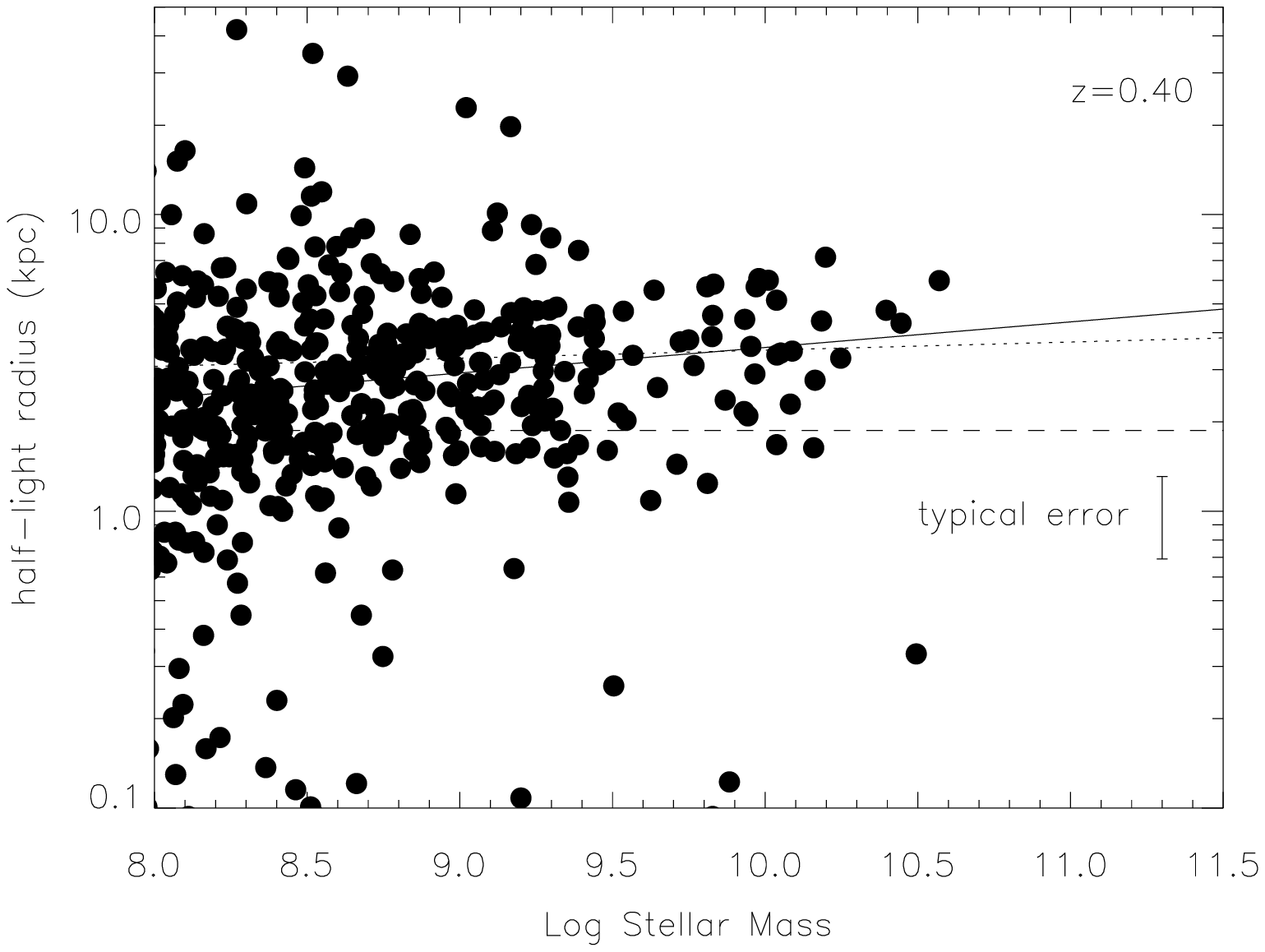} 
\includegraphics[scale=0.43,  trim=0 62 0 31, clip=true]{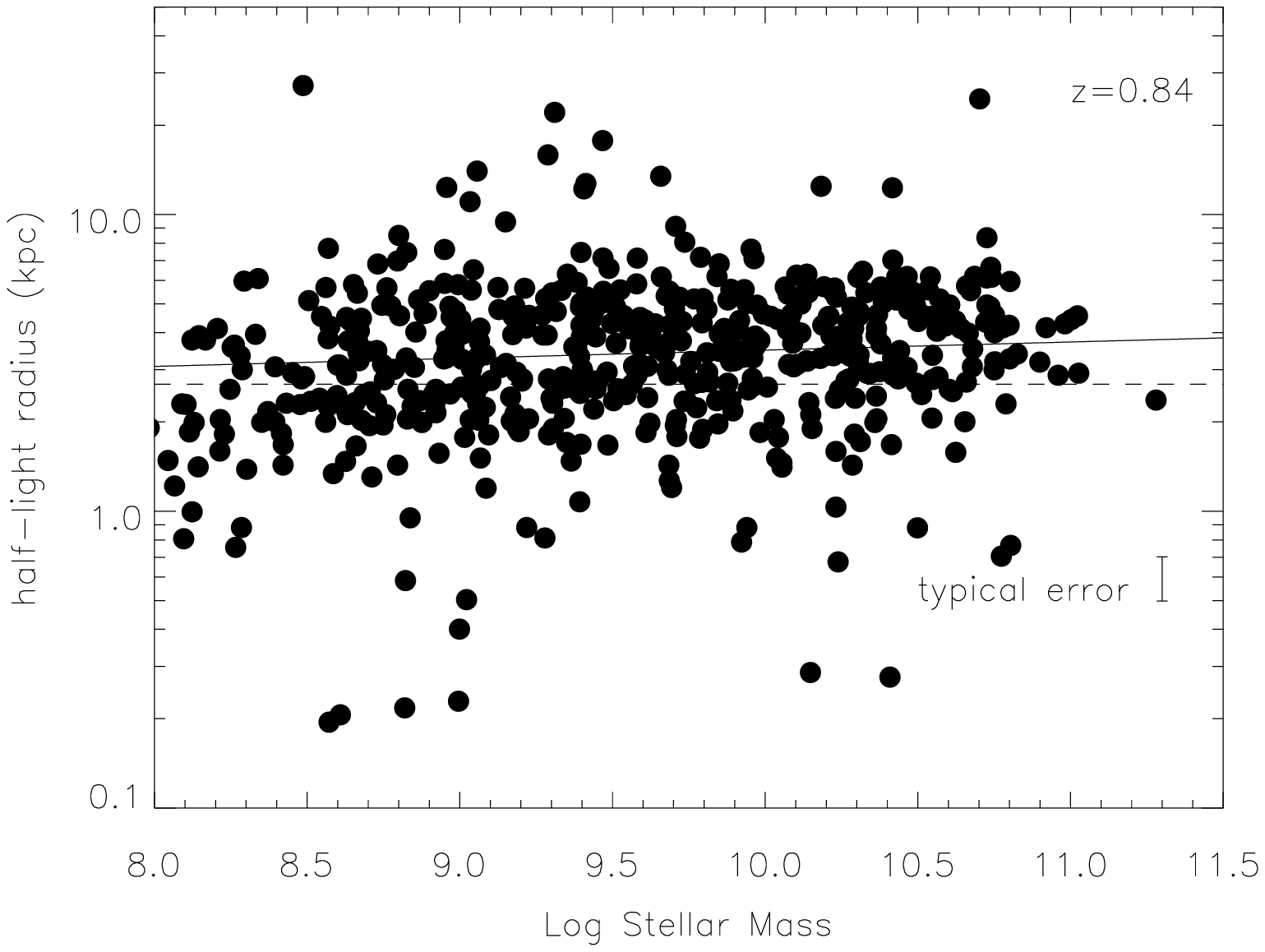} 
\includegraphics[scale=0.43,  trim=0 62 0 31, clip=true]{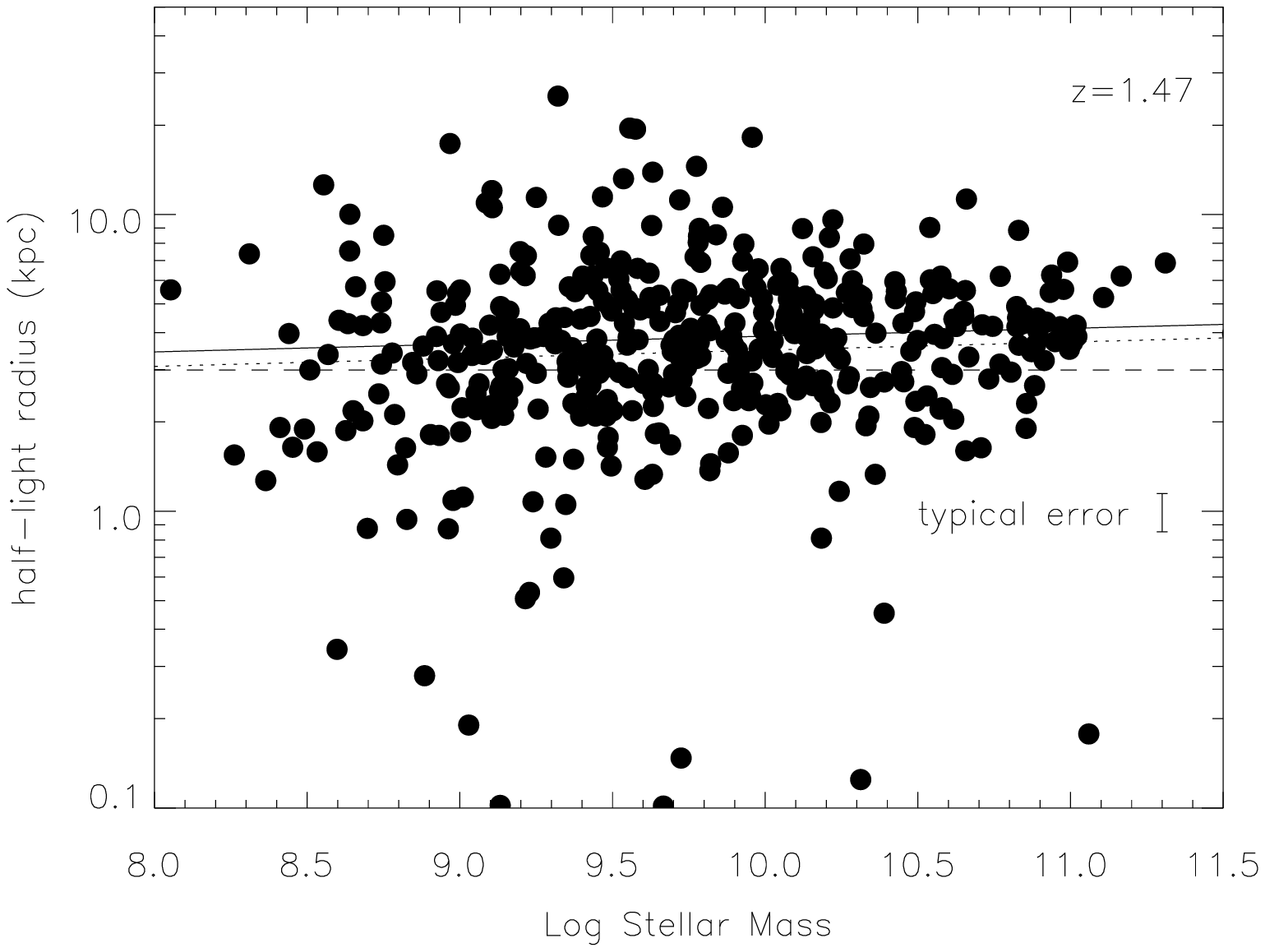} 
\includegraphics[scale=0.43, trim=0 0 0 31, clip=true]{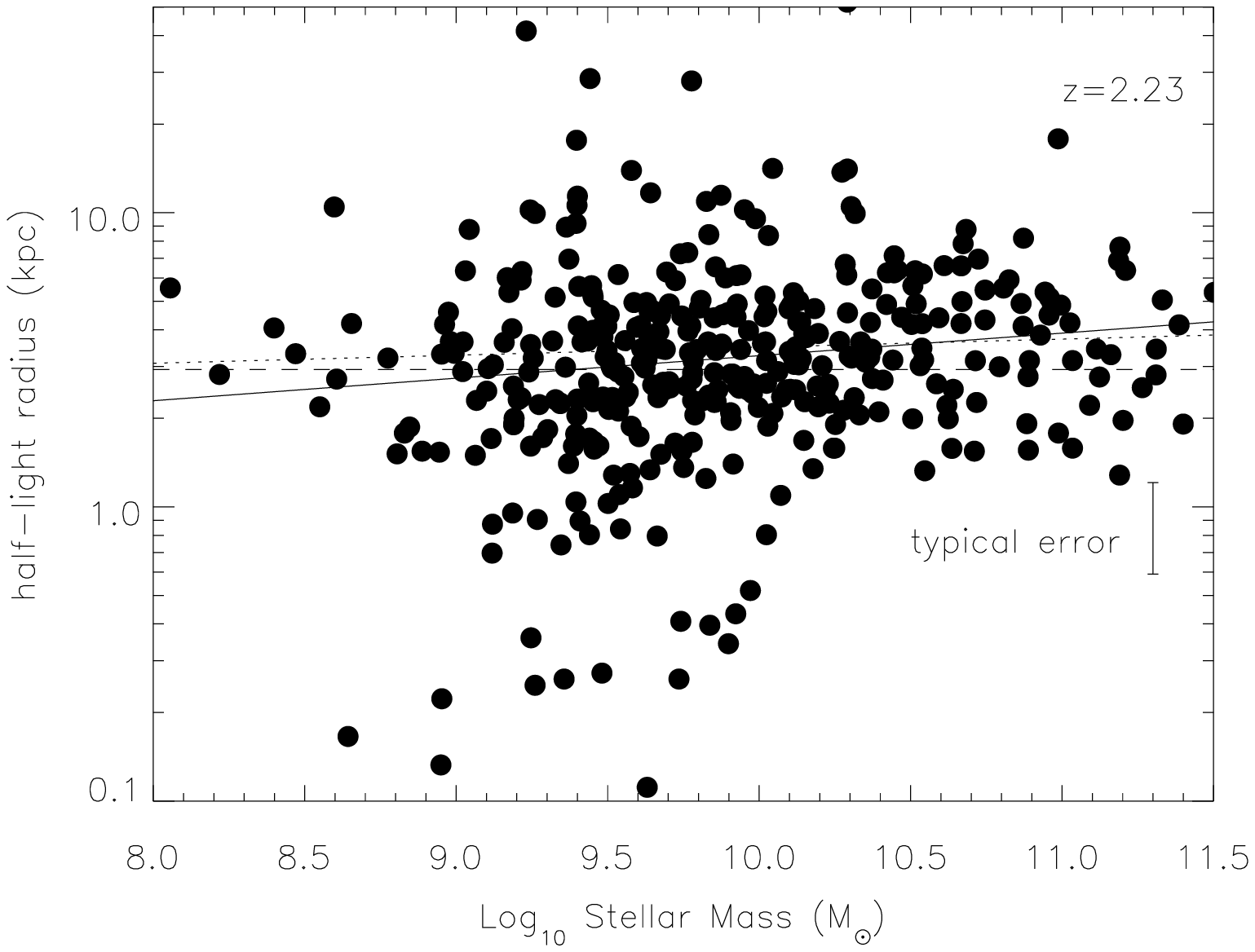} 
\caption[]{The half-light radius plotted against stellar mass for the $z=0.4$ (top), $z=0.84$ (upper middle), $z=1.47$ (lower middle) and $z=2.23$ (lower). The solid lines are linear fits to the relations with the dotted line the $z=0.84$ fit for reference. The dashed lines represent the PSF HWHM. The slope of the size--mass relation is found to be broadly constant.}
   \label{fig:sizemass}
\end{figure}

Figure \ref{fig:sizemass} shows the size--mass relations at each redshift slice. We perform linear fits to this relation of the form $\log_{10}\,r_{e}=a\,(\log_{10}\,(M_{\star}) - 10) + b$, where $r_{e}$ and $M_{\star}$ are in units of kpc and $\rm M_{\odot}$ respectively and we normalise the fits to $M_{\star}=10^{10} \rm M_{\odot}$. Table \ref{tab:size} contains the results of these fits at the four redshift slices considered. From these fits we find the surprising result that the typical size of a star-forming galaxy with $\log_{10}\,M_{\star}=10$ does not evolve significantly out to $z=2.23$, with $r_{e}=3.6\pm0.3\,\rm kpc$ on average. These results are in good agreement with the trends of \cite{barden2005,ichikawa2012} who also find little evidence of an evolution in this relation or the typical size of star forming galaxies. In a related study, \cite{kanwar2008} find no evolution in the shape of the size function of disc galaxies between $0.1<z<1.0$ with just an evolution in the number density of discs. However, other groups have found evidence for a stronger size evolution for the most massive ($M_{\star}>10^{10} \rm M_{\odot}$) disc-like galaxies, with a $2-4$ fold increase in size since $z\sim2$ \citep{trujillo2007,mosleh2011}.

By analysing the S\'{e}rsic index, $n$, which we obtain from the fitting process we divide our sample into disc-like and bulge-like galaxies where we define the former as having $0.5\leq n<2.5$ and the latter as $2.5\leq n<5.0$. From this we find that the fraction of disc-like galaxies is $>80\%$ in each redshift slice with no evidence for an evolution, which is not unexpected as star-forming galaxies such as those selected by HiZELS are in general found to be discs, consistent with \cite{sobral2009}. We note that this disc fraction also has no trend with SFR or stellar mass.

\begin{table}
\begin{center}
\caption[]{The size--mass relations at each redshift slice, of the form $\log_{10}\,r_{e}=a\,(\log_{10}\,(M_{\star}) - 10)+b$. Where $r_{e}$ and $M_{\star}$ are in units of kpc and $\rm M_{\odot}$ respectively.}

\label{tab:size}
\small\begin{tabular}{cccc}
\hline
$z$ & $a$  & $b$ & $r_{e}$ at $\log_{10}\,(M_{\star})=10$ \\
&&&(kpc)\\
\hline
0.40&0.08$\pm$0.02&0.55$\pm$0.03&3.6$\pm$0.2 \\
0.84&0.03$\pm$0.02&0.54$\pm$0.01&3.5$\pm$0.1 \\
1.47&0.03$\pm$0.02&0.59$\pm$0.01&3.9$\pm$0.2 \\
2.23&0.08$\pm$0.03&0.51$\pm$0.02&3.3$\pm$0.2 \\
\hline
\end{tabular}
\end{center}
\end{table}

\subsection{Morphologies}
\label{sec:morph}
\subsubsection{Quantifying and calibrating morphology}
\label{sec:morphcal}
To quantify the morphologies of the galaxies in this study we choose to use a combination of Gini and $\rm M_{20}$ coefficients first proposed by \cite{lotz2004}. The Gini coefficient, developed by statistician Corrado Gini, measures the inequality among values of a frequency distribution. It was first applied to studies of galaxy morphology by \cite{abraham2003}. A Gini coefficient of zero expresses an equality where all values are the same (i.e. a galaxy with uniform surface brightness). A Gini coefficient of 1 expresses maximal inequality among values (i.e. where all of the flux is in one pixel/element). The $\rm M_{20}$ coefficient, describes the second-order moment of the brightest 20\% of pixels in the galaxy and is sensitive to merger signatures such as multiple nuclei \citep{lotz2004}. The combination of Gini and $\rm M_{20}$ can differentiate between `normal' star forming galaxies and Ultra Luminous Infrared Galaxies (ULIRGs), as well as single galaxies and merging systems. However, there are some differences in the boundaries chosen to delineate these populations (e.g. see \citealt{lotz2006,lotz2008}) and therefore we choose to perform our own tests and calibrate the Gini and $\rm M_{20}$ coefficients by visual inspection.

The Gini and $\rm M_{20}$ coefficients are calculated using the Gini and $\rm M_{20}$ components of the galVSM software \citep{huertas2008}. This software requires a segmentation map which tells galVSM which pixels are associated with the galaxy. We first cutout 10$''$ postage stamp images, taken from the CANDELS mosaic, around each galaxy and generate a segmentation map using {\sc sextractor} \citep{bertin1996}. The Gini and $\rm M_{20}$ codes are then run on the postage stamps and the corresponding segmentation maps.

The sample we choose to run the initial visual inspection calibration analysis on is that of 167 star forming galaxies in the redshift range $1.4<z<2.5$ selected via the $BzK$ method \citep{daddi2004} which lie within the CANDELS survey region in the UDS. The F160W mosaic provides high resolution rest frame optical imaging of these galaxies. We choose this sample over the HiZELS narrow-band sample as it should consist of similar star forming galaxies but has a higher surface density and so a larger sample falls within the high resolution CANDELS imaging, key to testing the morphological classifications.

From visual inspection of the $BzK$ galaxy morphologies, the {\sc sextractor} parameters {\sc deblend\_mincont}=0.1 and {\sc detect\_minarea}=5 and {\sc detect\_thresh}=1 $\sigma$ are found to be relaxed enough to associate clear merging components of the same `galaxy' with one segmentation map but still stringent enough so as to not produce clear false positives. We note that having a {\sc deblend\_mincont} set too high means that unrelated galaxies would be considered as mergers whereas when set to a low value separate features within the same galaxy separate into distinct objects and therefore this parameter has the most effect on the $\rm M_{20}$ coefficient (see Appendix \ref{sec:sim} for a discussion of this parameter). Setting the detection threshold to low sigma values includes fainter `sky' pixels in the segmentation map and thus increases inequality, raising the Gini coefficient. In this way one can see that the way in which the segmentation map is created is the most important factor in determining the Gini and $\rm M_{20}$ coefficients and differences between how this is done in different studies are the reason why we choose to calibrate our own definitions of mergers and non-mergers. 

\begin{figure}
   \centering
\includegraphics[scale=0.45, trim=0 62 0 0, clip=true]{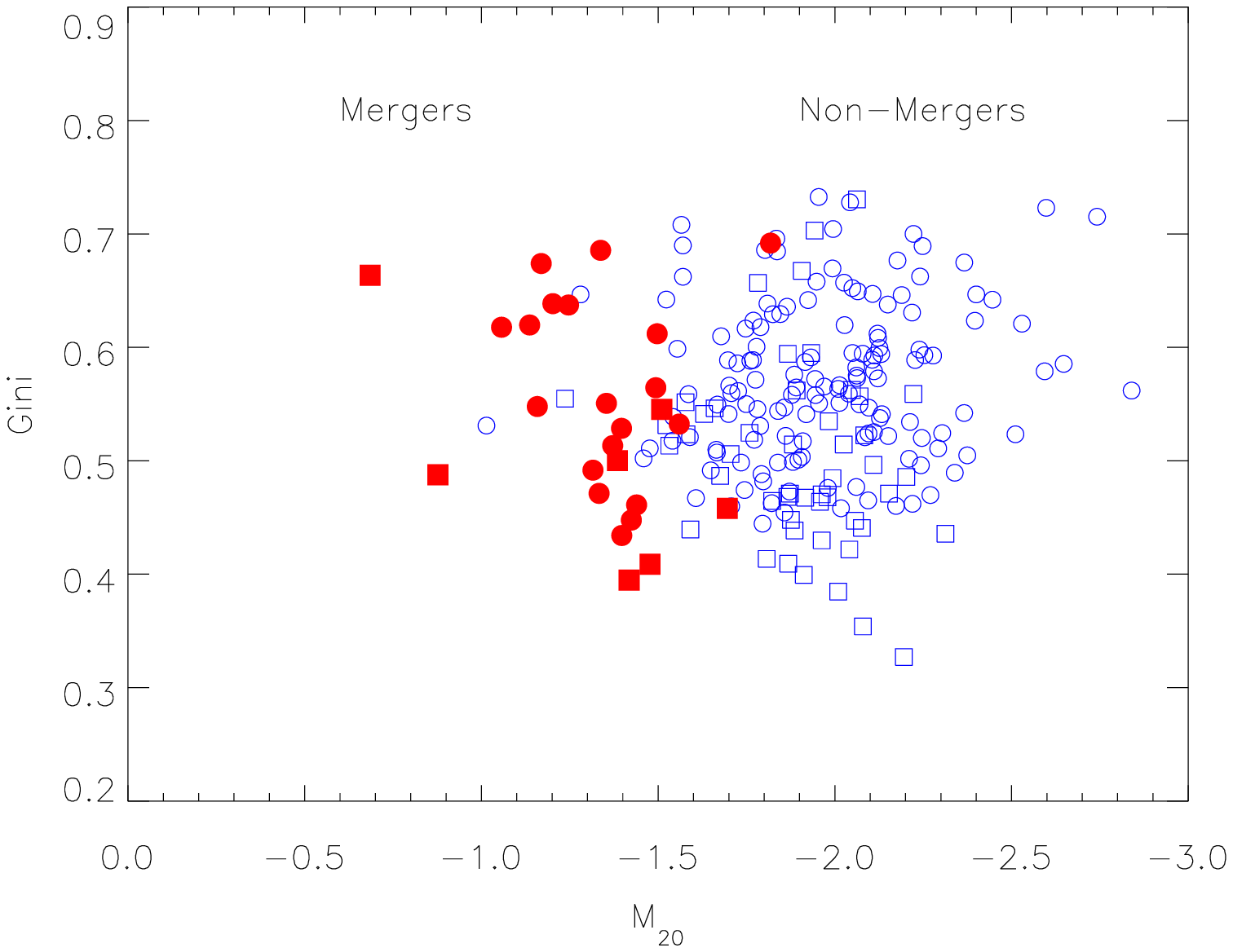}
\includegraphics[scale=0.45, trim=0 0 0 31, clip=true]{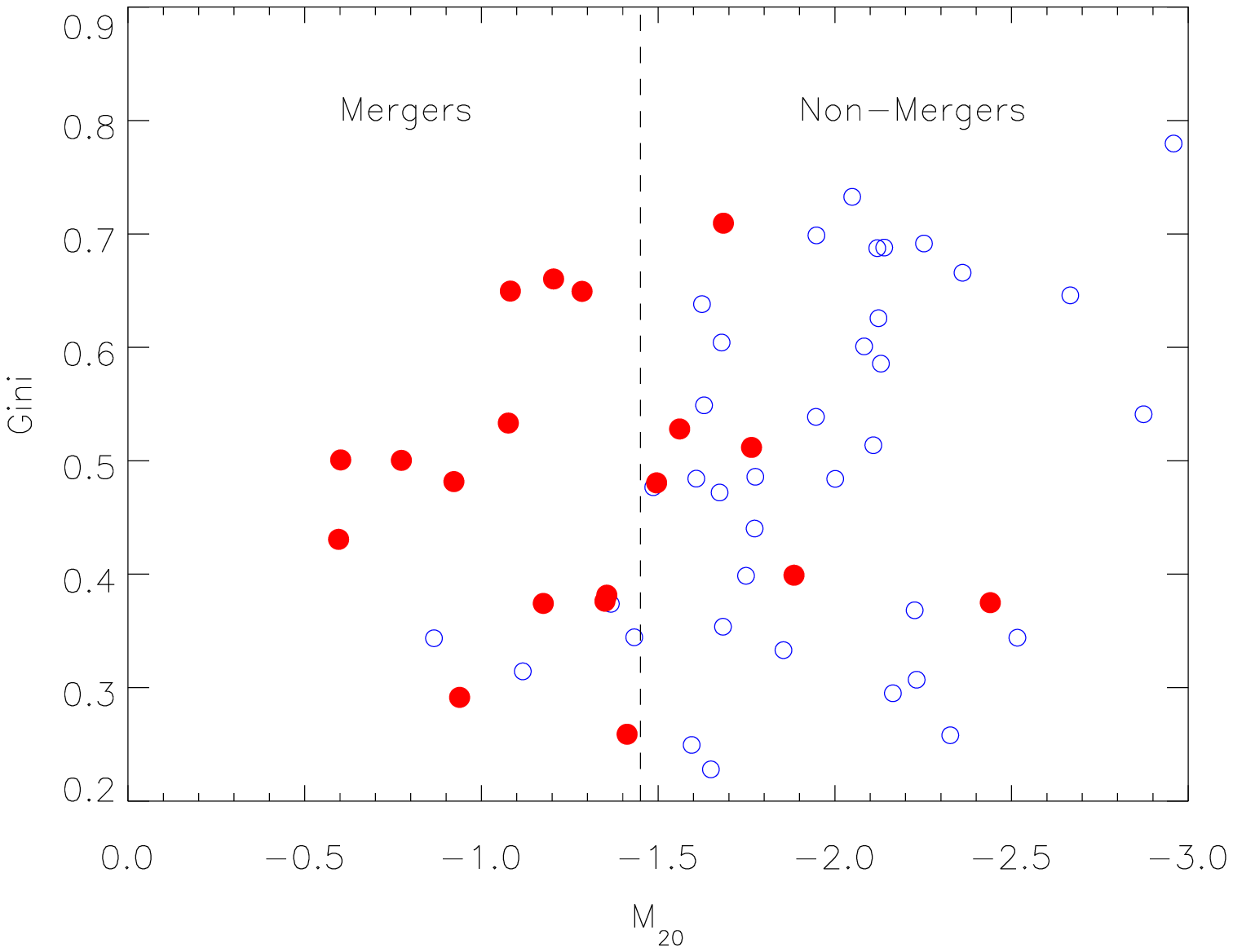}  
\caption[]{{\it Upper}: The Gini coefficient plotted against the $\rm M_{20}$ value for the $z\sim1.4-2.5$ $BzK$ population and a photometric redshift sample with $z\sim0.4$ from the F160W CANDELS imaging data in the UDS field. The filled red and open blue symbols are those classified as mergers and non-mergers respectively by visual inspection of the CANDELS imaging with circles representing mergers and non-mergers for the $BzK$ population and squares for the $z\sim0.4$. From visual inspection $\rm M_{20}\sim-1.5$ appears to be an excellent delineation between mergers and non-mergers. This demonstrates that the for our particular analysis the key parameter for determining whether a galaxy has a merger-like morphology is the $\rm M_{20}$ parameter and not the Gini coefficient. {\it Lower}: The Gini coefficient plotted against the $\rm M_{20}$ coefficient for HiZELS galaxies at all redshifts, as measured from the UDS K band imaging and calibrated using equations \ref{eq:gini} and \ref{eq:m20} but with morphologies visually identified from the CANDELS F160W image. The filled red and open blue circles are those visually classified as mergers and non-mergers respectively. The vertical line at $\rm M_{20}=-1.45$ is the value we now choose from visual inspection to delineate the mergers and non-mergers. This demonstrates that the calibrated ground-based near-infrared imaging can be used to derive $\rm M_{20}$ values that differentiate between mergers and non-mergers.}
   \label{fig:bzkginim}
\end{figure}

Using the above method, fixing the {\sc sextractor} parameters to those found to give the best performance, the Gini and $\rm M_{20}$ codes are run on the CANDELS imaging with the results for the $BzK$ sample are displayed in Figure \ref{fig:bzkginim} ({\it upper}). Also included is a $0.35<z<0.45$ photometric redshift sample with a similar magnitude range to the HiZELS $z=0.40$ sample sourced from \cite{williams2009} to demonstrate that this classification technique is not affected by redshift. 

By visually assigning the galaxies into two categories `mergers' and `non-mergers' with the former classification based on evidence of merging components either creating disturbed morphologies or very close potential mergers (on-sky separation $\lesssim2''$). This information is included in Figure \ref{fig:bzkginim}, with the delineation between mergers and non-mergers found to occur at an $\rm M_{20}\sim-1.5$ for both high and low redshift regimes and thus the Gini coefficient does not seem to add any information. Using this method there is a contamination of $\sim10\%$ non-mergers in the mergers and $<5\%$ mergers in the non-mergers. The simulations performed for Appendix \ref{sec:sim} demonstrate that the $\rm M_{20}$ coefficient is sensitive to merging components down to a luminosity (mass) ratio of $\sim1:10$  (in agreement with the simulations of  \cite{lotz2010}). As such we note that our analysis throughout this paper is a measure of major mergers only.

\begin{figure*}
   \centering
\includegraphics[scale=0.45, trim=0 0 0 0, clip=true]{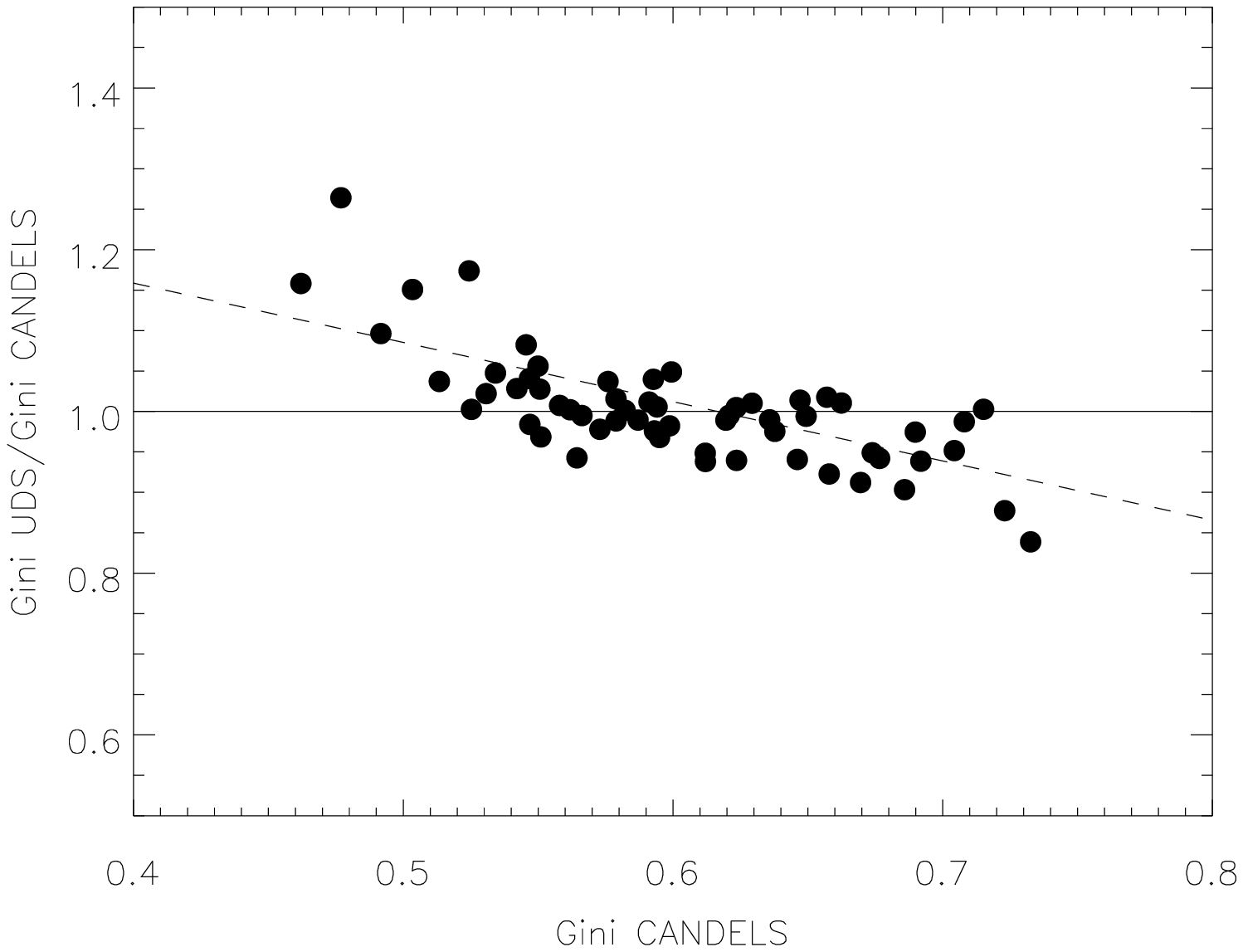} \includegraphics[scale=0.45, trim=0 0 0 0, clip=true]{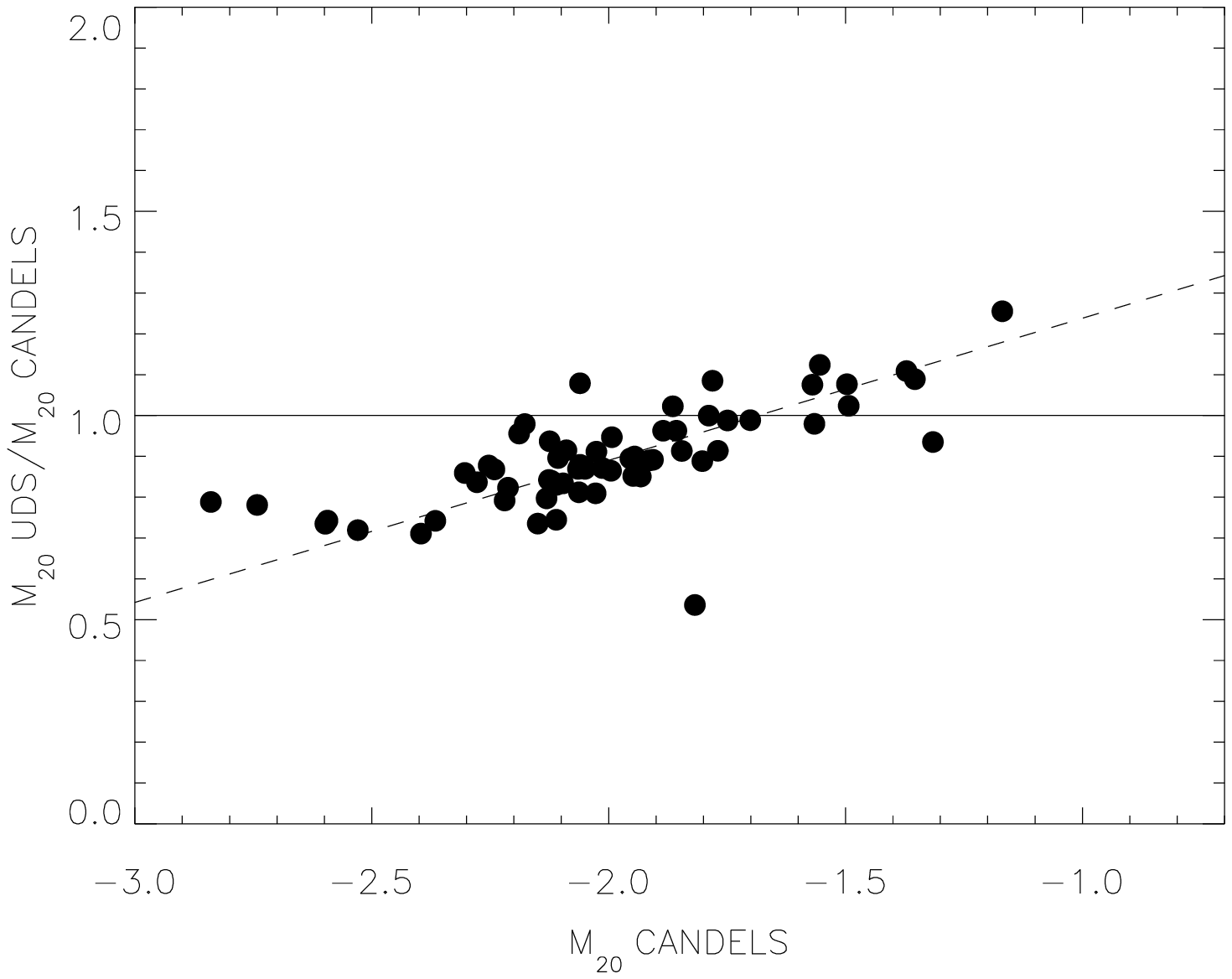}
\caption[]{{\it Left}: The ratio of the Gini coefficient for $BzK$ galaxies measured from the UKIDSS UDS ground based $K$-band imaging to that measured from the CANDELS {\it HST} F160W imaging plotted against the Gini coefficient measured from the CANDELS F160W imaging. The solid line is the 1-to-1 relation and the dashed line is a fit to the observed trend. {\it Right}: The ratio of the $\rm M_{20}$ coefficient for $BzK$ galaxies measured from the UKIDSS UDS ground based $K$-band imaging to to that measured from the CANDELS F160W imaging plotted against the $\rm M_{20}$ coefficient measured from the CANDELS F160W imaging. The solid line is the 1-to-1 locuss and the dashed line is a fit to the relation. From these plots we can see that it is possible to calibrate the values of Gini and $\rm M_{20}$ derived from ground-based imaging to those from {\it HST} imaging.}
   \label{fig:ginimatch}
\end{figure*}

The morphology codes are then run on the same galaxies but using the deep $K$-band ground-based UKIDSS UDS so we can compare the two independent measurements. We expect the higher resolution CANDELS imaging to be a truer reflection of a galaxy's intrinsic morphology. We also note here that it is difficult to measure the morphologies of the lowest luminosity galaxies in our sample as they tend to be smaller (see size--mass relations in \S\ref{sec:size}) and are thus more affected by the seeing of the ground-based near-infrared imaging. By performing tests we find that setting {\sc deblend\_mincont}=0.03 ensures that the $\rm M_{20}$ parameter selects the same type of mergers in the ground-based data as that derived from the {\it HST} data (again see Appendix \ref{sec:sim}). The ground-based versus {\it HST} Gini and $\rm M_{20}$ values are plotted in Figure \ref{fig:ginimatch}. By performing linear fits to these relations we can calibrate the ground-based Gini and $\rm M_{20}$ values to those derived from {\it HST}. These fits are: 

\begin{equation}
\label{eq:gini}
\rm Gini_{\,CANDELS}=0.78\,Gini_{\,UDS} +0.13
\end{equation}

\begin{equation}
\label{eq:m20}
\rm M_{20, CANDELS}=0.68\,M_{20, UDS} -0.39
\end{equation}

and will now be applied to the HiZELS morphologies derived from the ground-based near-infrared imaging. 

One potential problem with measuring the morphologies of galaxies at different epochs, using the same near-infrared imaging, is that of morphological $k$ correction. Galaxies look smoother at longer wavelengths, meaning that the lowest redshift galaxies in our sample may artificially appear less disturbed than those at high redshift. When we analyse the {\it HST} Advanced Camera for Surveys (ACS) F814W imaging data available in COSMOS, many of the galaxies are very low surface brightness and therefore it is difficult to assess whether the morphological classifications given by the $\rm M_{20}$ coefficient are reliable. However, for the galaxies in the $z=0.4$ sample with $K_{AB}<22.5$, the same classifications as those derived from  the near-infrared CANDELS imaging are recovered in $\sim90\%$ of the cases, so we conclude that our results are not significantly affected by this.

An additional concern is that some disc galaxies at high redshift are found to contain large star-forming clumps (e.g. \citealt{swinbank2010a, swinbank2012}). There may therefore be a degeneracy between what we classify as `mergers' and those galaxies that contain a small number of large star-forming clumps. It is practically impossible to differentiate between these two populations without dynamical information and thus we note with caution that so called `clumpy disc' galaxies may make a up some fraction of our `merger' sample, if the clumps are on scales of $\gtrsim4\,\rm kpc$. In fact when we run the sub-sample of nine HiZELS galaxies which, from dynamical analysis of integrated field unit data, are all found to contain clumps (see \citealt{swinbank2012} for a description of this sample) all of them have $\rm M_{20}\gtrsim-1.5$ and thus we would classify them as `mergers'. We note here that when visually classified not all of these clumpy galaxies appear as clear mergers which may explain the non-merger interlopers with $\rm M_{20}\gtrsim-1.5$ in Figure \ref{fig:bzkginim}.

\subsubsection{HiZELS morphologies}

\begin{figure*}
   \centering
%
\includegraphics[scale=0.14, trim=77 0 77 0, clip=true]{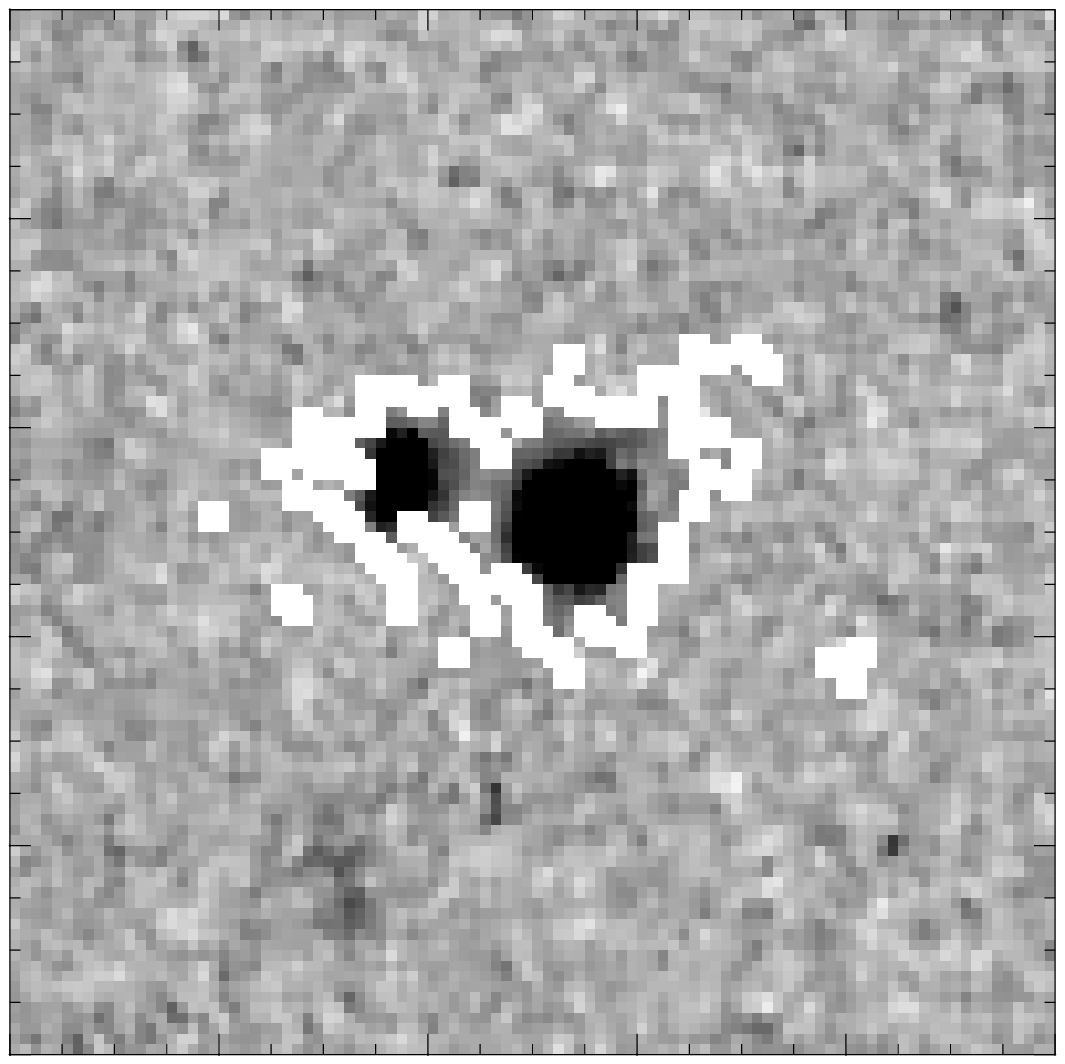} \includegraphics[scale=0.14, trim=77 0 77 0, clip=true]{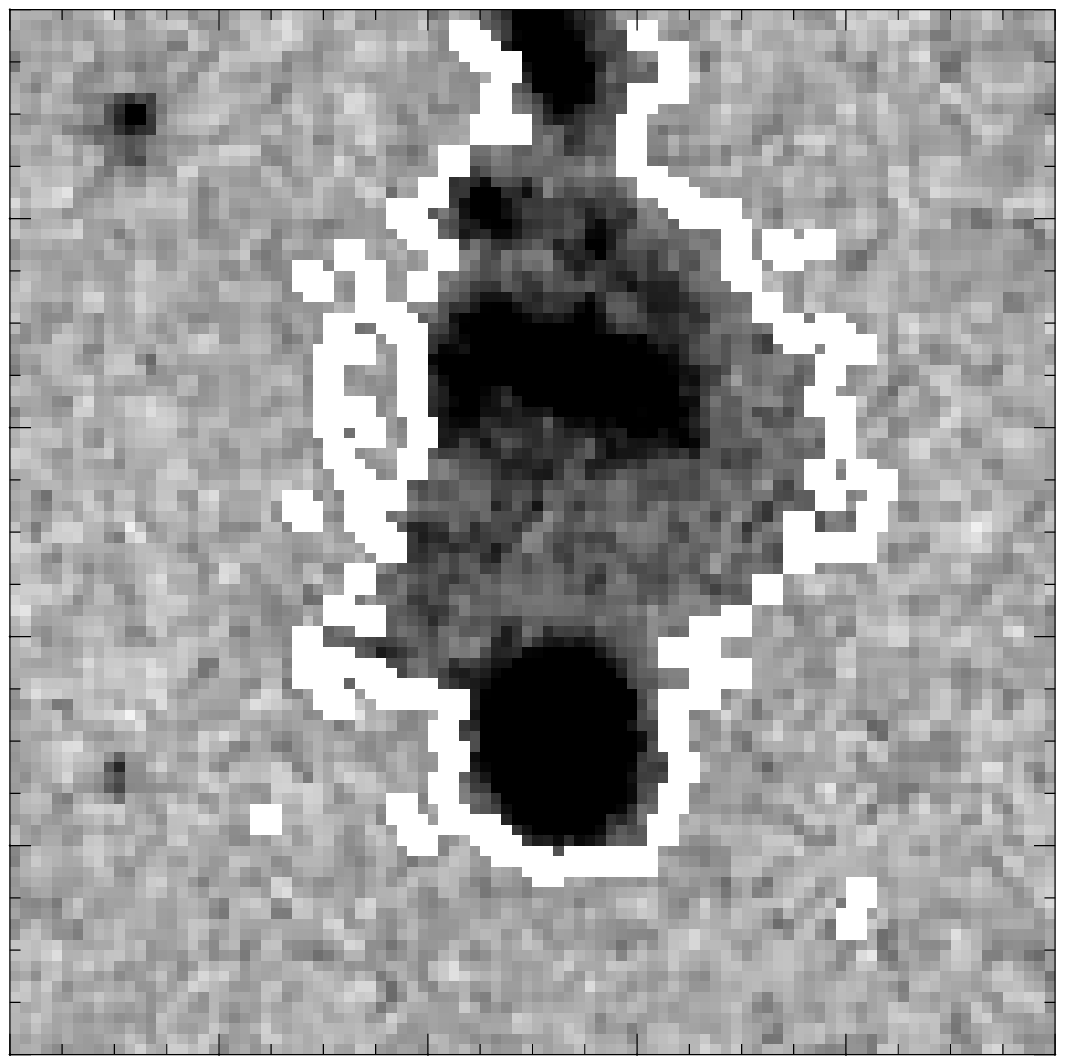} \includegraphics[scale=0.14, trim=77 0 77 0, clip=true]{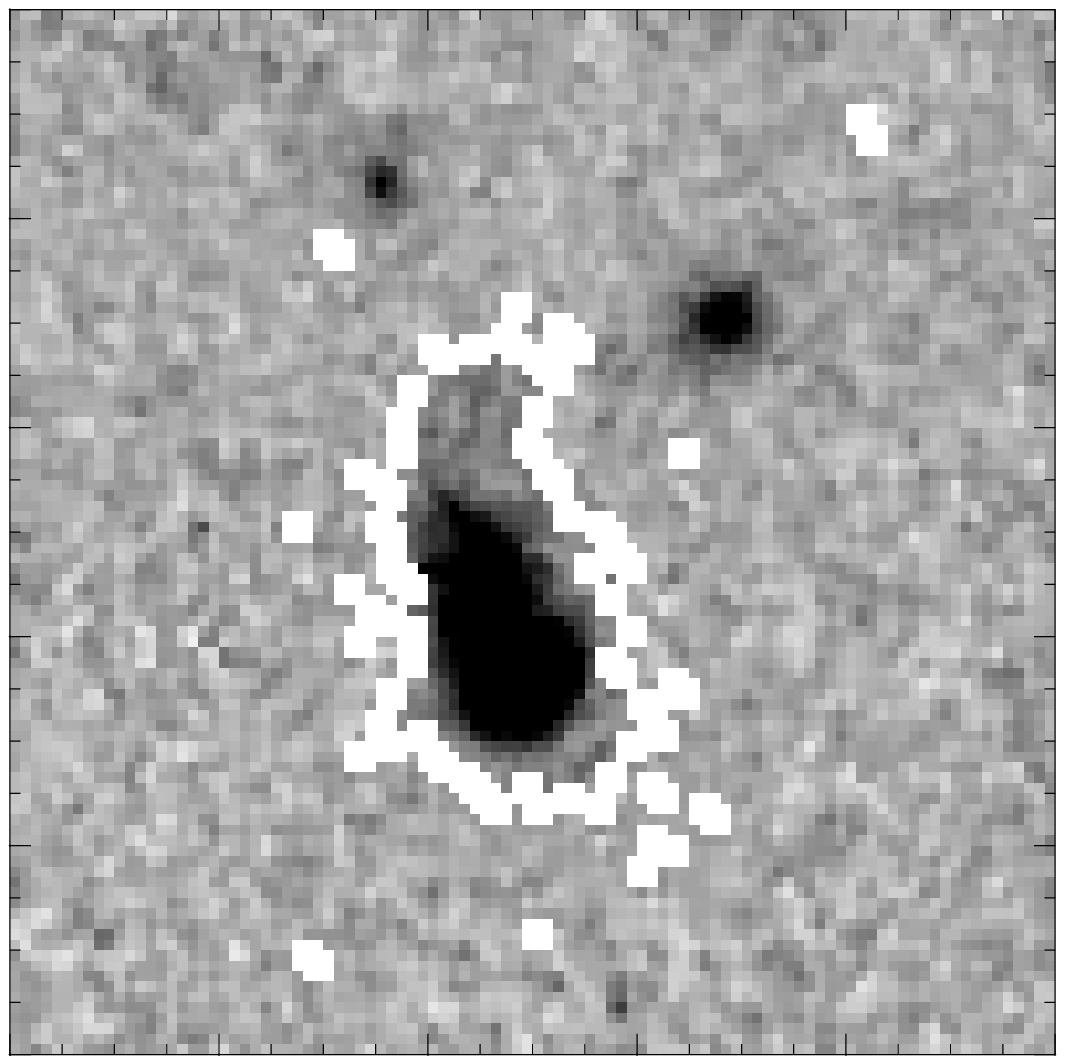} \includegraphics[scale=0.14, trim=77 0 77 0, clip=true]{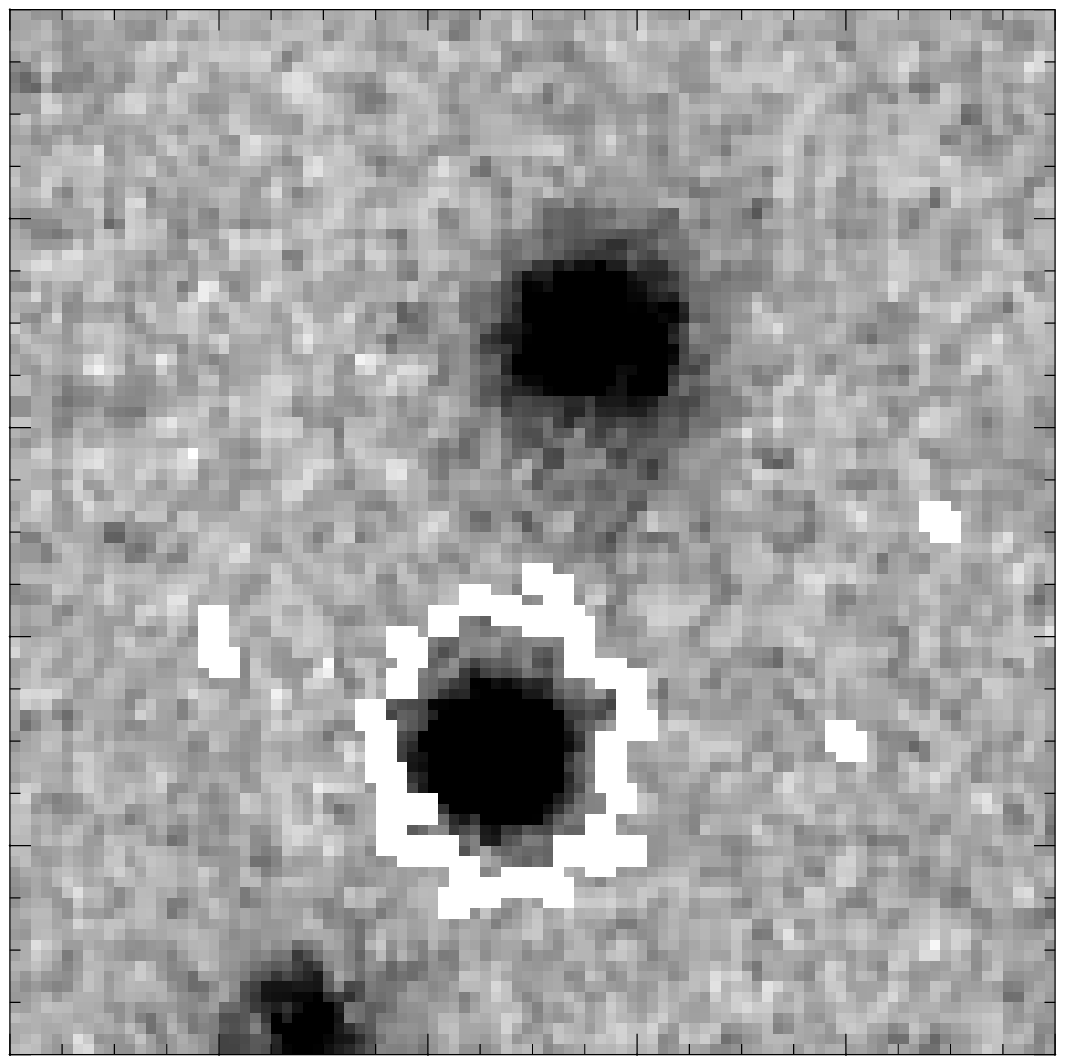} \includegraphics[scale=0.14, trim=77 0 77 0, clip=true]{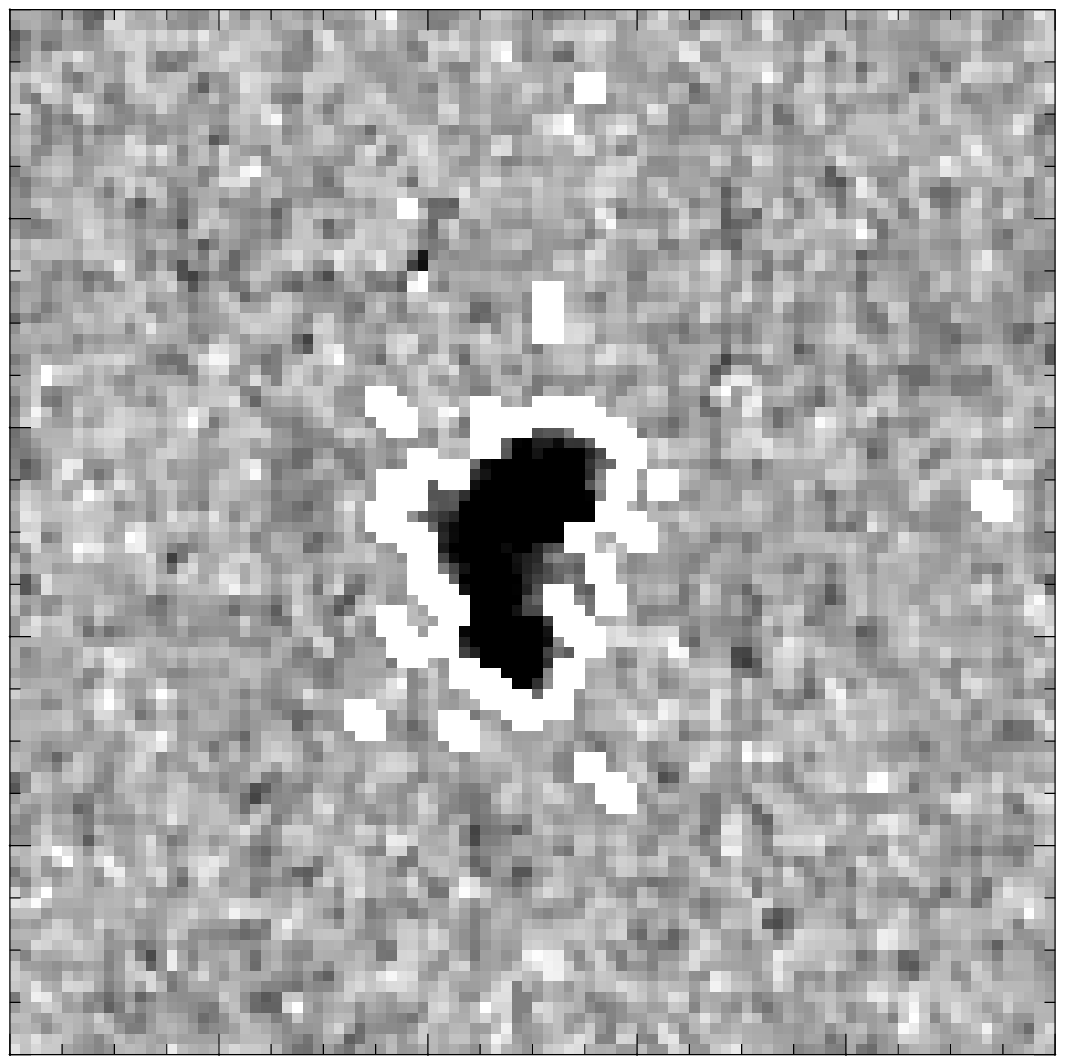}  
\hspace*{2em}
\includegraphics[scale=0.14, trim=77 0 77 0, clip=true]{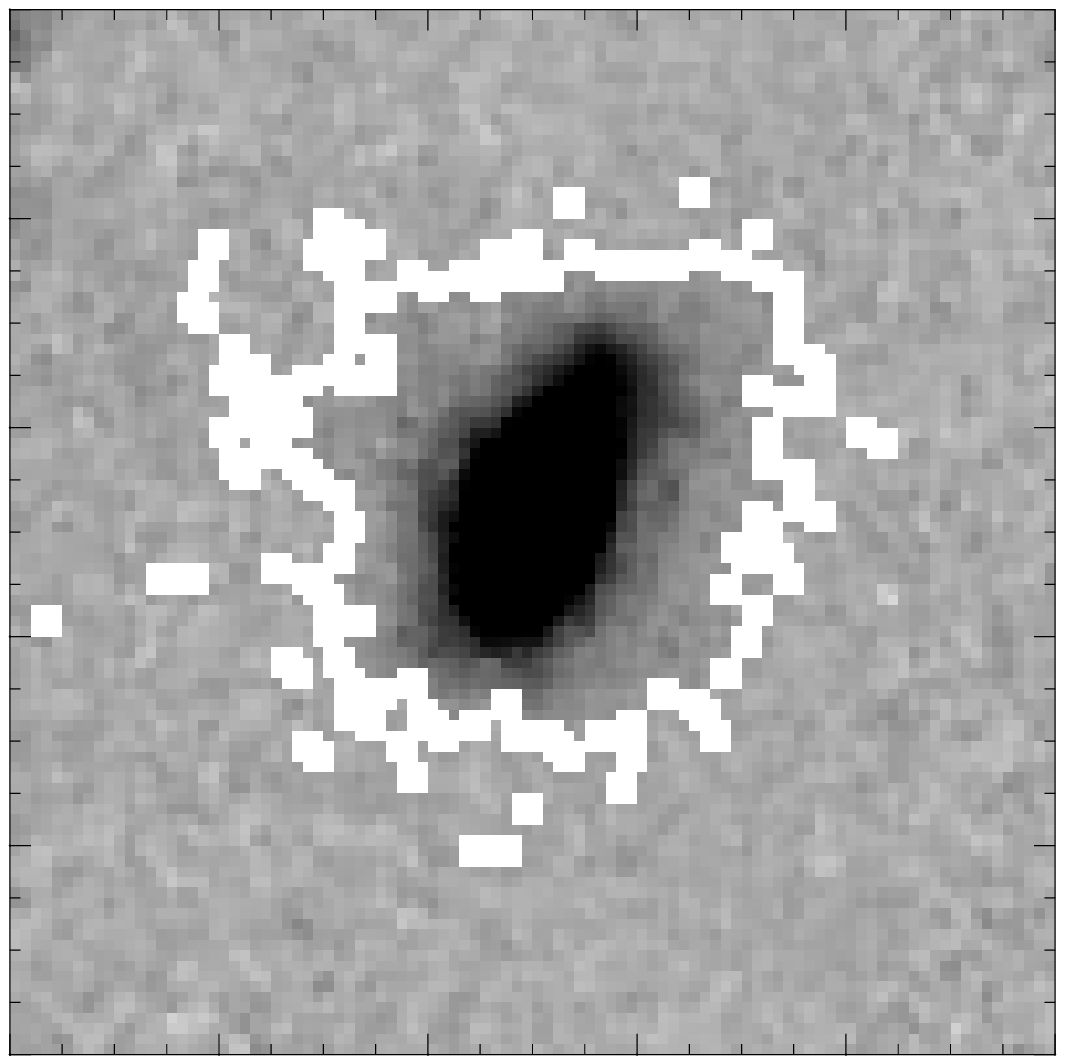} \includegraphics[scale=0.14, trim=77 0 77 0, clip=true]{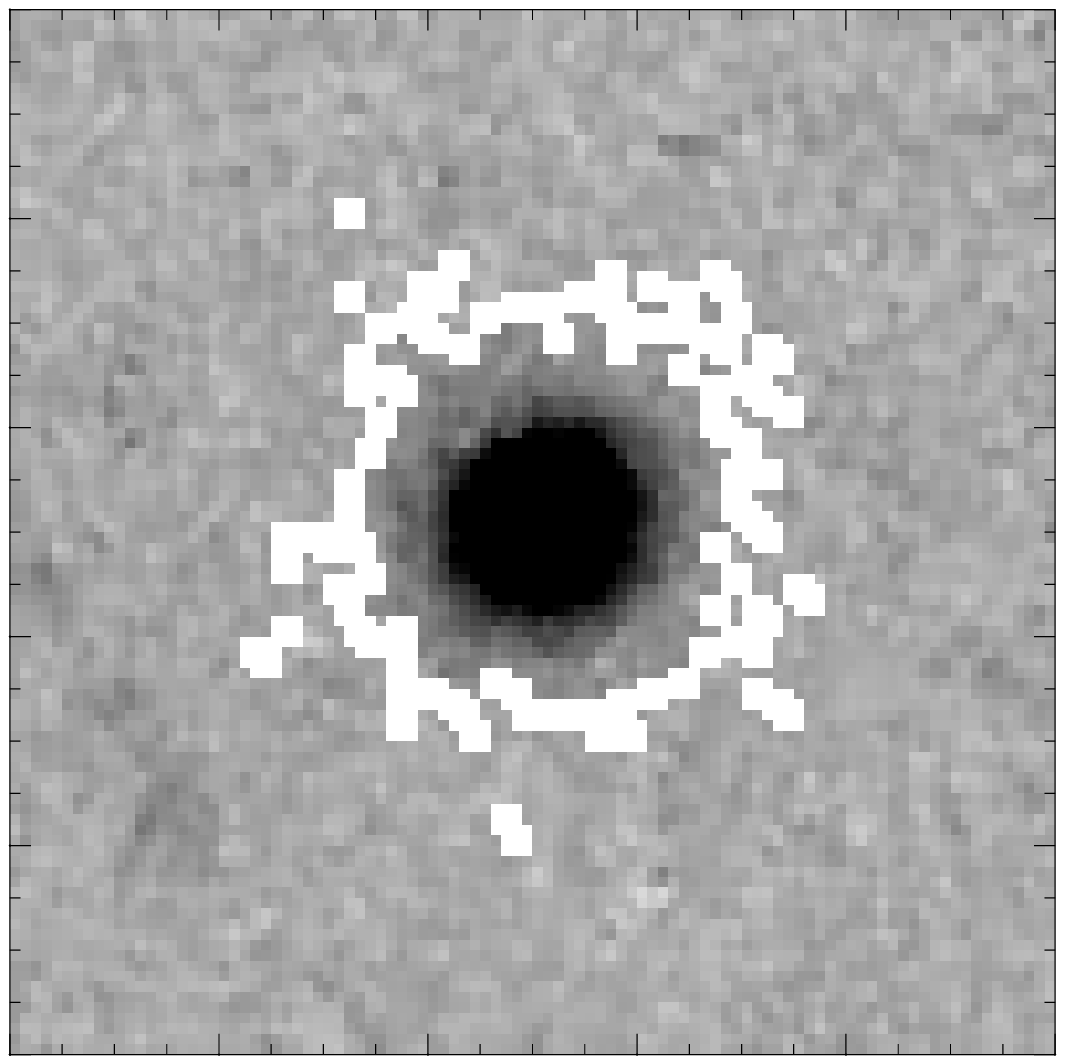} \includegraphics[scale=0.14, trim=77 0 77 0, clip=true]{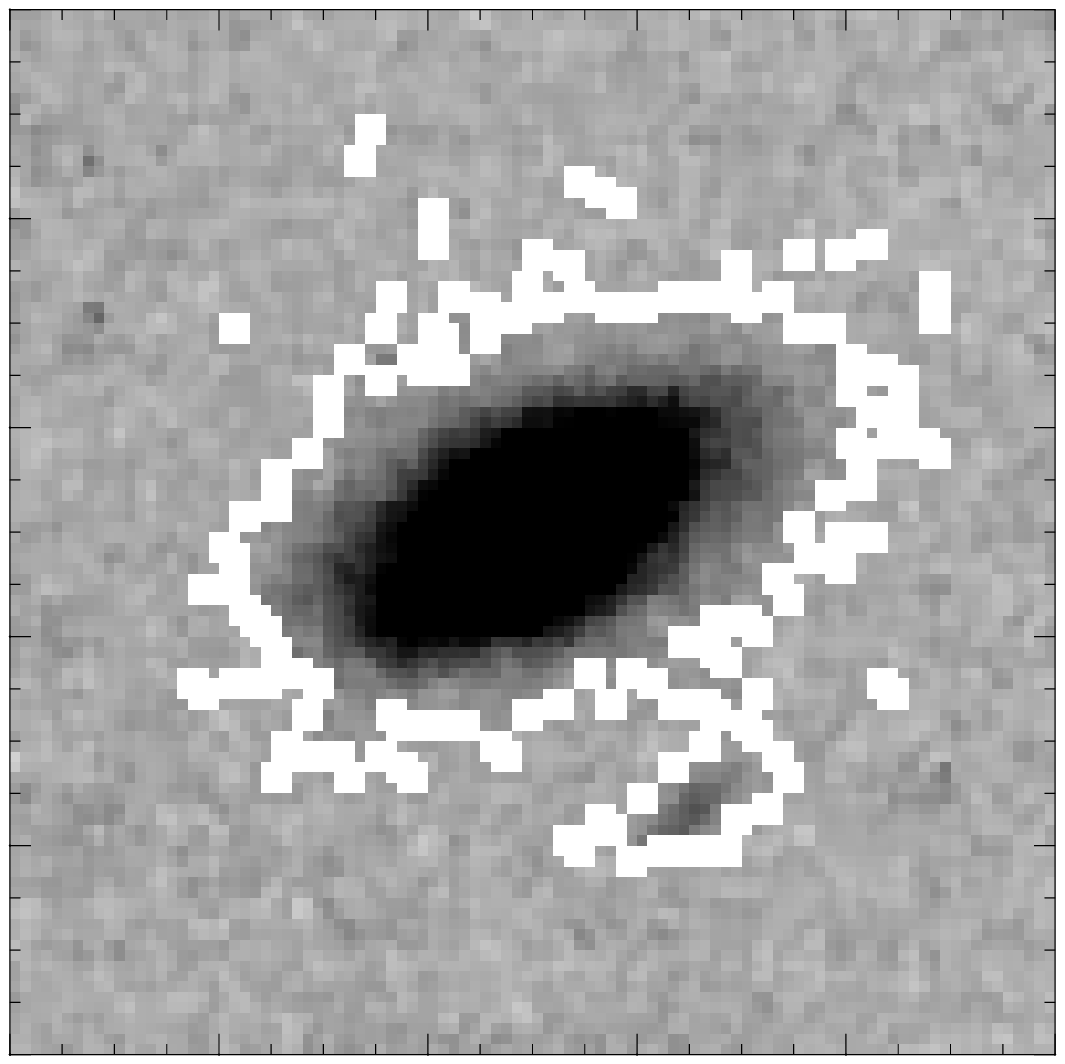} \includegraphics[scale=0.14, trim=77 0 77 0, clip=true]{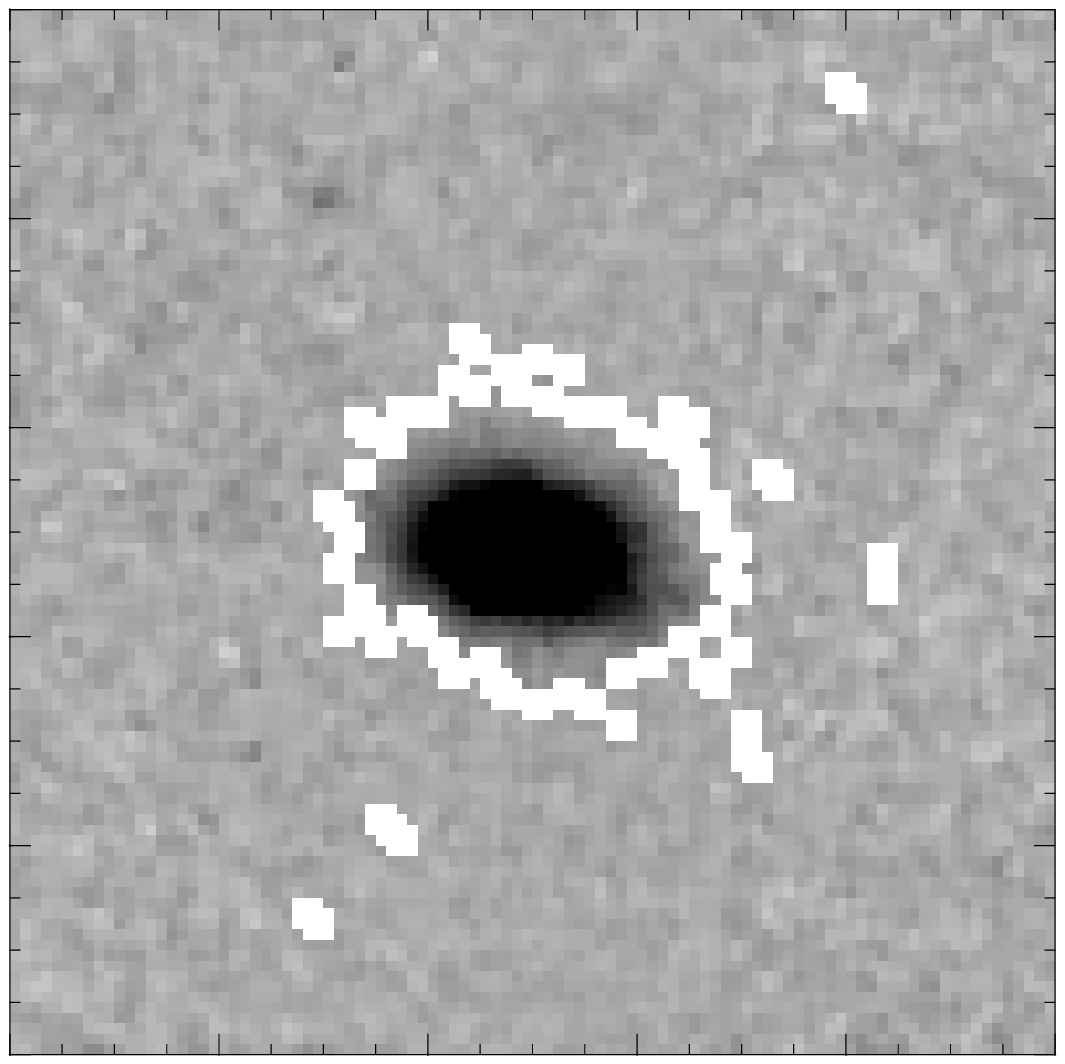} \includegraphics[scale=0.14, trim=77 0 77 0, clip=true]{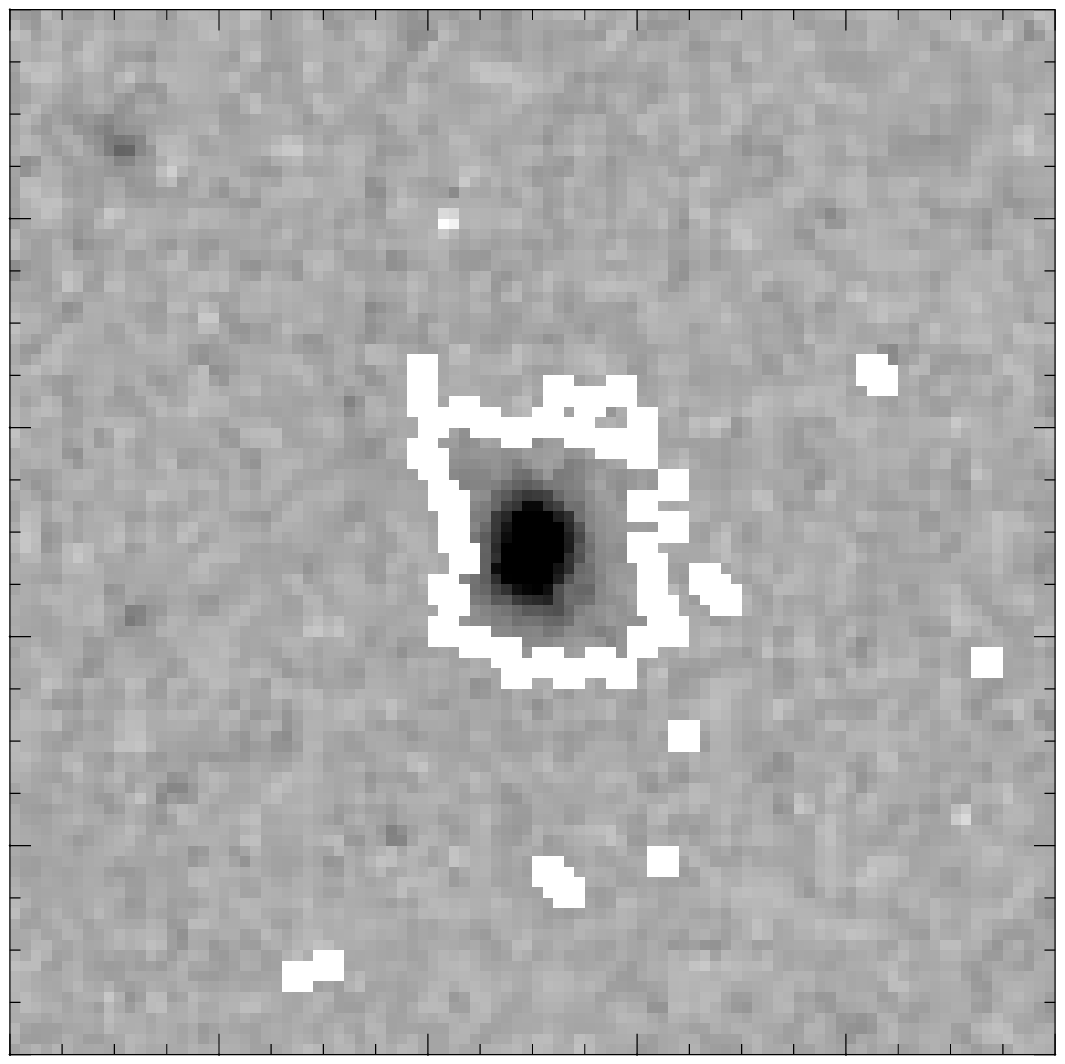}
\includegraphics[scale=0.14, trim=77 0 77 0, clip=true]{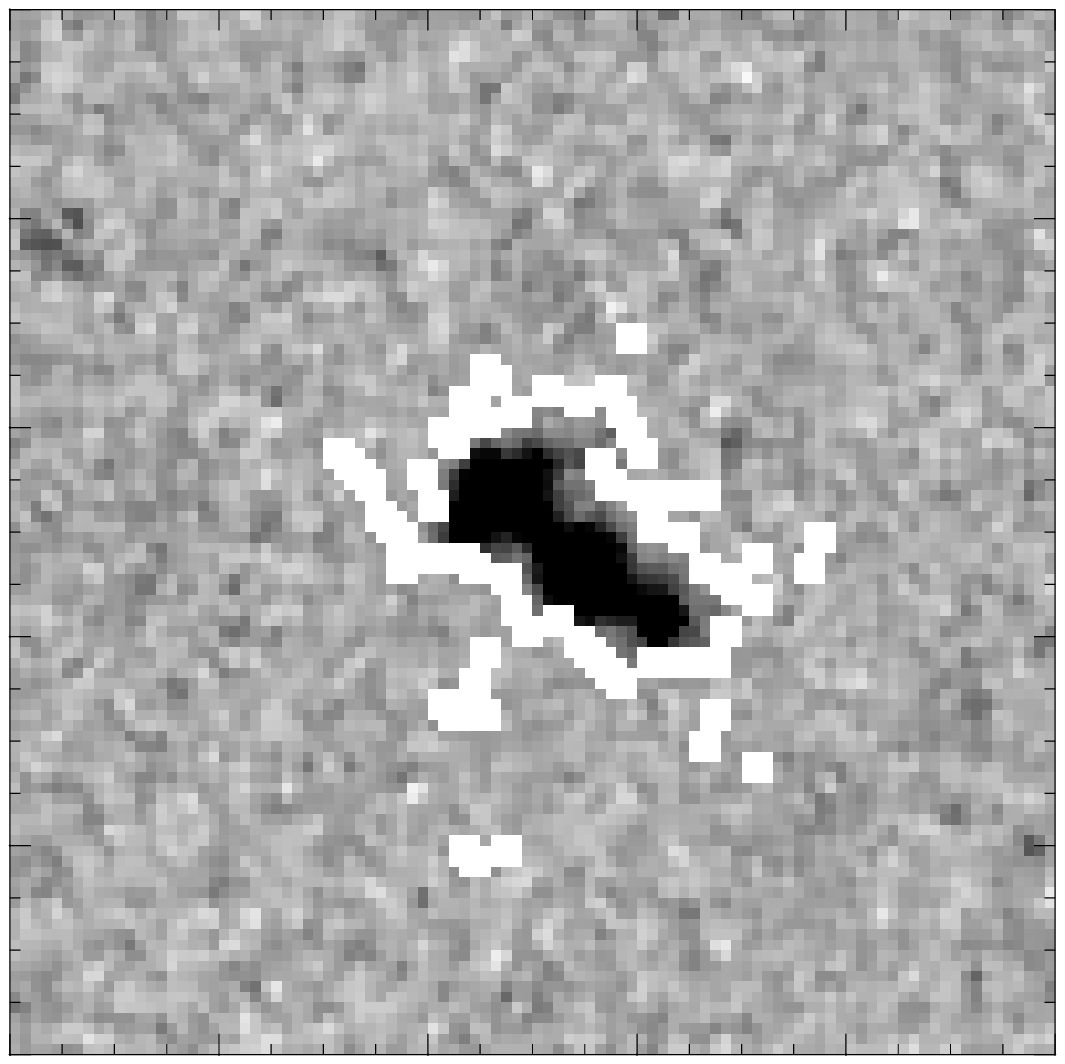} \includegraphics[scale=0.14, trim=77 0 77 0, clip=true]{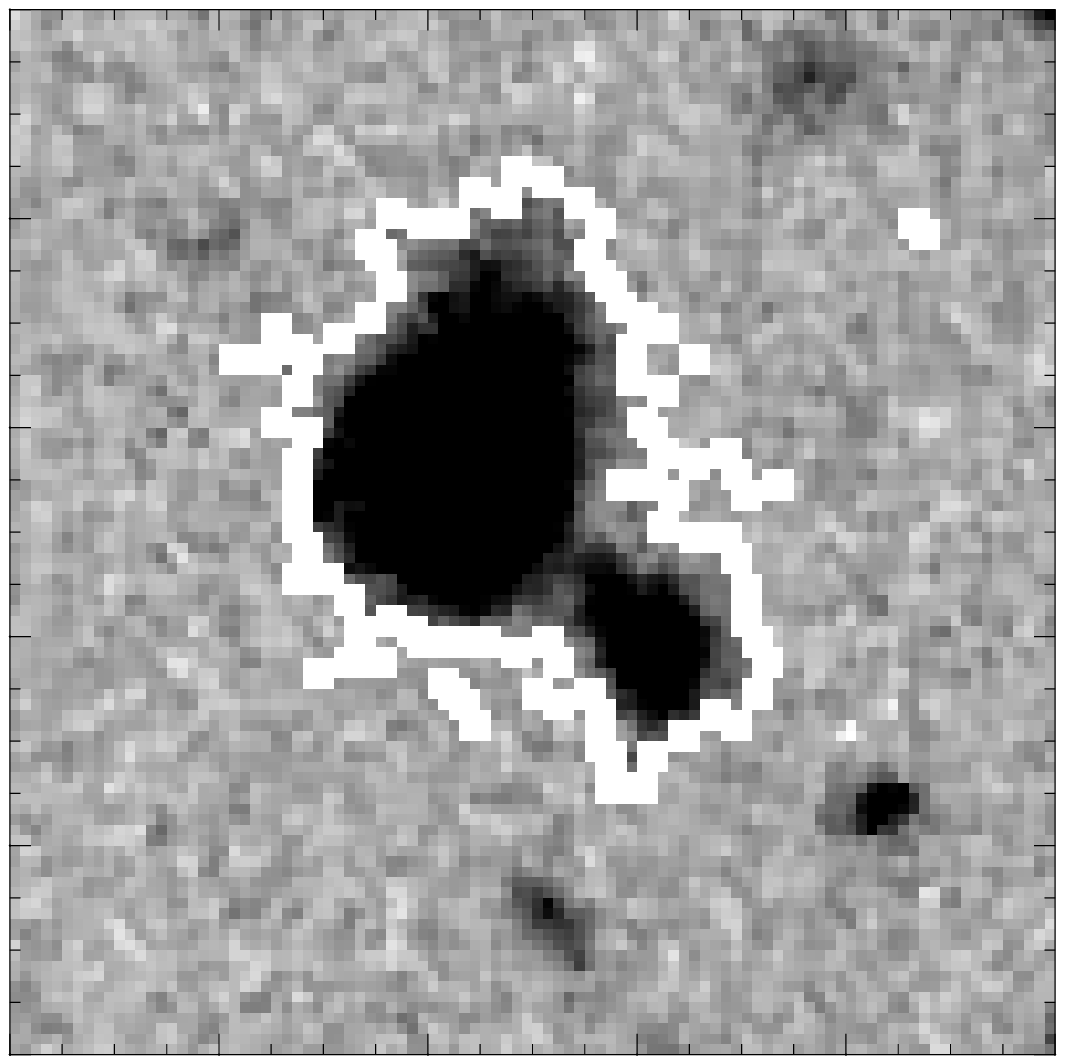} \includegraphics[scale=0.14, trim=77 0 77 0, clip=true]{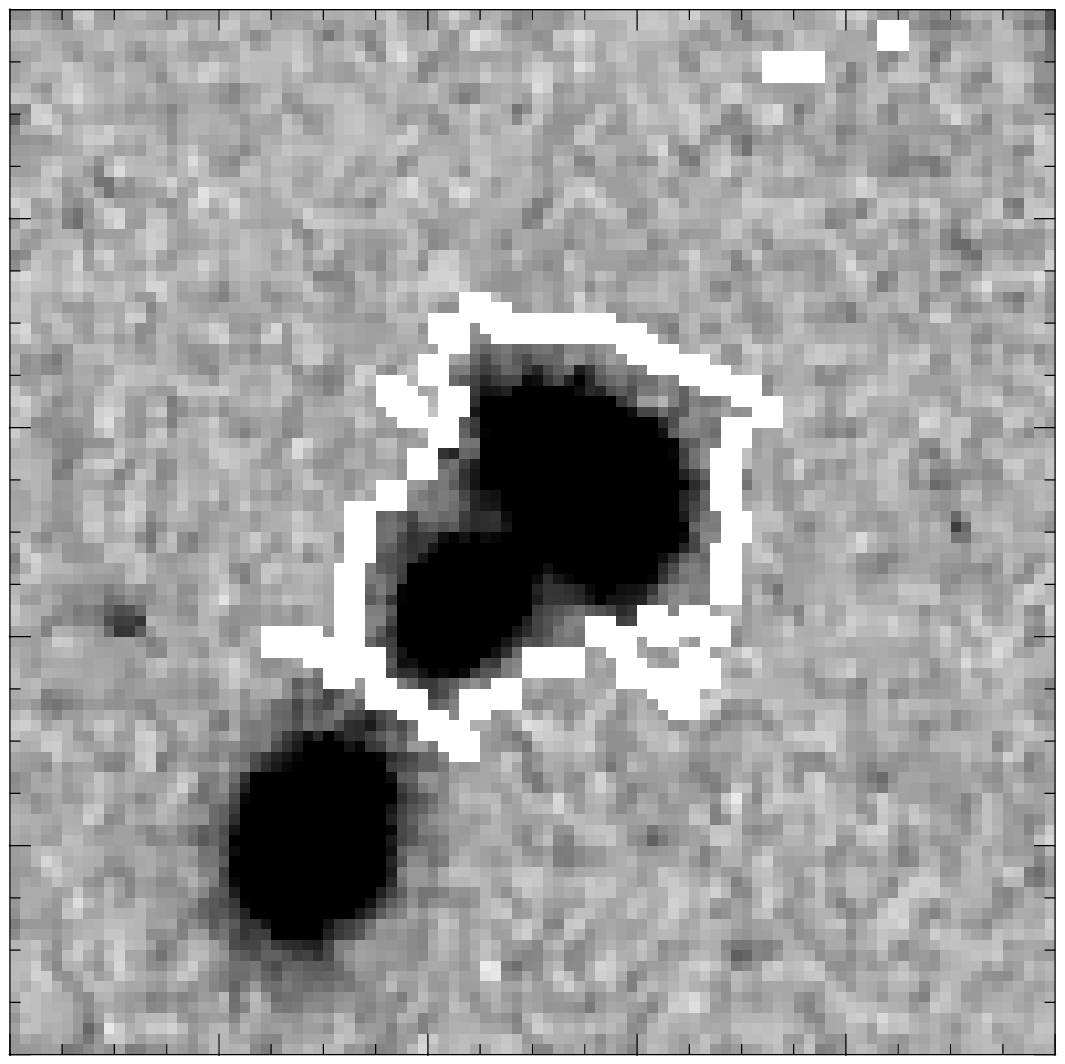} \includegraphics[scale=0.14, trim=77 0 77 0, clip=true]{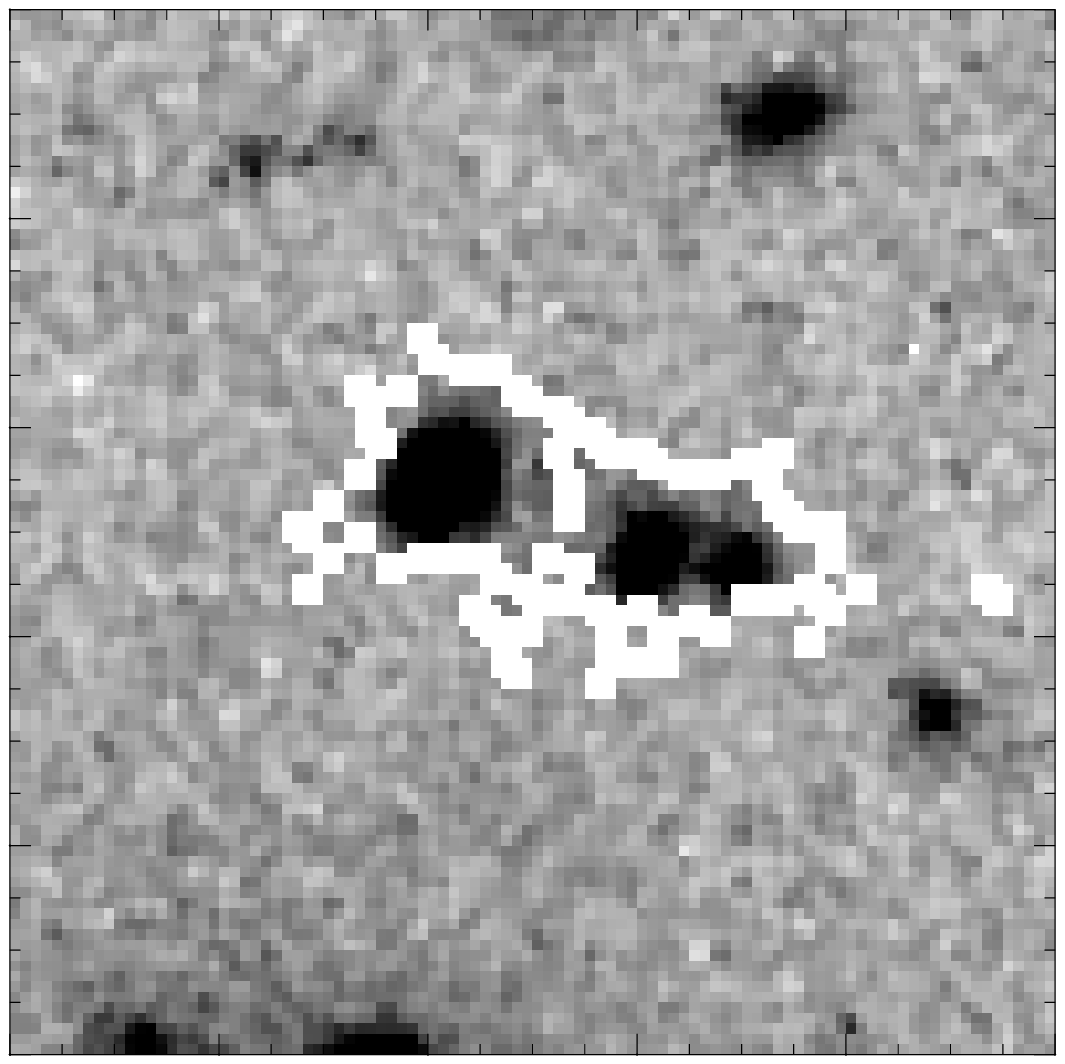} \includegraphics[scale=0.14, trim=77 0 77 0, clip=true]{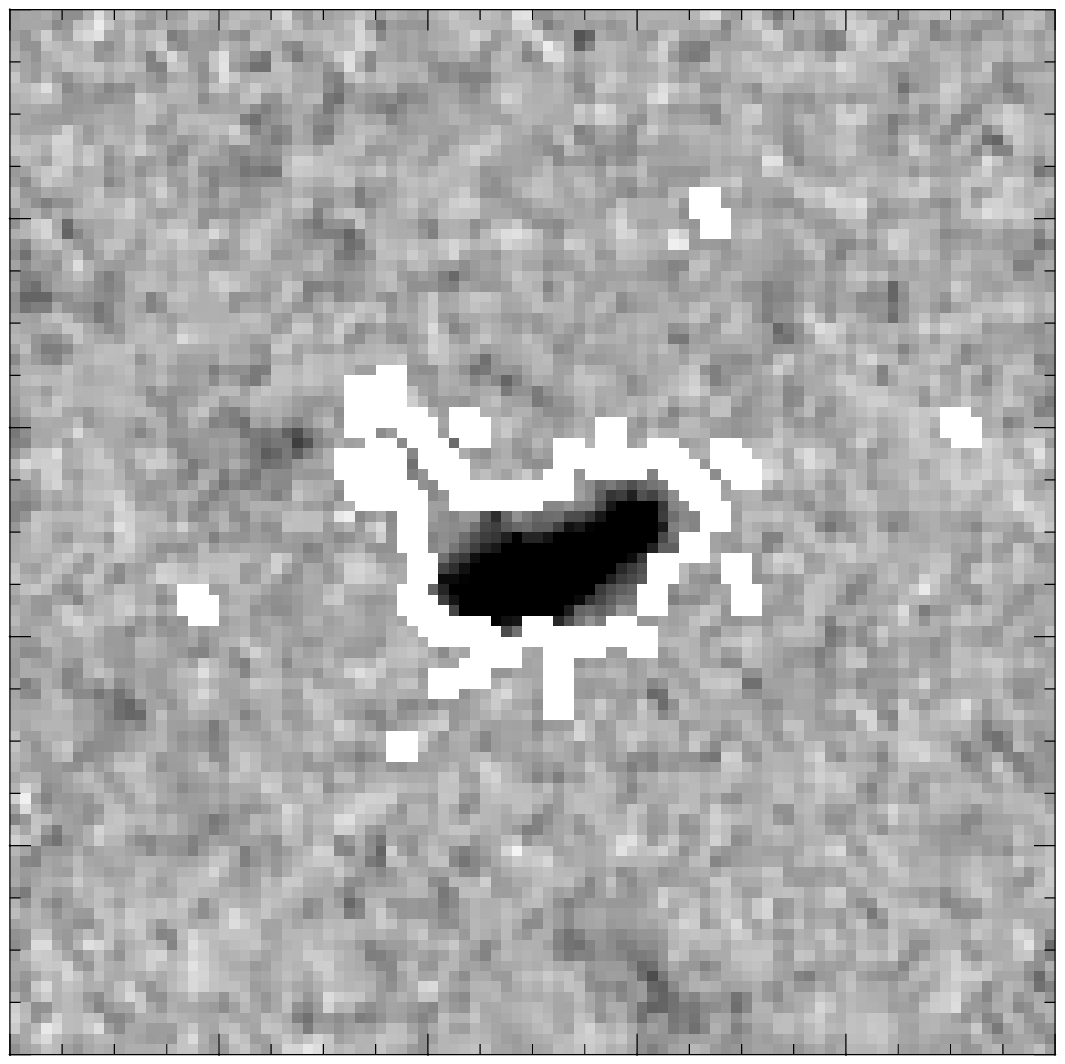}  
\hspace*{2em}
\includegraphics[scale=0.14, trim=77 0 77 0, clip=true]{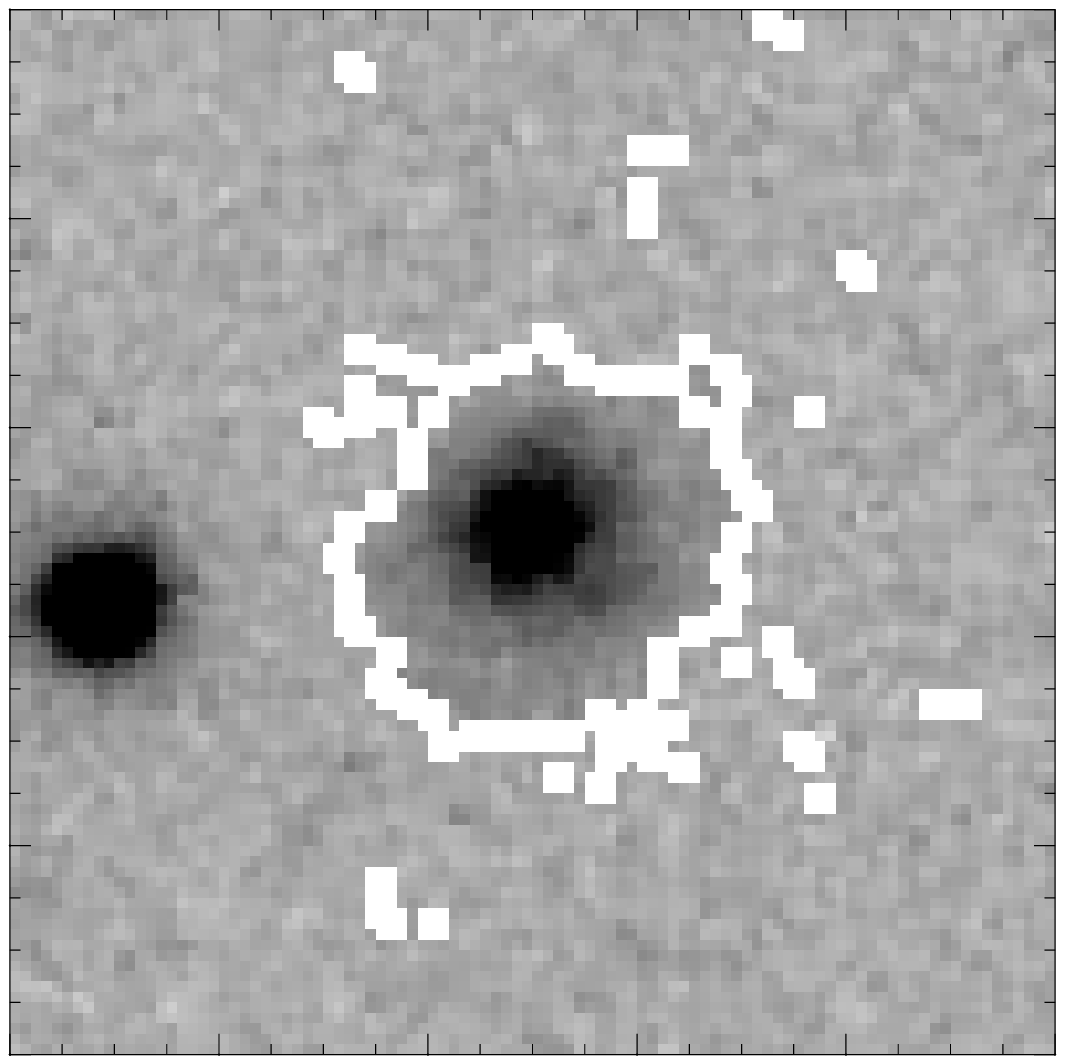} \includegraphics[scale=0.14, trim=77 0 77 0, clip=true]{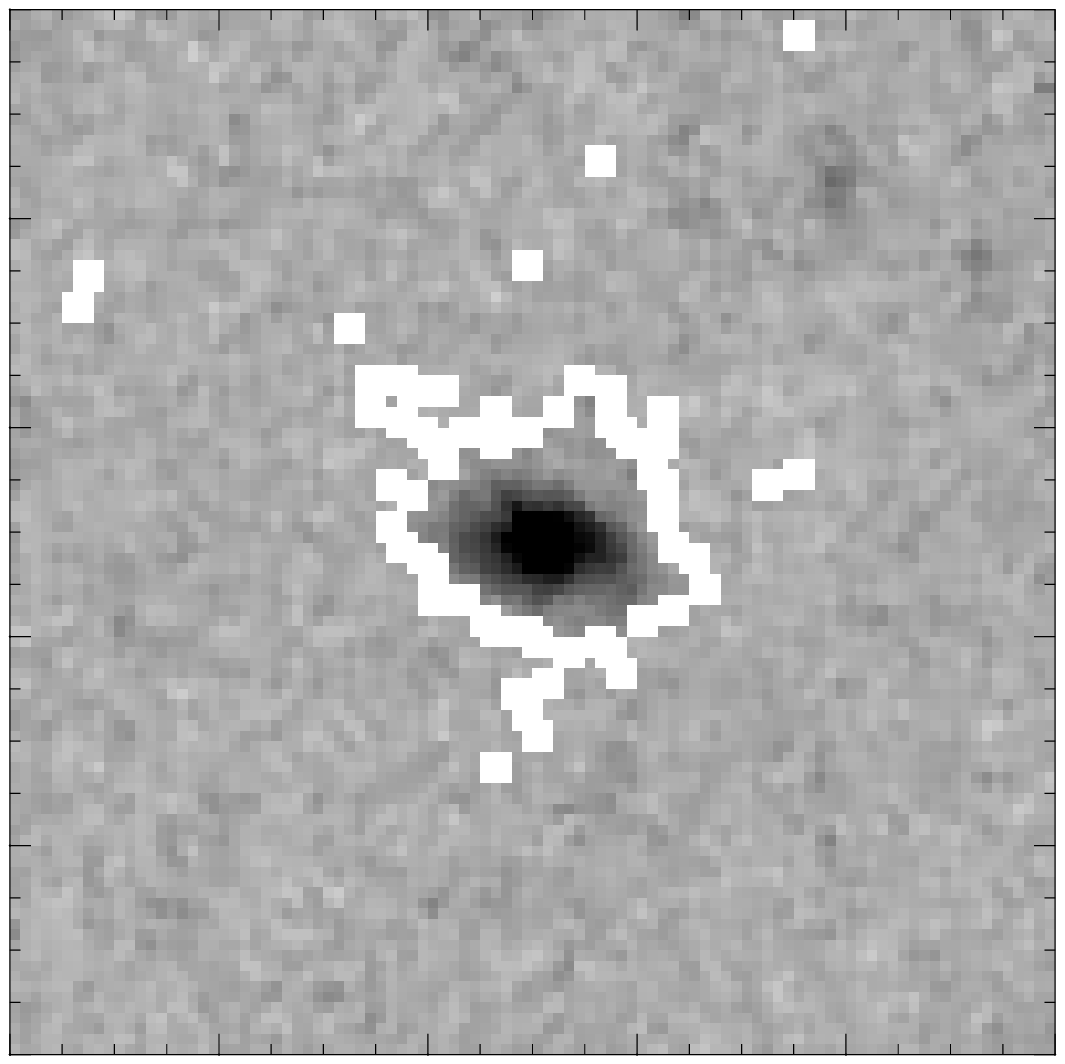} \includegraphics[scale=0.14, trim=77 0 77 0, clip=true]{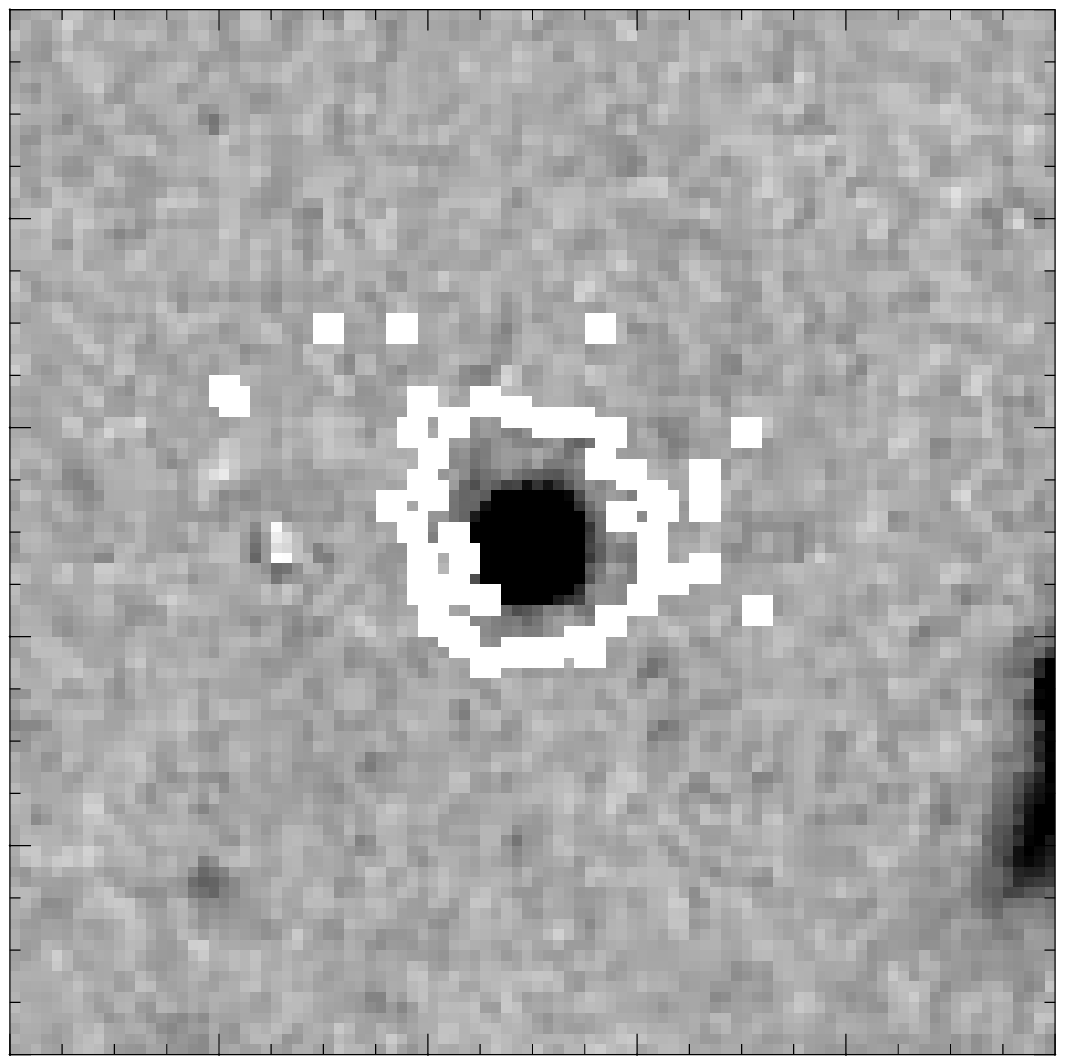} \includegraphics[scale=0.14, trim=77 0 77 0, clip=true]{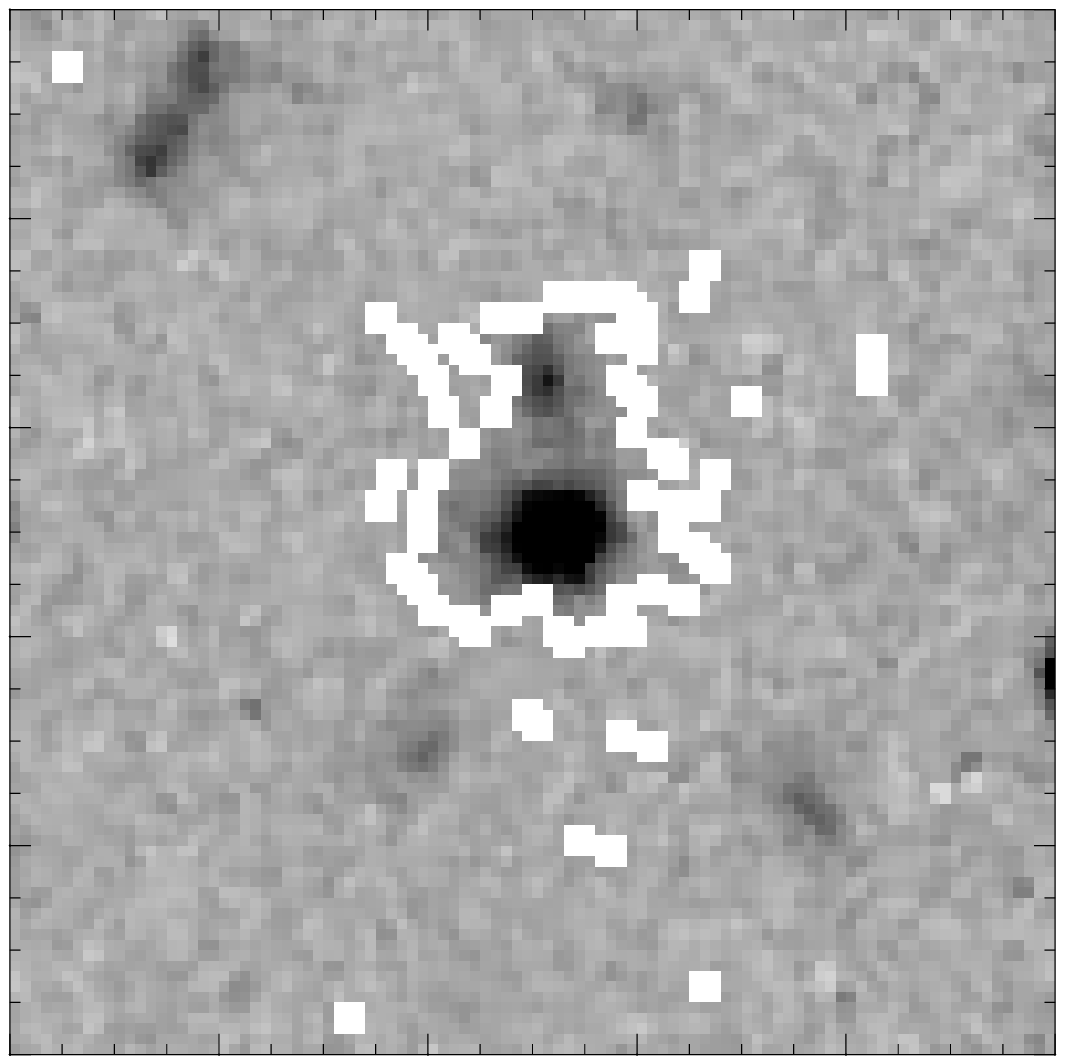} \includegraphics[scale=0.14, trim=77 0 77 0, clip=true]{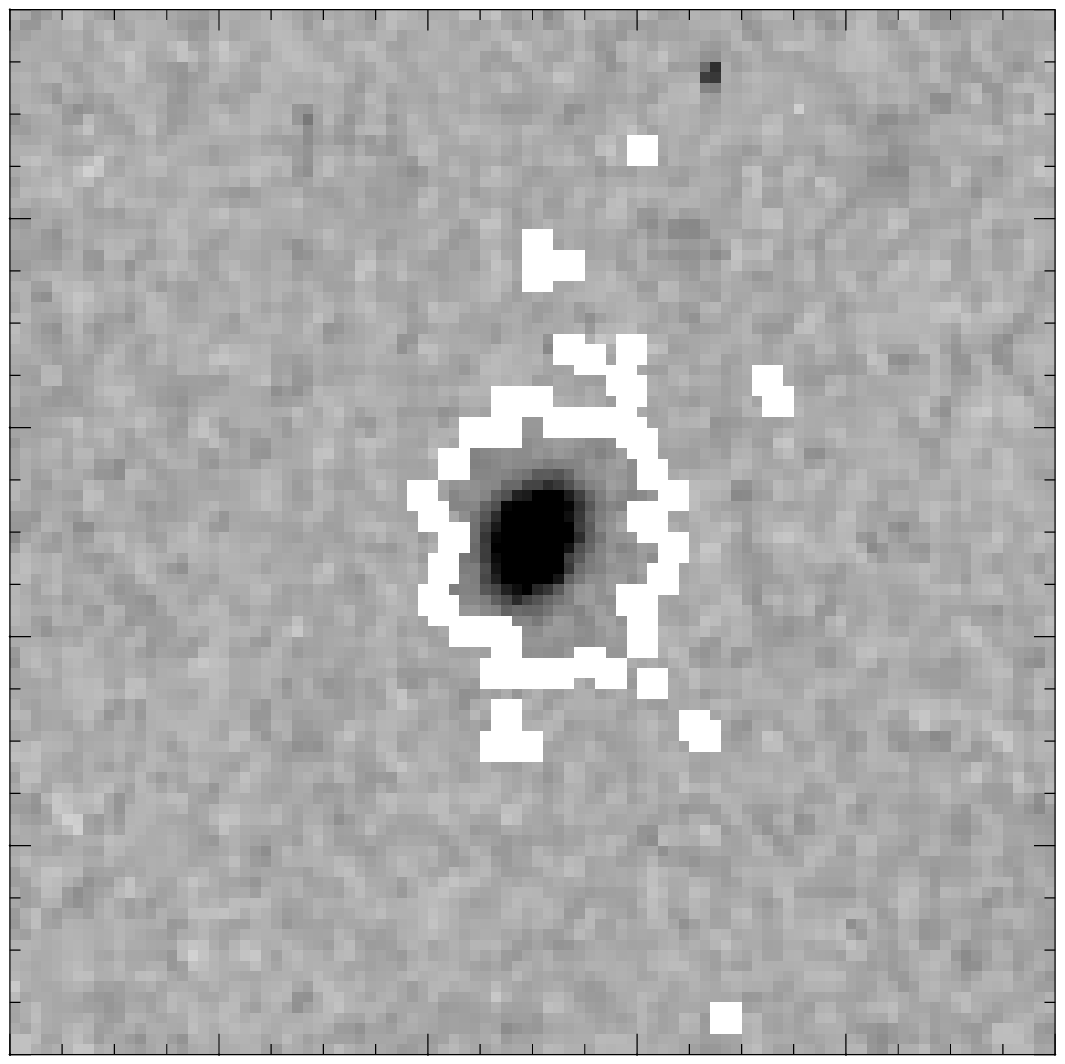}\vspace{0.25cm}
%
%
\includegraphics[scale=0.14, trim=77 0 77 0, clip=true]{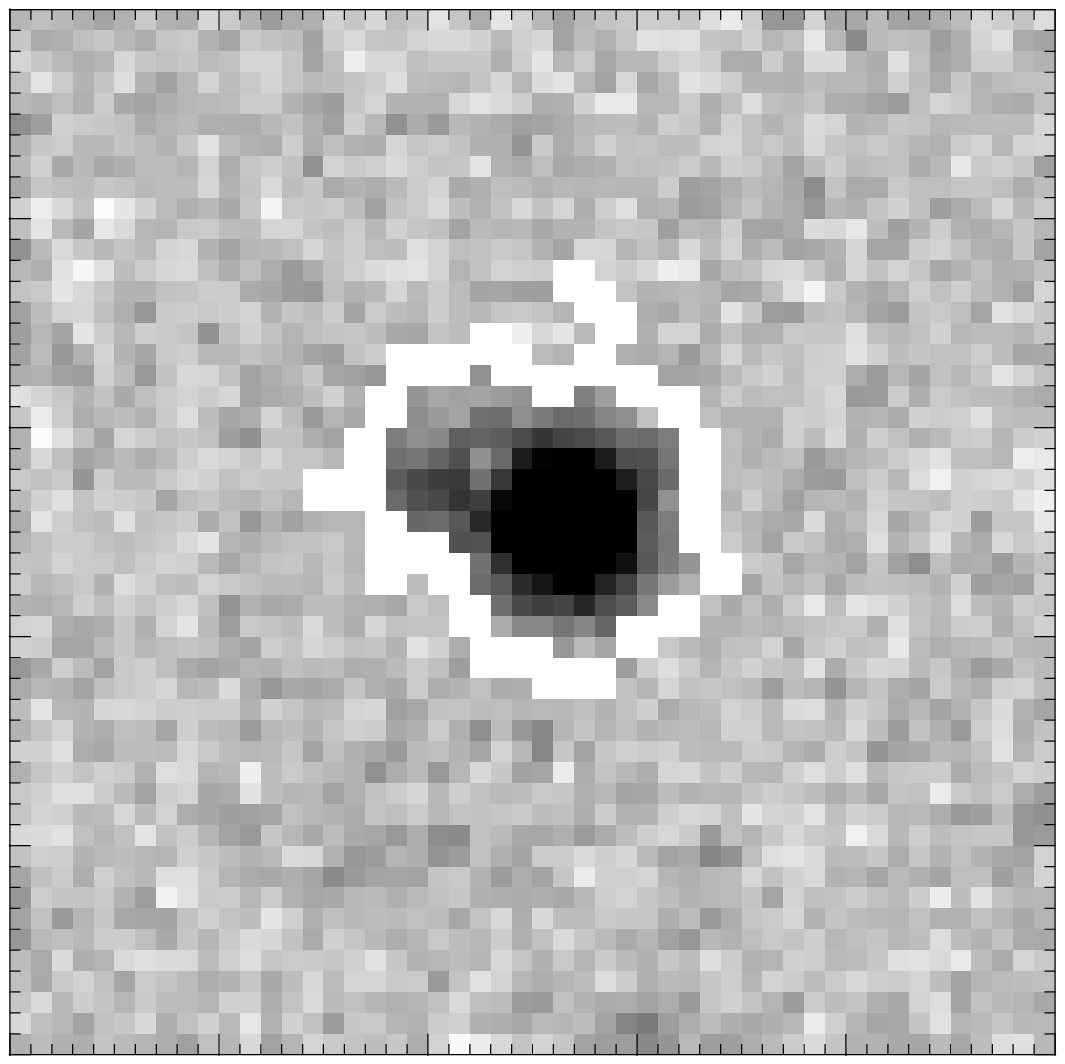} \includegraphics[scale=0.14, trim=77 0 77 0, clip=true]{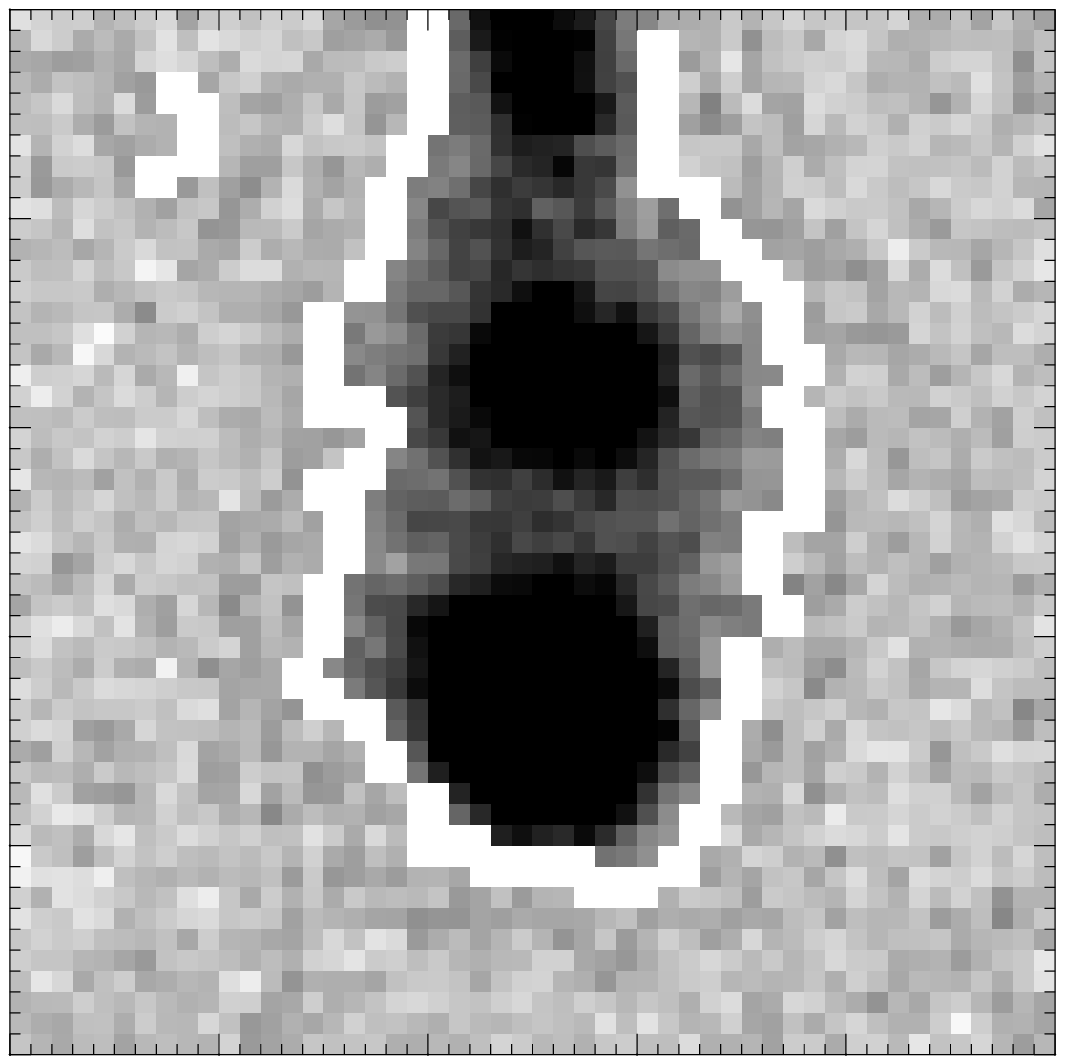} \includegraphics[scale=0.14, trim=77 0 77 0, clip=true]{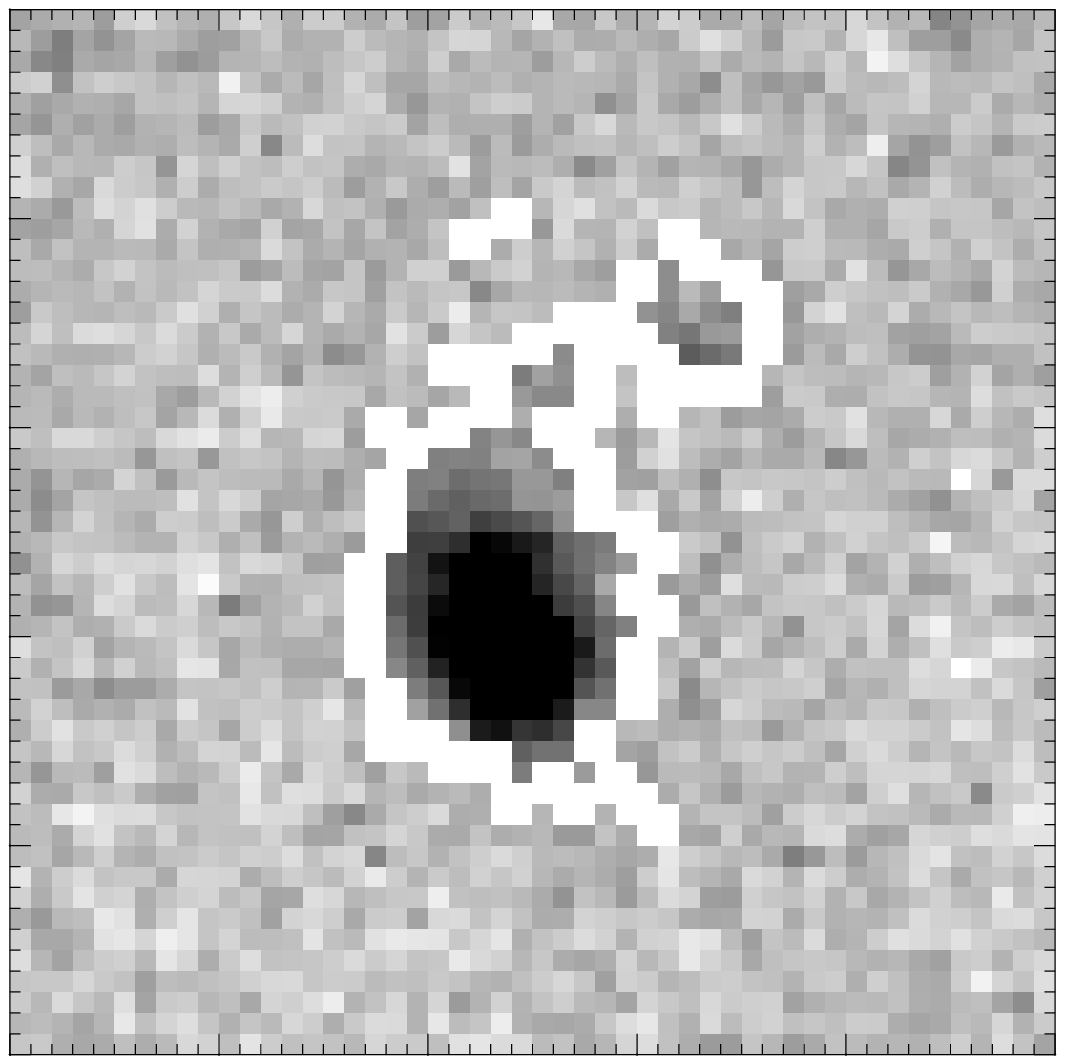} \includegraphics[scale=0.14, trim=77 0 77 0, clip=true]{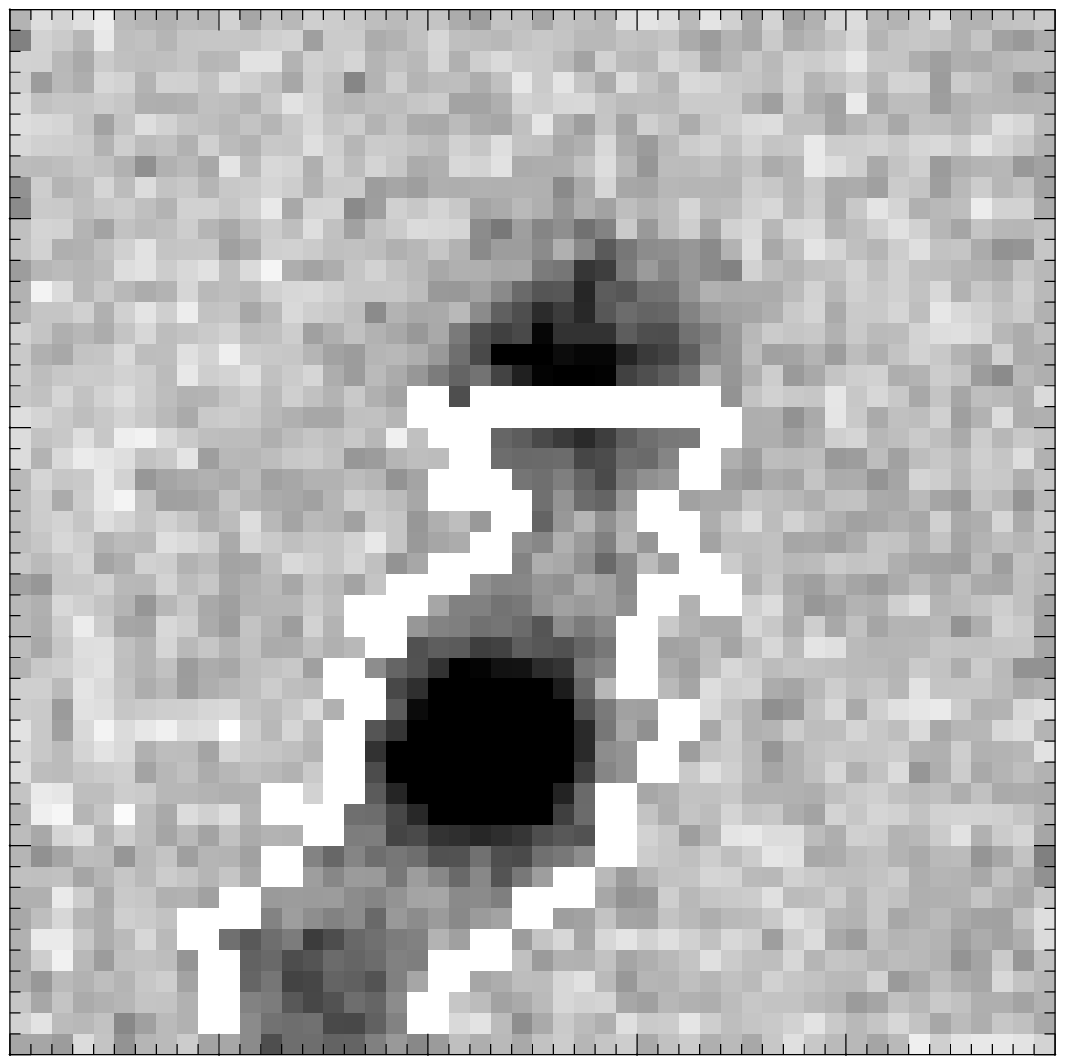} \includegraphics[scale=0.14, trim=77 0 77 0, clip=true]{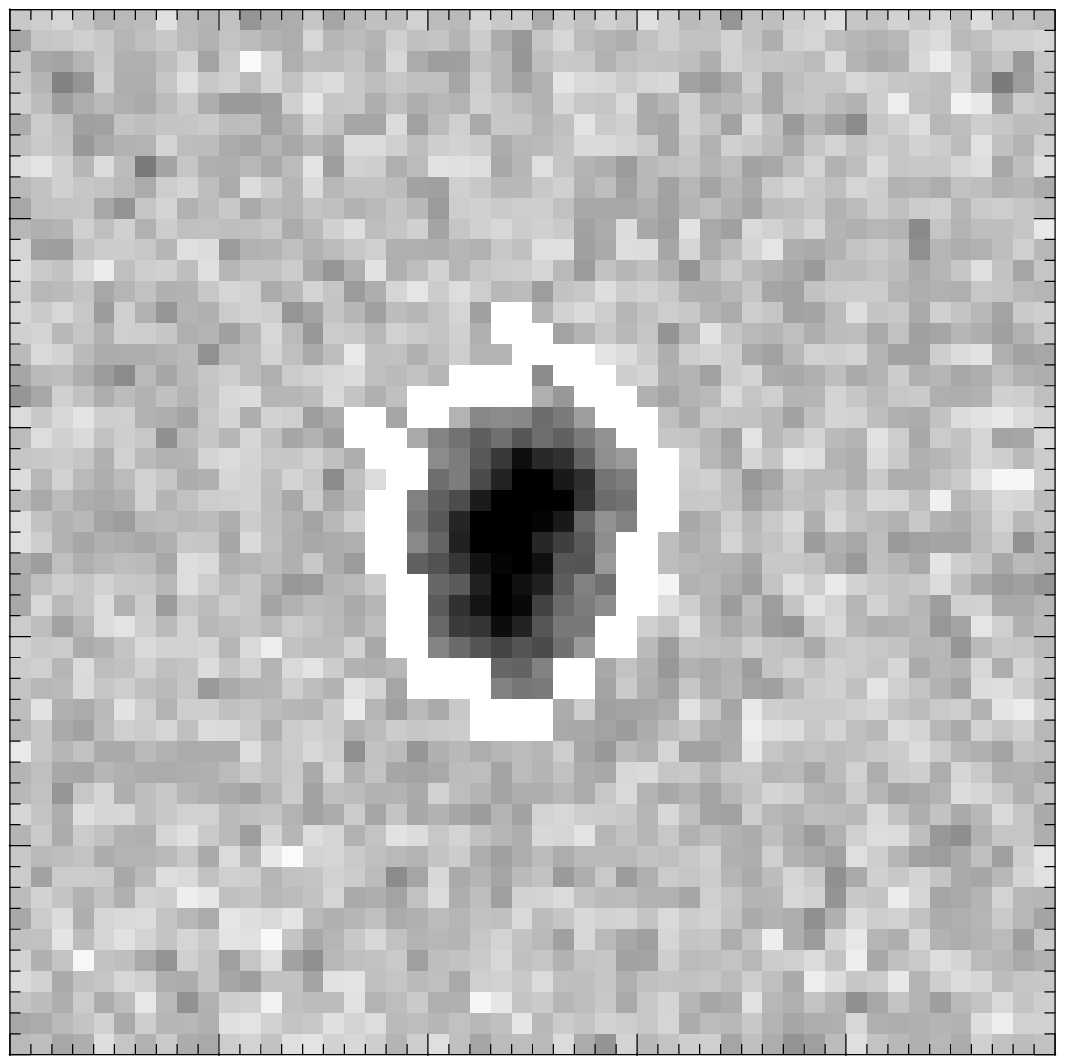}
\hspace*{2em}
\includegraphics[scale=0.14, trim=77 0 77 0, clip=true]{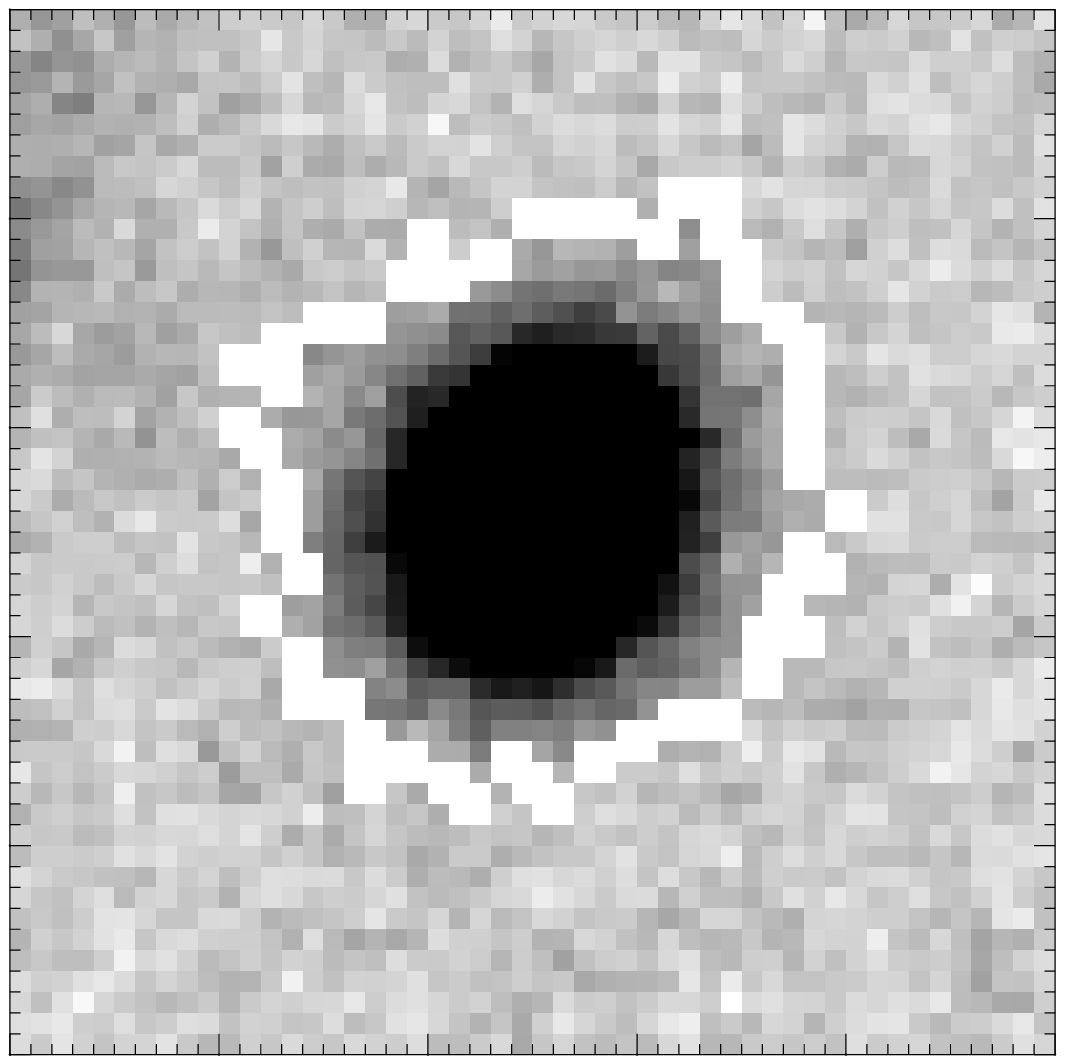} \includegraphics[scale=0.14, trim=77 0 77 0, clip=true]{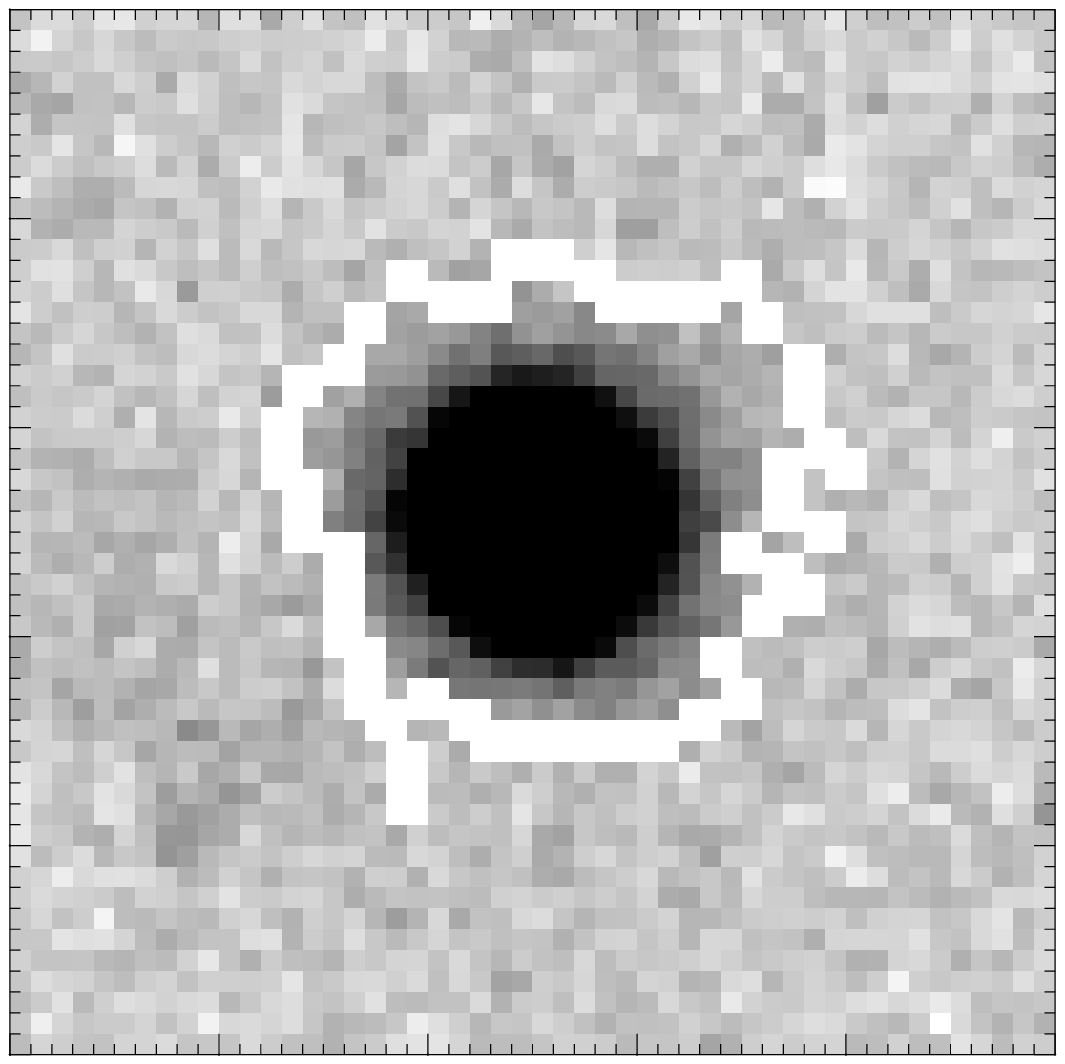} \includegraphics[scale=0.14, trim=77 0 77 0, clip=true]{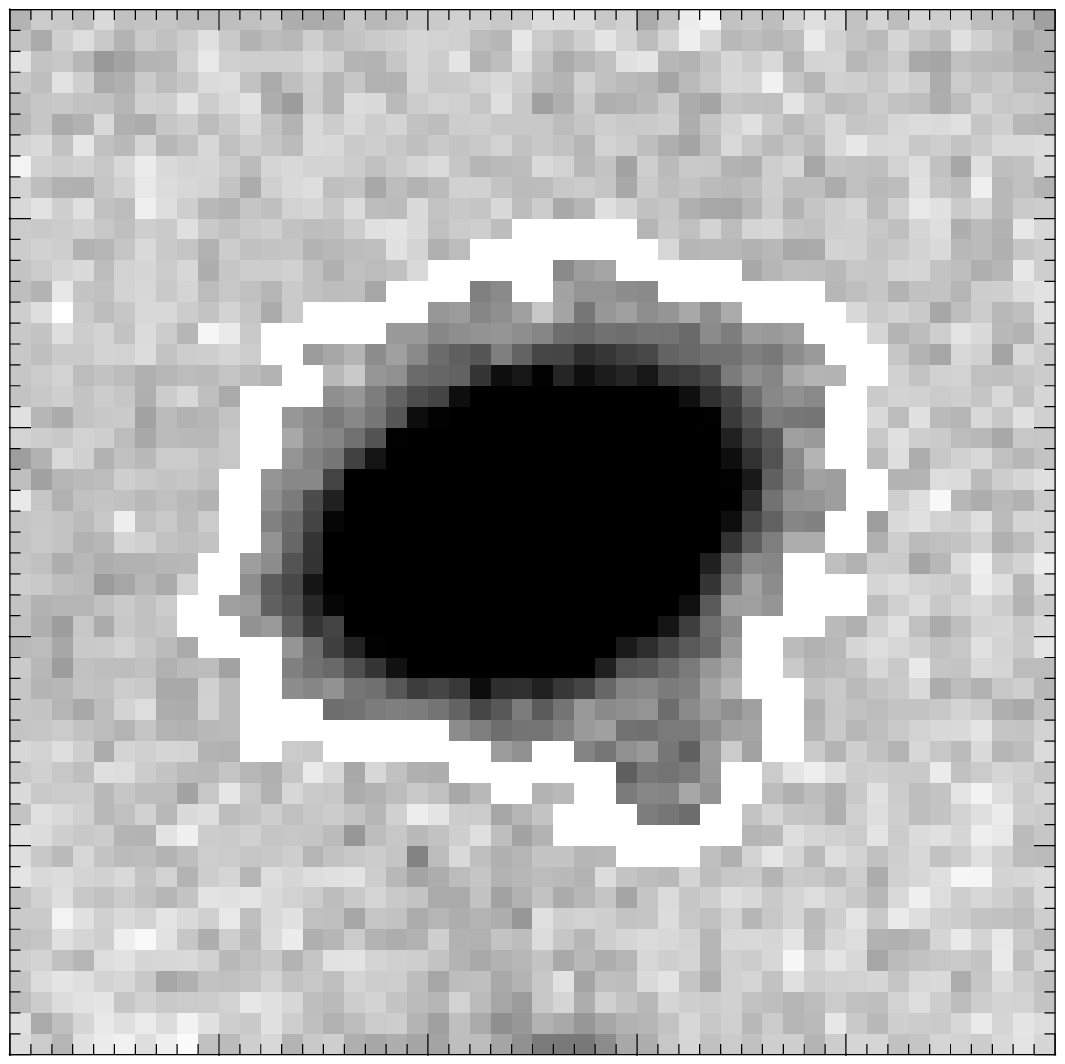} \includegraphics[scale=0.14, trim=77 0 77 0, clip=true]{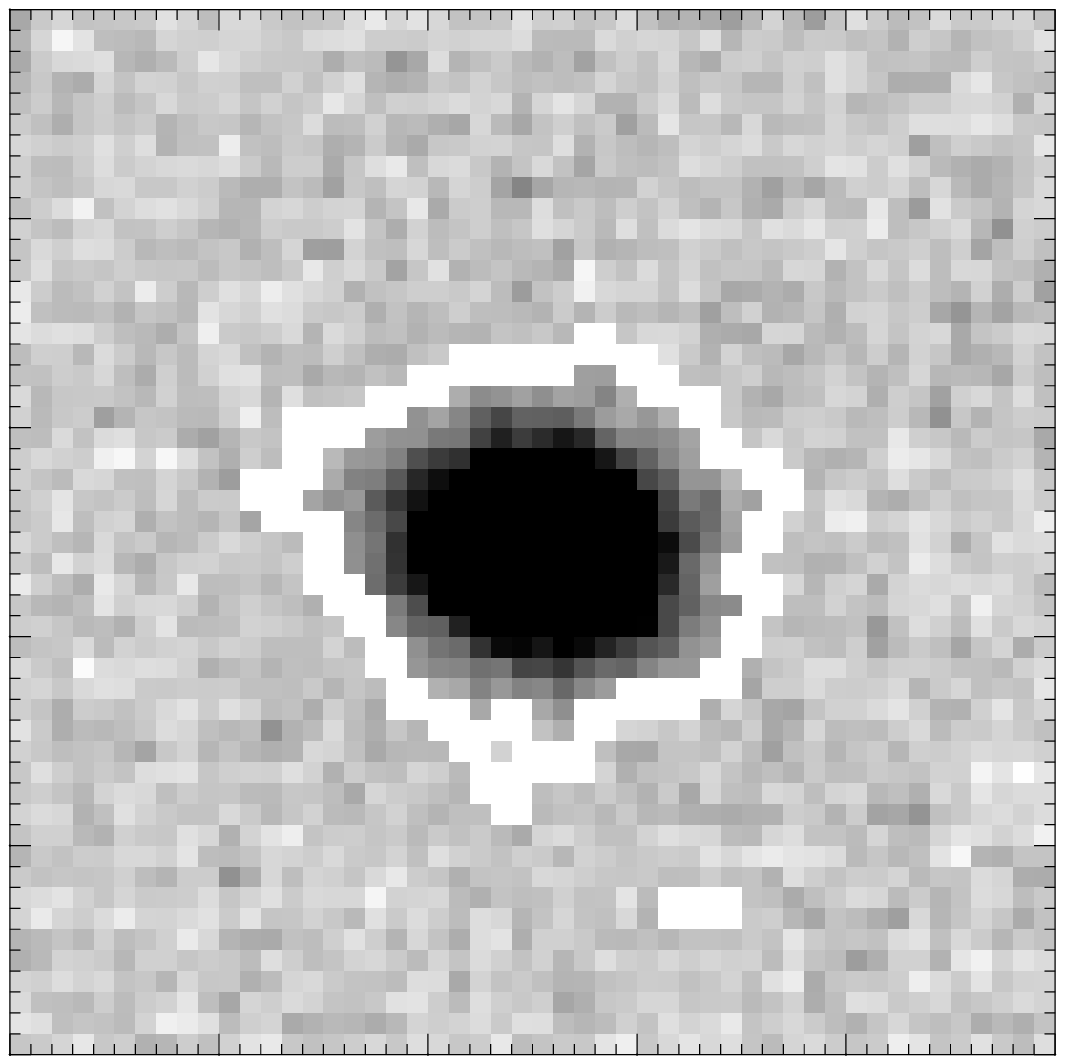} \includegraphics[scale=0.14, trim=77 0 77 0, clip=true]{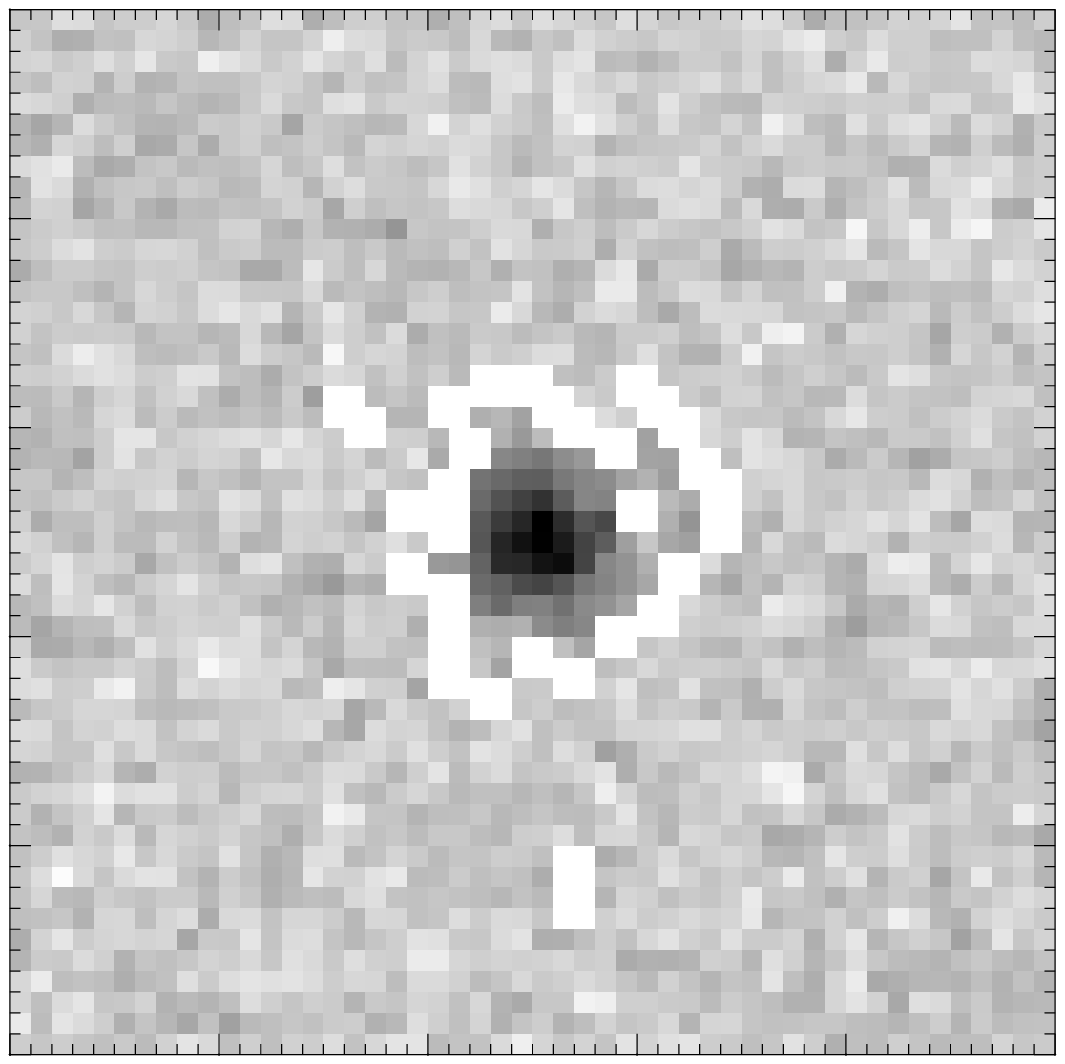}
%
%
\includegraphics[scale=0.14, trim=77 0 77 0, clip=true]{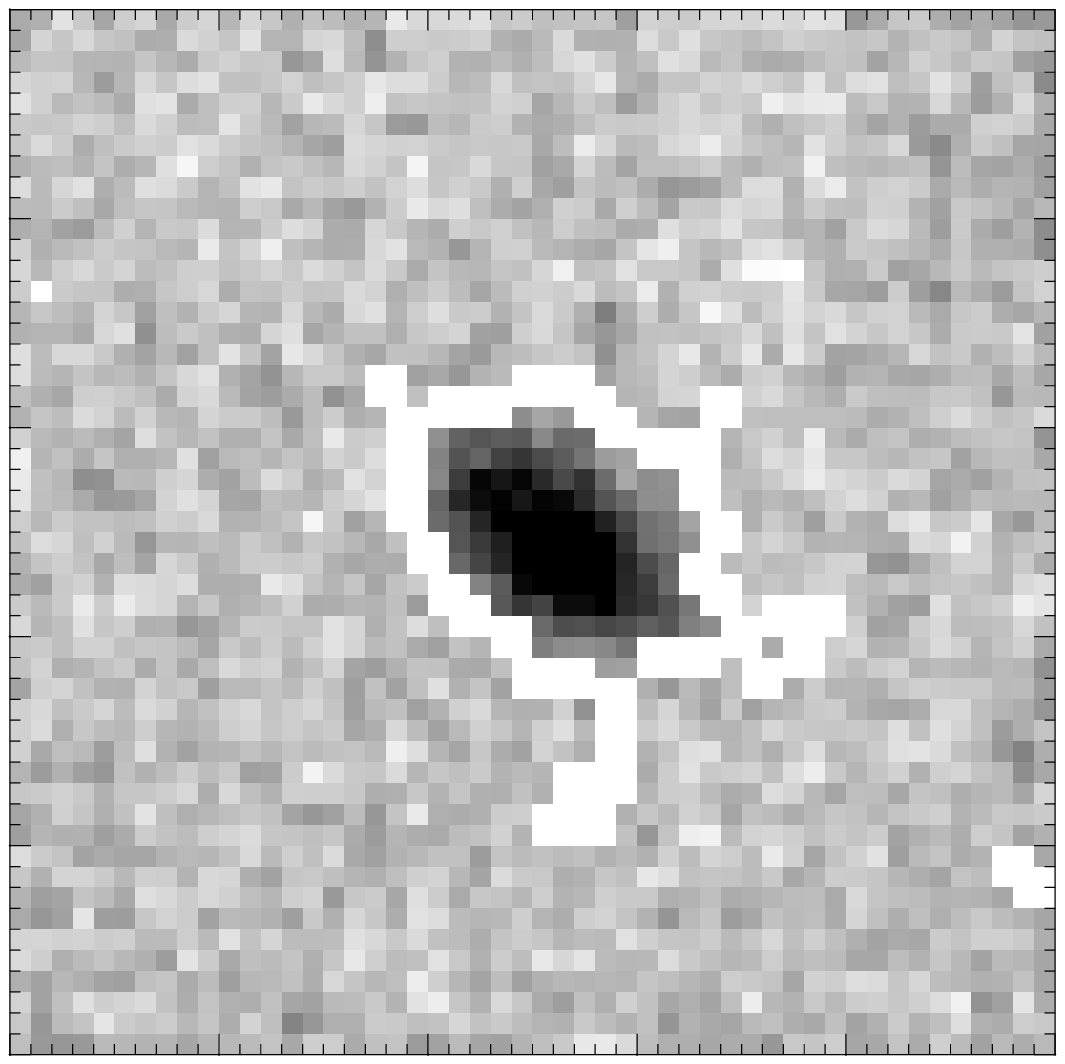} \includegraphics[scale=0.14, trim=77 0 77 0, clip=true]{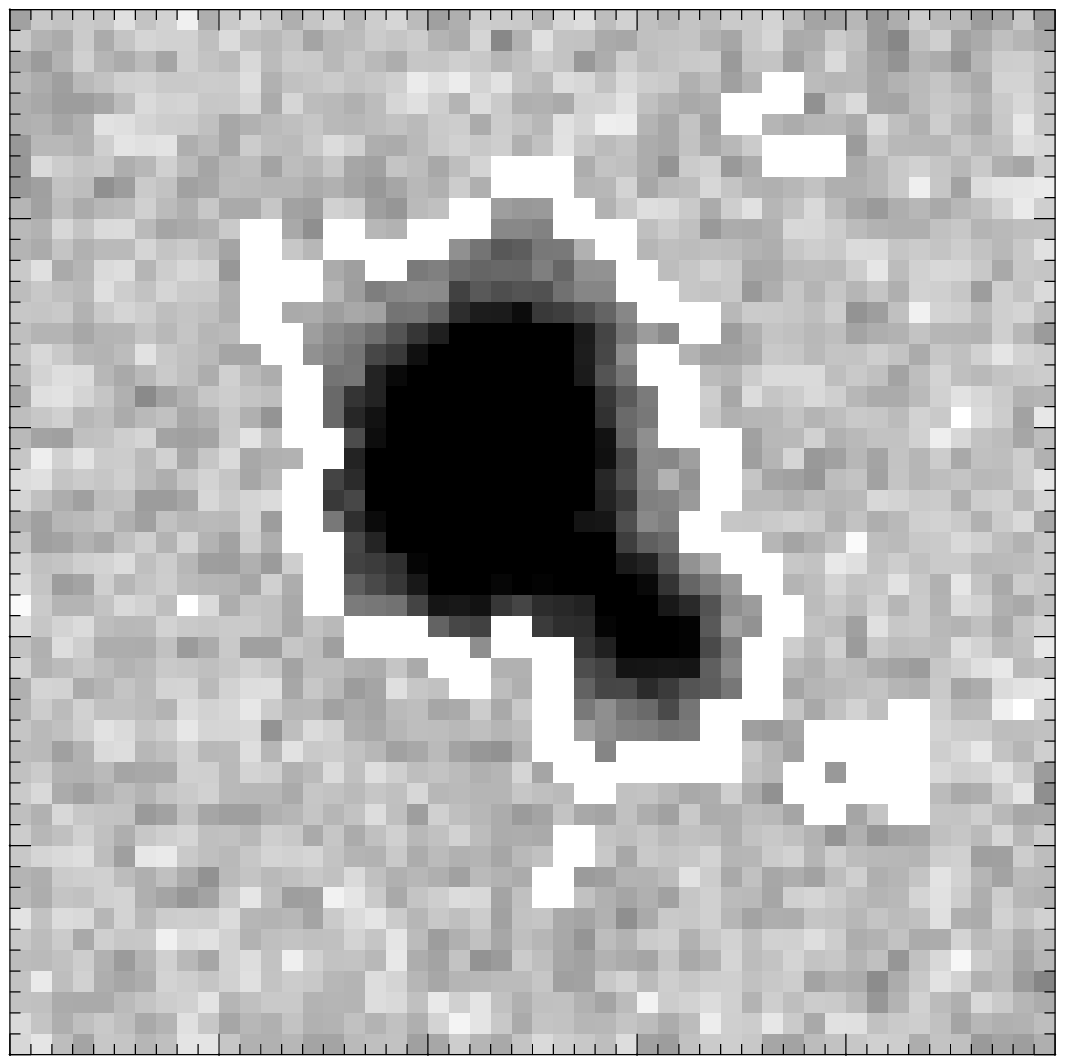} \includegraphics[scale=0.14, trim=77 0 77 0, clip=true]{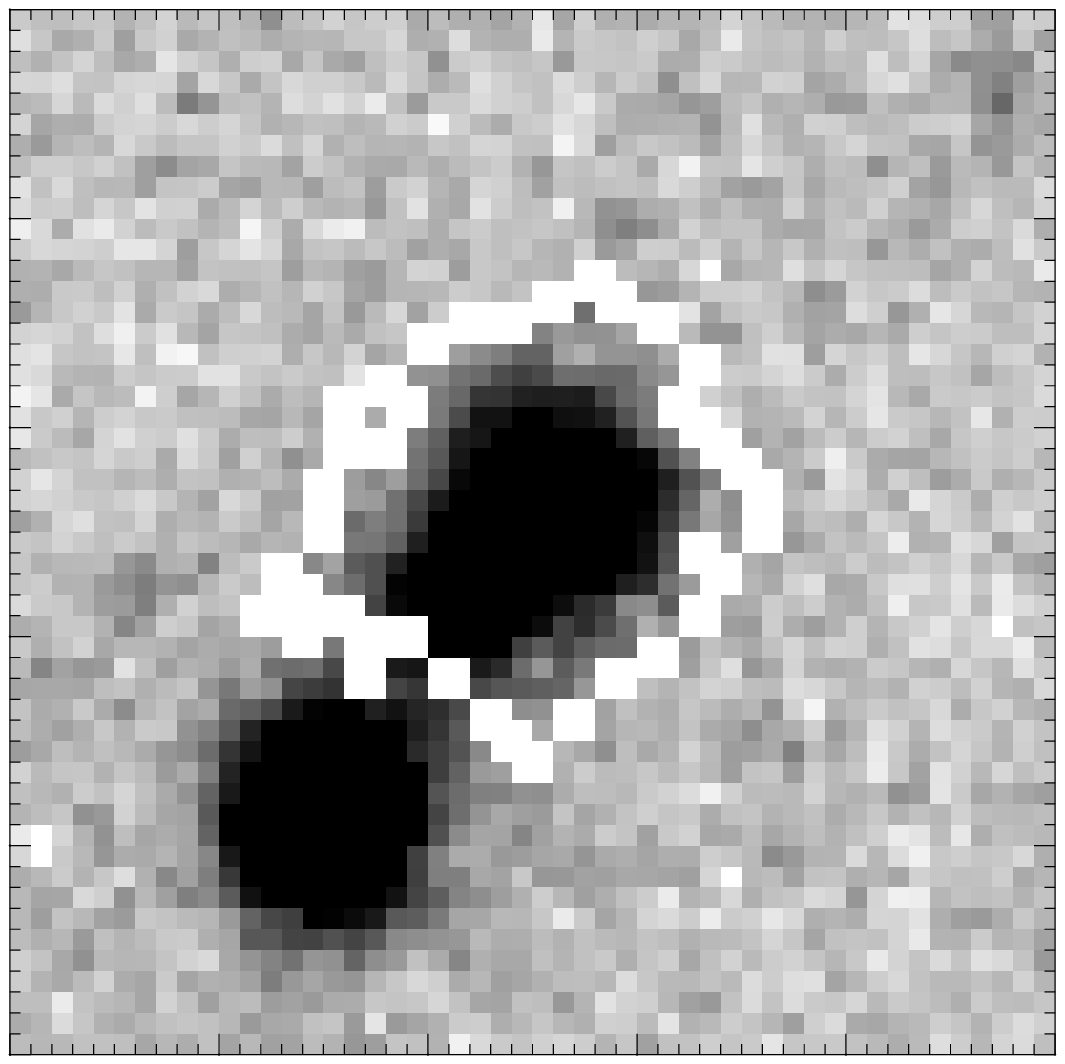} \includegraphics[scale=0.14, trim=77 0 77 0, clip=true]{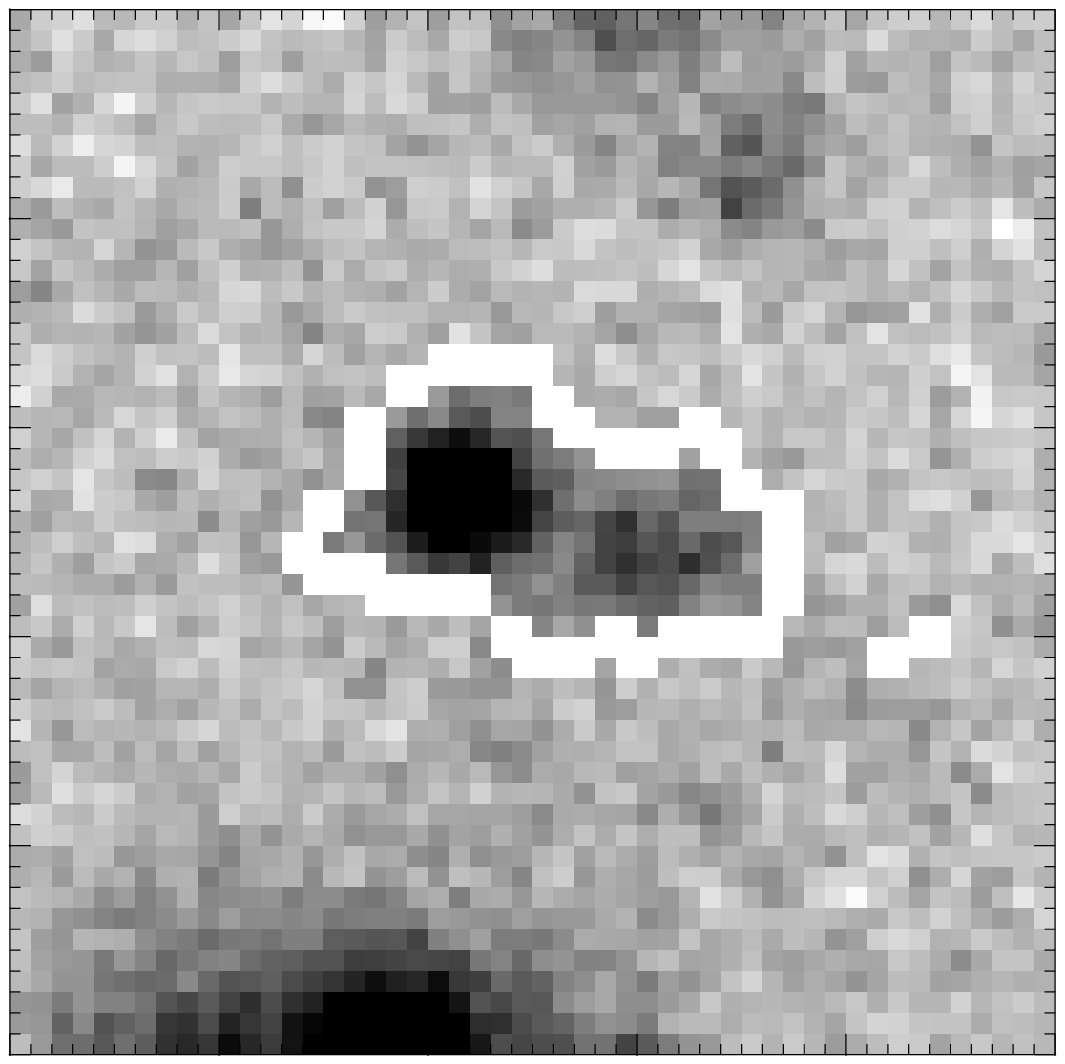} \includegraphics[scale=0.14, trim=77 0 77 0, clip=true]{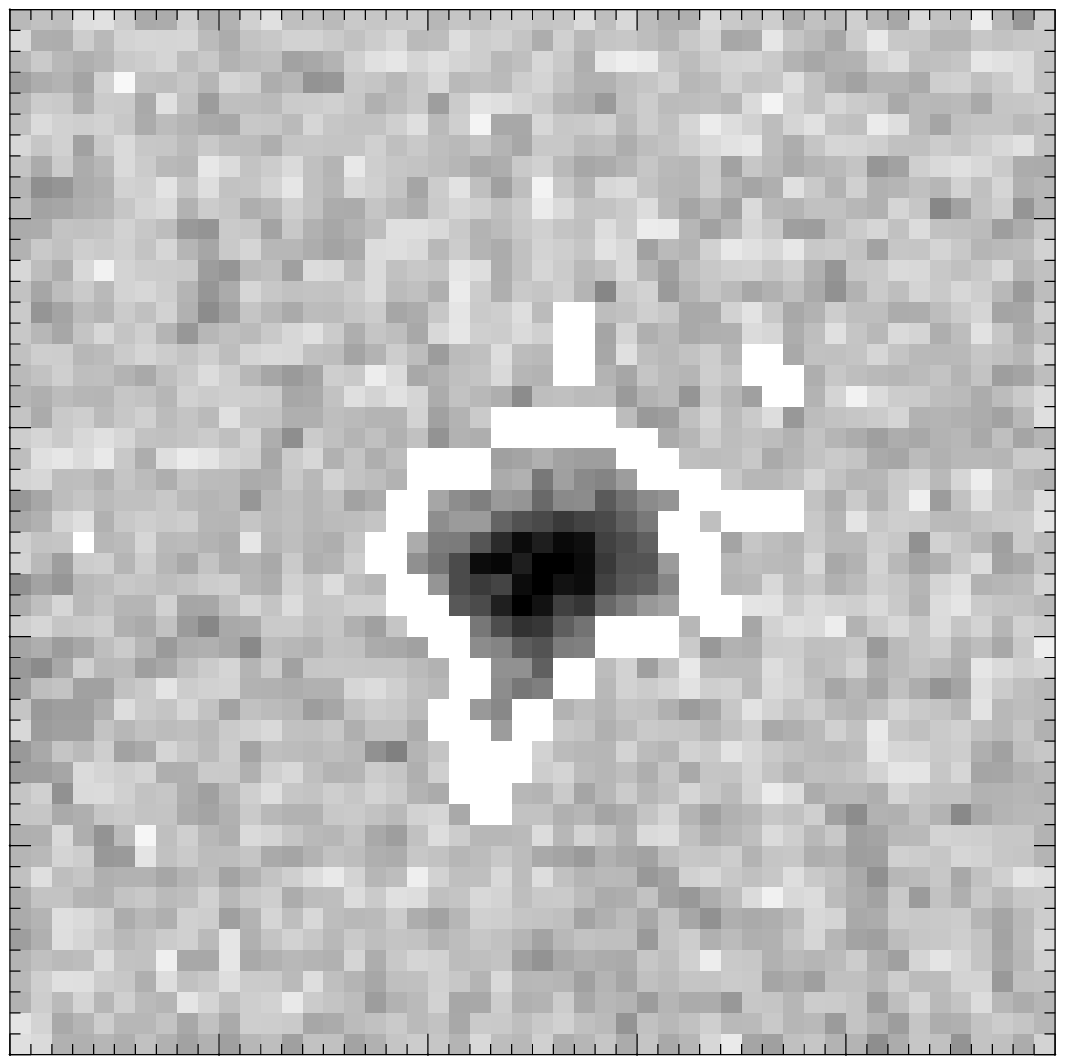}
\hspace*{2em}
%
%
\includegraphics[scale=0.14, trim=77 0 77 0, clip=true]{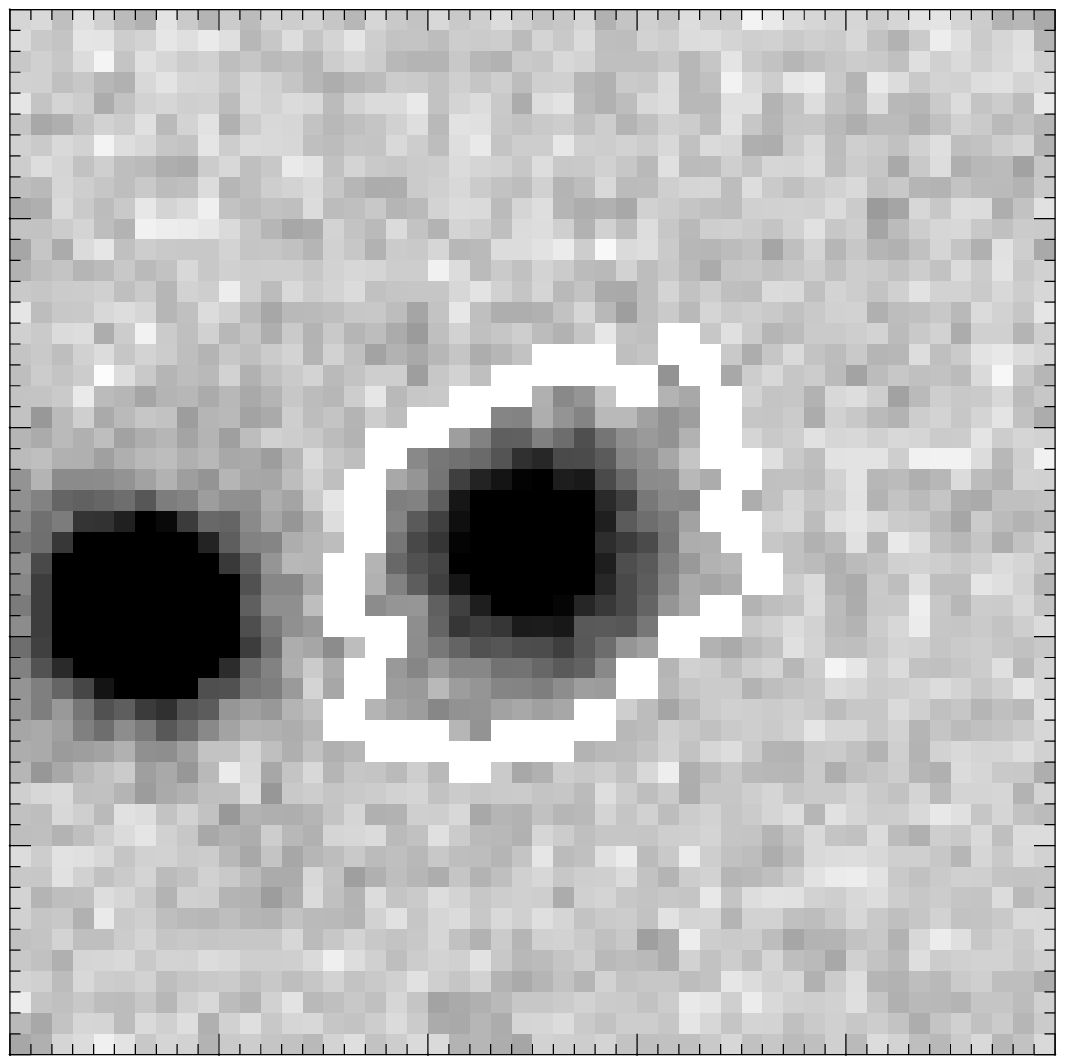} \includegraphics[scale=0.14, trim=77 0 77 0, clip=true]{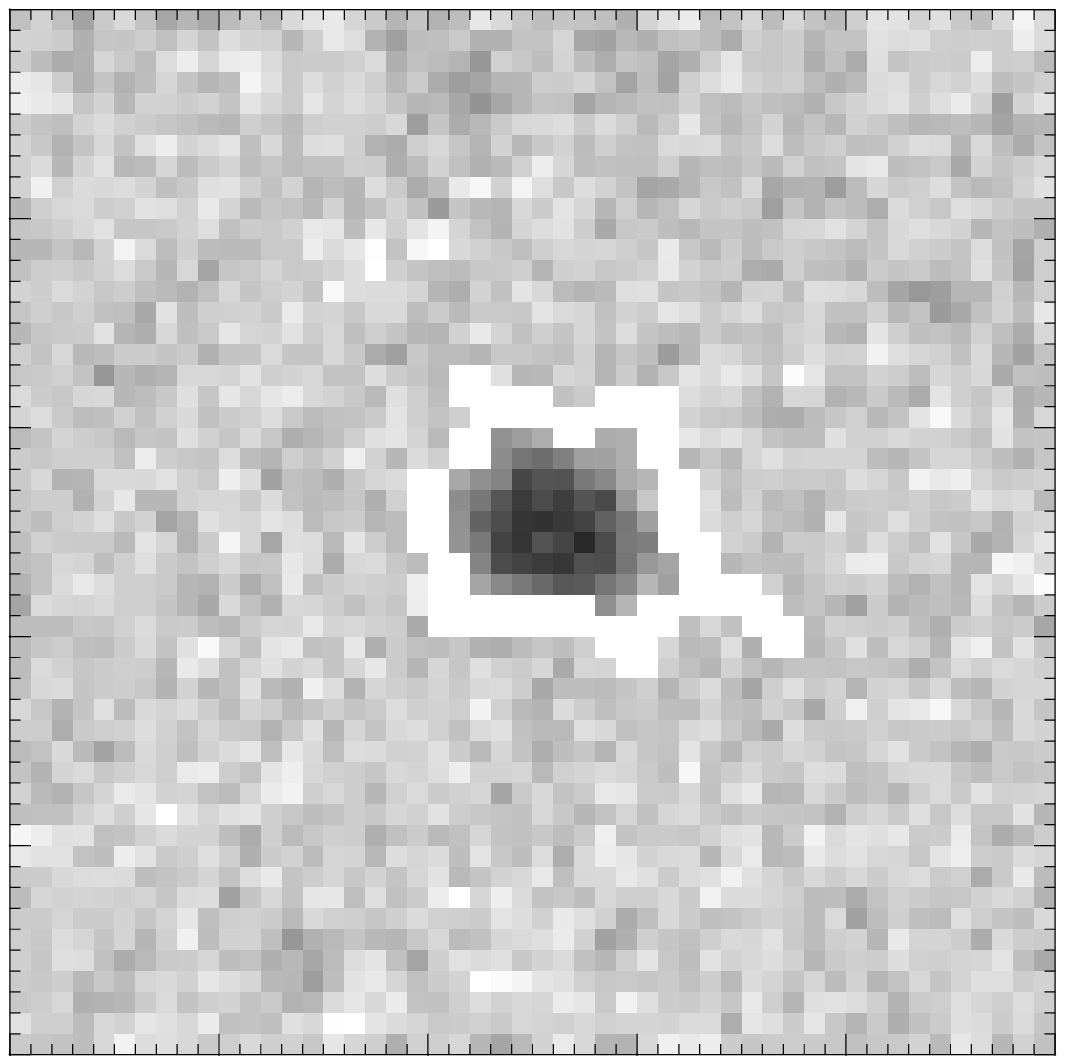} \includegraphics[scale=0.14, trim=77 0 77 0, clip=true]{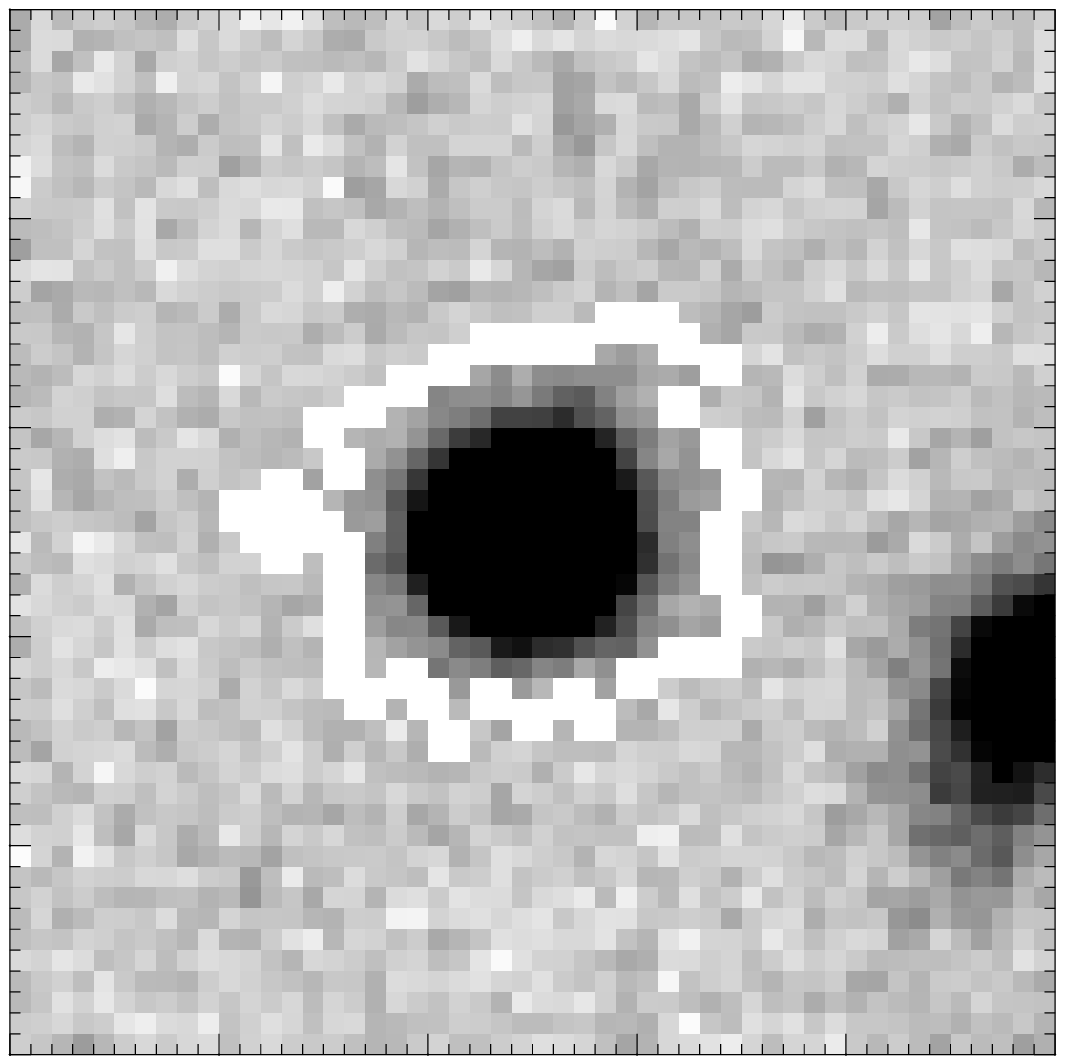} \includegraphics[scale=0.14, trim=77 0 77 0, clip=true]{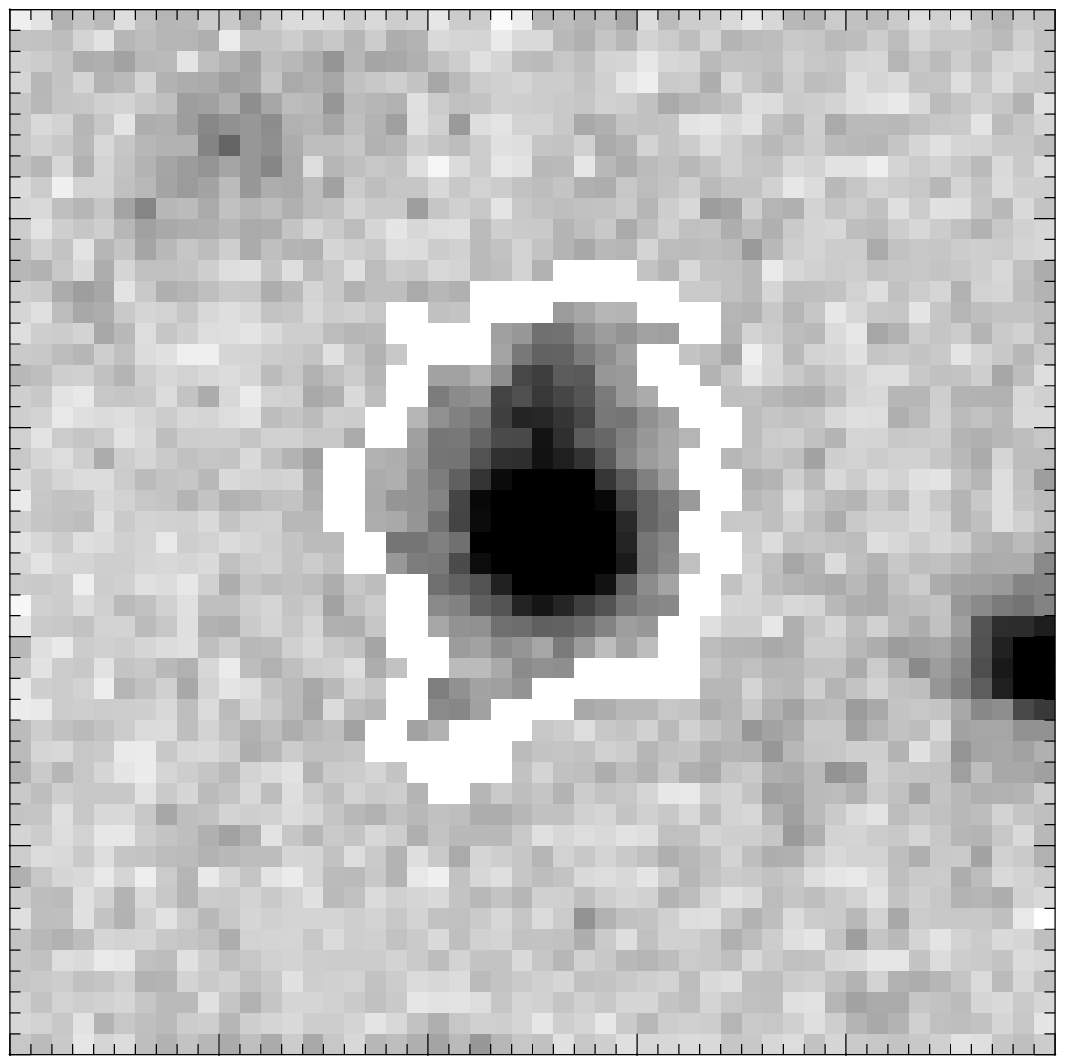} \includegraphics[scale=0.14, trim=77 0 77 0, clip=true]{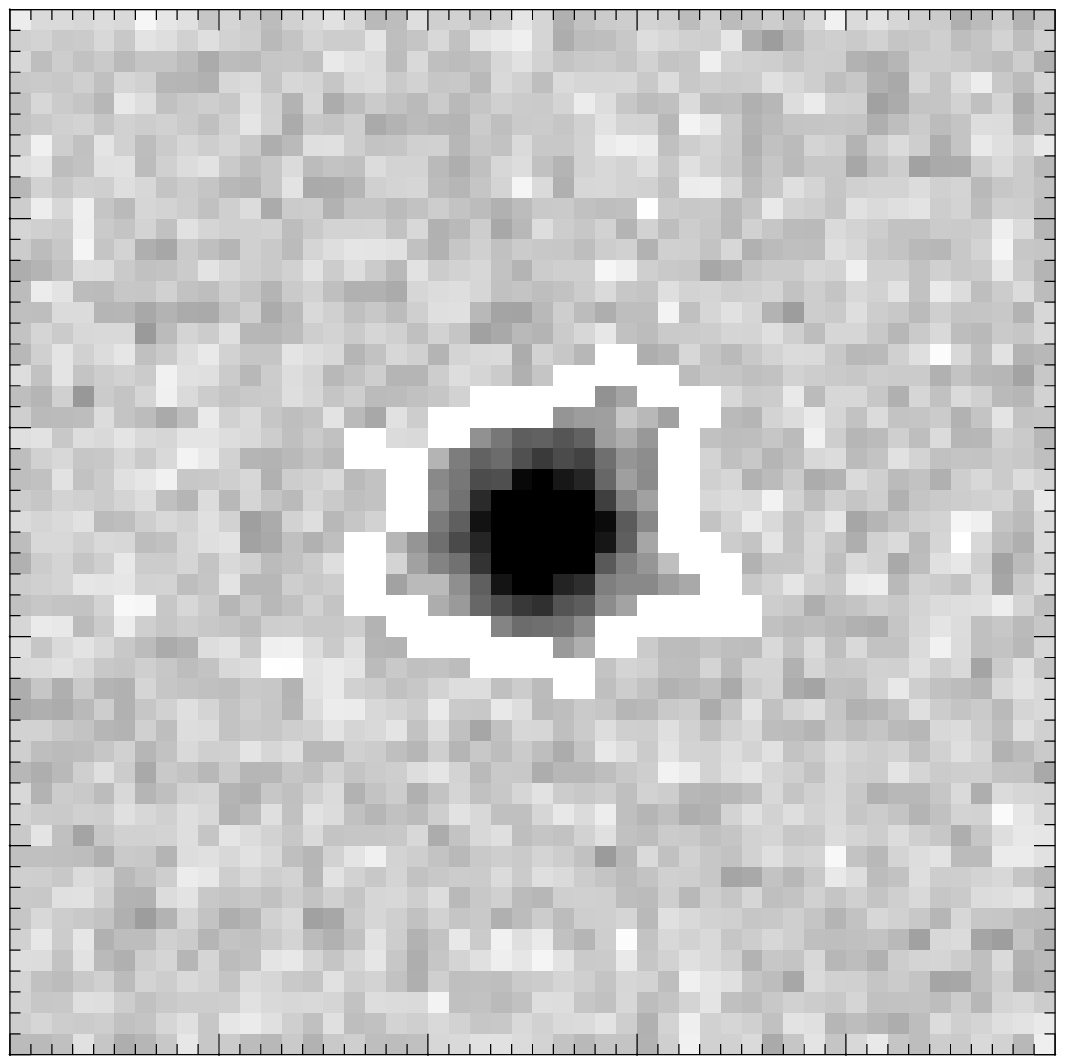}
\caption[]{{\it Left}: Postage stamp images ($10''\times10''$) from both CANDELS {\it HST} F160W ({\it upper ten}) and UDS $K$ ({\it lower ten}) of the same HiZELS galaxies classified by $\rm M_{20}$ as major mergers. {\it Right}: Postage stamp images ($10''\times10''$) from both CANDELS F160W ({\it upper ten}) and UDS $K$ ({\it lower ten}) of the same HiZELS galaxies classified by $\rm M_{20}$ value as non-mergers. It is clear from this plot that mergers are well separated from non-mergers in this morphological classification system and that it is possible to identify mergers from the ground-based imaging. The white outlines represent the {\sc sextractor} segmentation maps used for the morphological analysis.}
   \label{fig:post}
\end{figure*}

\begin{figure}
   \centering
\includegraphics[scale=0.5]{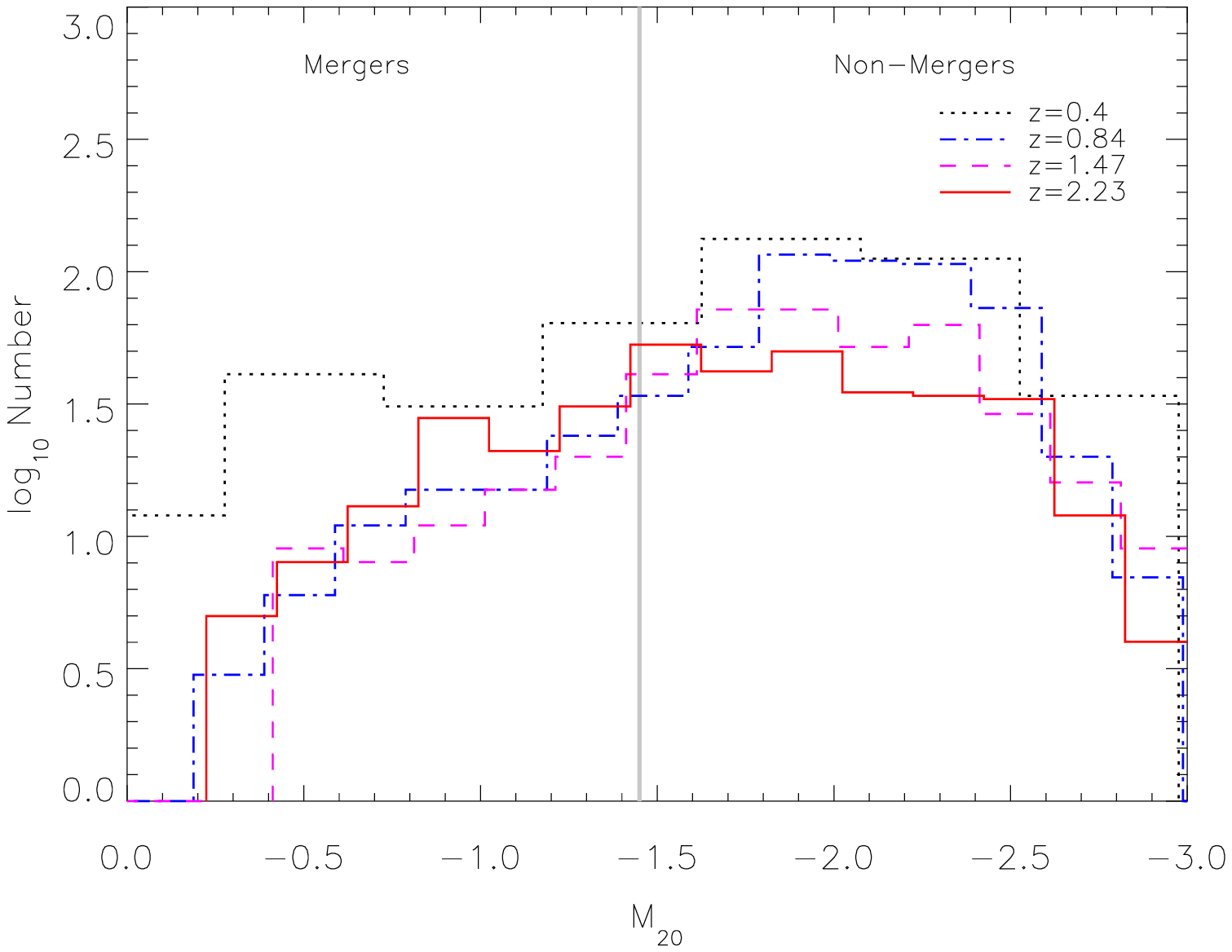} 

\caption[]{A histogram of $\rm M_{20}$ values for the four HiZELS redshift slices. The vertical line at $\rm M_{20}=-1.45$ delineates mergers from non-mergers.}
   \label{fig:m20hist}
\end{figure}

The number densities of galaxies in the HiZELS samples are lower than the $BzK$ morphology calibration sample used in \S\ref{sec:morphcal} and therefore do not have the same level of overlap with the CANDELS imaging region in the UDS. We instead run the morphology codes on the CANDELS, UKIDSS UDS and COSMOS WIRDS imaging for the HiZELS samples at each of the four redshifts. The output Gini and $\rm M_{20}$ values for the ground-based near-infrared imaging are calibrated to the CANDELS values using the fits found for the $BzK$ sample in \S\ref{sec:morphcal}. As a confirmation of the calibration of the ground-based morphologies to those derived from the {\it HST} data the UKIDSS UDS Gini and $\rm M_{20}$ coefficients for those that line in the CANDELS sub-region are plotted in Figure \ref{fig:bzkginim} ({\it lower}) but with the visual classifications derived from the CANDELS data indicated. The result of analysing the morphologies of these calibrated data is that we now choose to delineate the difference between mergers and non-mergers at $\rm M_{20}=-1.45$ which minimises the visual contamination to $22\pm12\%$ non-mergers in the merger region and $15\pm7\%$ mergers in the non-merger region. We note that some of the contamination of visually classified non-mergers to the merger fraction may in fact be due to galaxies with clumpy discs (see \S\ref{sec:morphcal}). Figure \ref{fig:post} displays a sub-sample of the HiZELS galaxies classified by the $\rm M_{20}$ parameter as mergers ({\it left}) and non-mergers ({\it right}) for both ground and space-based imaging, with their {\sc sextractor} segmentation maps over-plotted. As the Gini coefficient is found to add little information, when using our particular analysis methods, Figure \ref{fig:m20hist} presents a histogram of $\rm M_{20}$ values, as measured from the ground-based imaging of the HiZELS population at all redshift slices.

\section{Results}
\subsection{Merger fractions}
\label{sec:mfrac}
Here we define `merger fraction' as the number of galaxies with a merger-like morphology (regardless of how many galaxies actually make up this merger) divided by the total number of galaxies in the redshift slice. The total fraction of mergers for the HiZELS galaxies in the redshift bins $z=0.40, 0.84, 1.47, 2.23$ are 0.33, 0.13, 0.18 and 0.32 respectively (see Figure \ref{fig:m20hist}), however these are not comparable as they are measured for different stellar mass and SFR ranges at the different redshifts. 

For comparison \cite{sobral2009} find a higher merger fraction of 0.28 at $z=0.84$ using the morphological classifications of \cite{scarlata2007} and a visual classification that included mergers and close pairs (which explains the higher merger fraction), although this is from rest-frame $B-$band imaging. However, when we study the COSMOS {\it HST} ACS imaging used in that study we find that many of the galaxies appear as very low surface brightness meaning that their morphological classifications are more uncertain.

\begin{figure*}
   \centering

\includegraphics[scale=0.43, trim=30 0 20 0, clip=true]{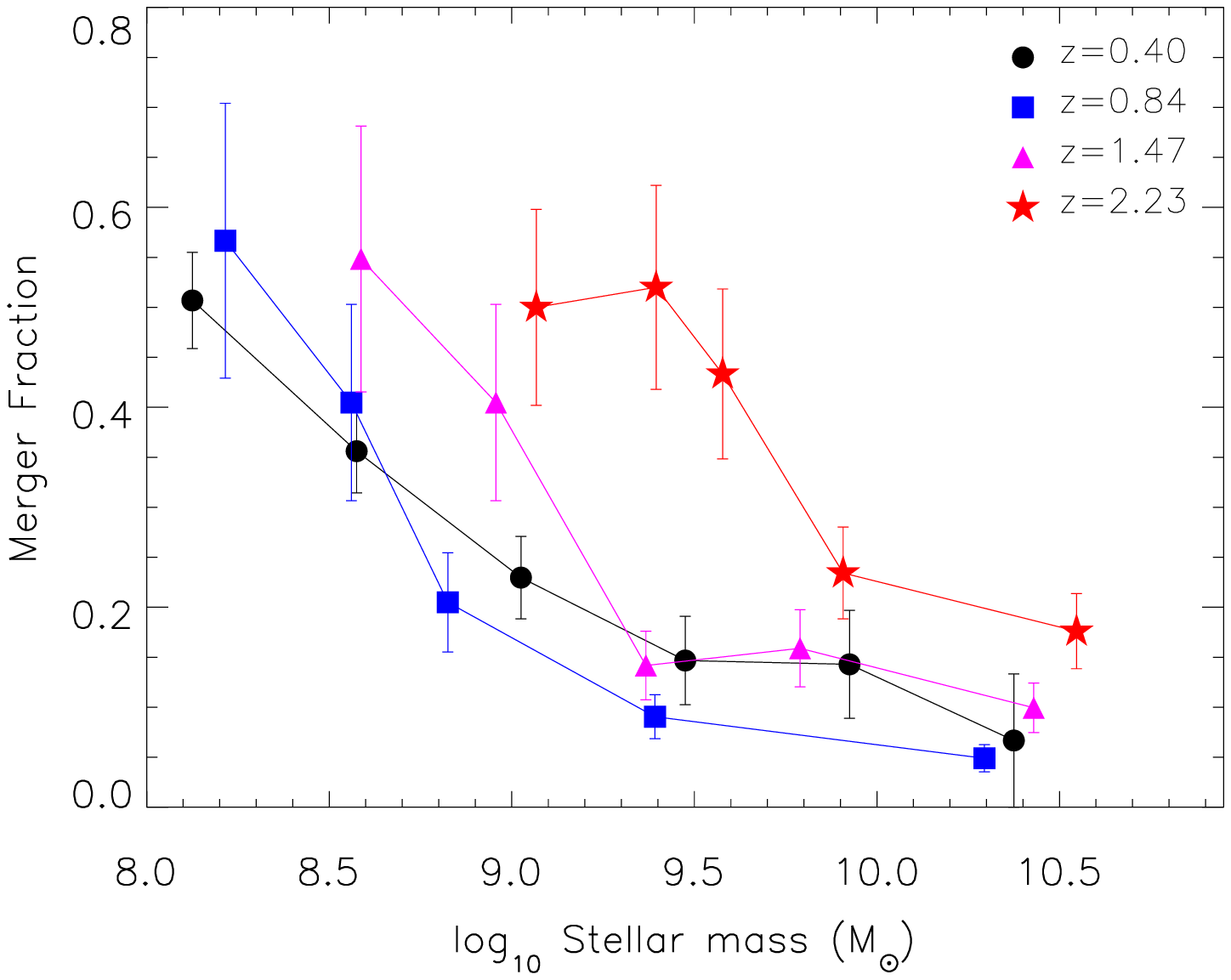}\includegraphics[scale=0.43, trim=77 0 20 0, clip=true]{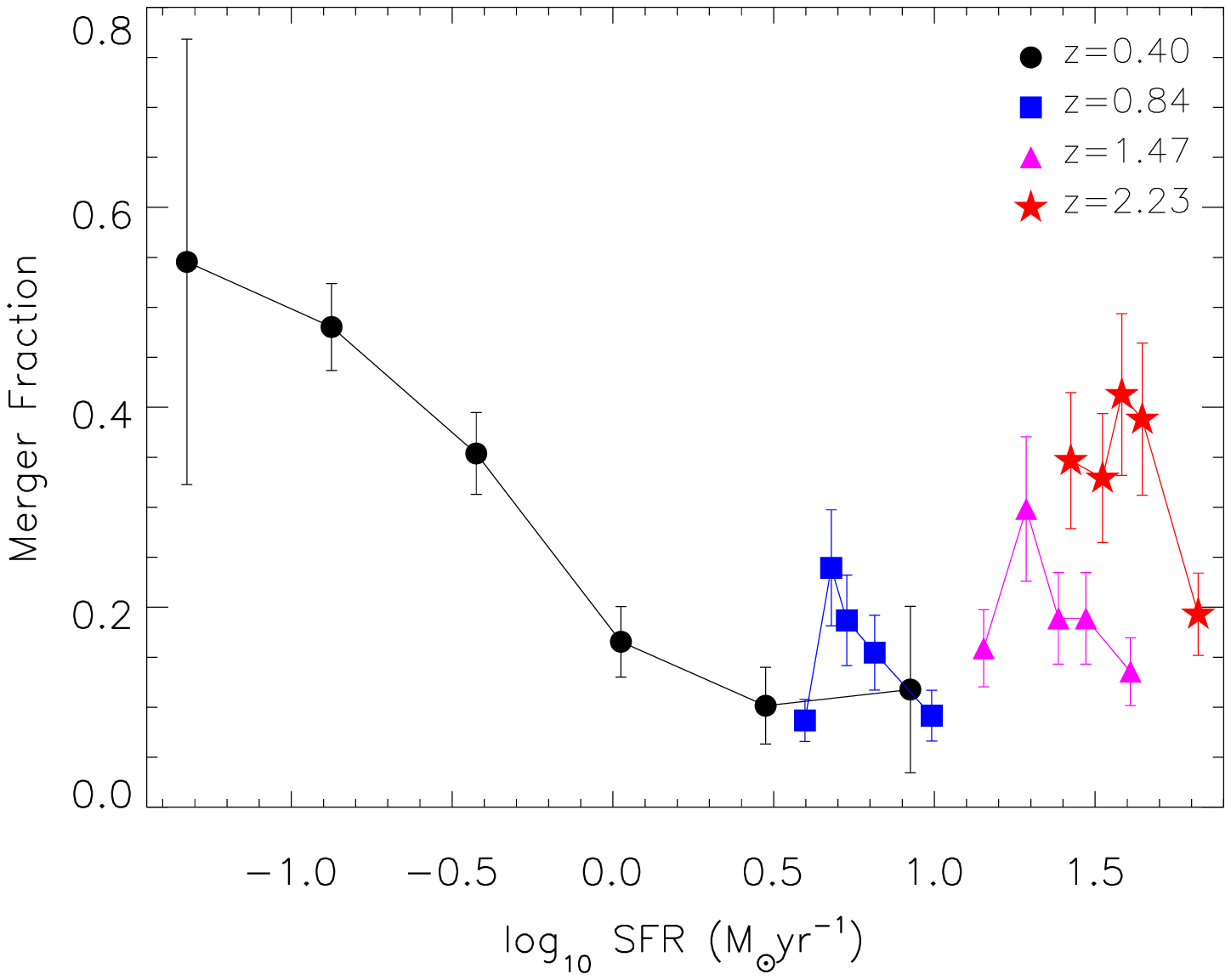}\includegraphics[scale=0.43, trim=77 0 20 0, clip=true]{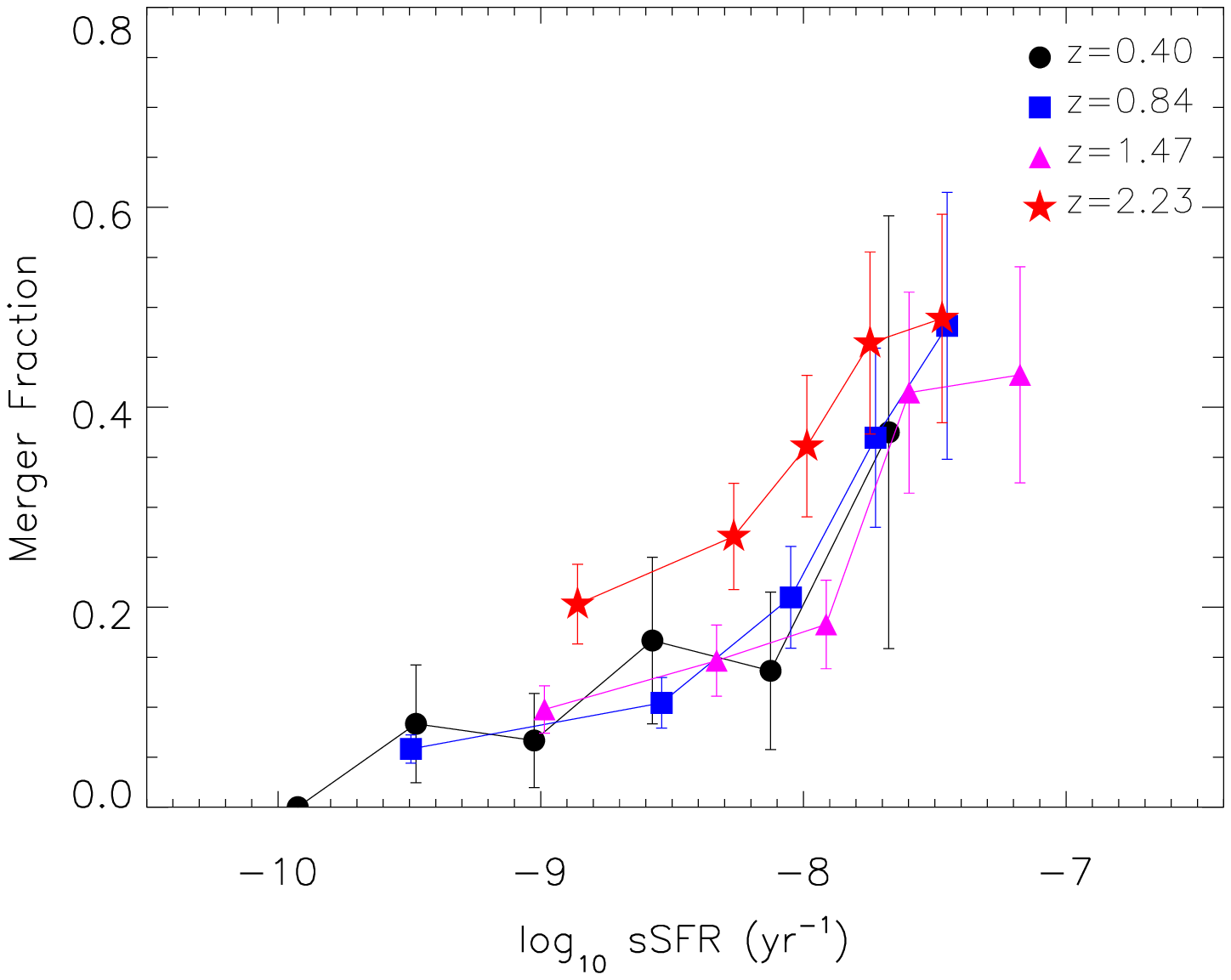}
\caption[]{{\it Left}: Fraction of $\rm M_{20}$ identified mergers versus stellar mass for the four HiZELS redshift slices. {\it Centre}: Fraction of $\rm M_{20}$ identified mergers versus SFR for the four HiZELS redshift slices. From these plots we can see that the merger fraction depends on mass and perhaps SFR with the most massive and most star-forming galaxies having the lowest merger fractions. {\it Right}: Fraction of $\rm M_{20}$ identified mergers versus sSFR for galaxies with $\rm ENSFR>0.2$ for the four HiZELS redshift slices. This suggests that major mergers can lead to galaxies having unusually high sSFR compared to the typical value at a given mass and redshift.}
   \label{fig:mfracmass}
\end{figure*}

In Figure \ref{fig:mfracmass} ({\it left}) the merger fraction is plotted against stellar mass, with the lowest mass galaxies progressively more likely to be classed as mergers with $\sim5-20\%$ of the star-forming population being mergers at the highest stellar masses in each of our redshift slices. The $z=0.4$, 0.84 and 1.47 trends are all remarkably similar and in agreement but there is an increase in merger fraction at all masses to $z=2.23$. However, the HiZELS selection is dependent on SFR, not mass and as described in \S\ref{sec:sample} the typical sSFR for galaxies increases with redshift and therefore we need to investigate these effects too.

A fraction of 10-20\% mergers is seen in the most strongly star-forming galaxies at each redshift (Figure \ref{fig:mfracmass}, {\it centre}). However, due to the flux-limited nature of the samples and the evolution of typical sSFR there is little overlap between different redshifts. In this figure the combined SFR data for all of the HiZELS redshift bins taken at face value may actually hint at a trend in merger fraction with SFR rather than any evolution with redshift (at least out to $z=1.47$). There is some evidence of an increase in merger rate at the same SFR when going from $z=1.47$ to $z=2.23$ but again this does not account for the evolution in typical sSFR. 

Combining the two results above we investigate the relative contribution of mergers to the range of sSFR covered by our sample, for galaxies with $\rm ENSFR>0.2$ to which we are complete at all redshifts, in Figure \ref{fig:mfracmass} ({\it right}). From this plot it is clear that the galaxies with the higher sSFR at all redshifts are increasingly more likely to have a merger-like morphology, with those with the highest sSFR having a merger fraction of $\sim40-50\%$. This strongly suggests that starbursts are more likely to be driven by major mergers when compared to the rest of the star-forming population. This is in agreement with the far infrared selected sample of \cite{kart2012} who find that major mergers have, on average, a high sSFR compared to typical star forming galaxies. It is also in broad agreement with results from the mass selected sample of \cite{kav2012} who also find that major mergers tend to have high sSFR compared to undisturbed galaxies.

We test whether the $A_{\rm H\alpha}=1.0$ dust correction we universally employ is reasonable and how it affects our results. This is by including both SED fit extinction values \citep{sobral2012b} and those derived from the relation between stellar mass and extinction from \cite{garn2010b}. We find that using these more sophisticated estimates makes little difference for the range of masses we study. The value of $A_{\rm H\alpha}$ is $\sim1$\,mag at a mass of $10^{10}\rm M_{\odot}$ with this value increasing/decreasing to higher/lower mass, with the typical range being $A_{\rm H\alpha}=0.5-2$\,mags. In fact when this more sophisticated treatment of dust obscuration is included it acts to strengthen our conclusions by smoothing the relations in Figure \ref{fig:mfracmass}. However, we choose to keep the extinction value at $A_{\rm H\alpha}=1.0$ as this is easier to compare to other works including the main results in \cite{sobral2012b} and to`epoch normalise' with the $\rm H\alpha$ luminosity function.

It is unlikely that HiZELS is missing a large population of `typical' high redshift star forming galaxies with high dust obscurations, as \cite{reddy2012} demonstrate that the dust content of typical star forming galaxies actually decreases with redshift. However, the HiZELS sample may miss extreme star forming and highly obscured galaxies such as sub-mm galaxies. Sub-mm galaxies are found to have a spread in morphologies which is indistinguishable from that of typical star forming galaxies at high redshift \citep{swinbank2010b} and are relatively rare objects (1-2$\times10^{-5} \rm Mpc^{-3}$, \citealt{wardlow2011}), thus their omission would not affect our conclusions.

\subsection{Merger rates}
\label{sec:mrate}
To calculate the merger rates (the number of mergers per Gpc$^3$ per Gyr) we follow the prescription outlined in \cite{lotz2011}: that the merger rate is simply the number of mergers per Gpc$^3$ divided by the average timescale over which the merger would be observed. In \cite{lotz2011} this observed merger timescale is found, from simulations, to be $\sim0.2$\,Gyr, when the Gini/$\rm M_{20}$ method is employed. We adopt this value for consistency with that study and with the data from other groups recalculated and used there. We note that as HiZELS is a narrow-band survey the volumes covered at each redshift slice are well defined with values of $\sim1-7\times10^{-4}\rm \,Gpc^3$ \citep{sobral2012b}. We now also assume that there are on average two galaxies per merger for consistency with other studies.  

\begin{figure*}
   \centering
\includegraphics[scale=0.5]{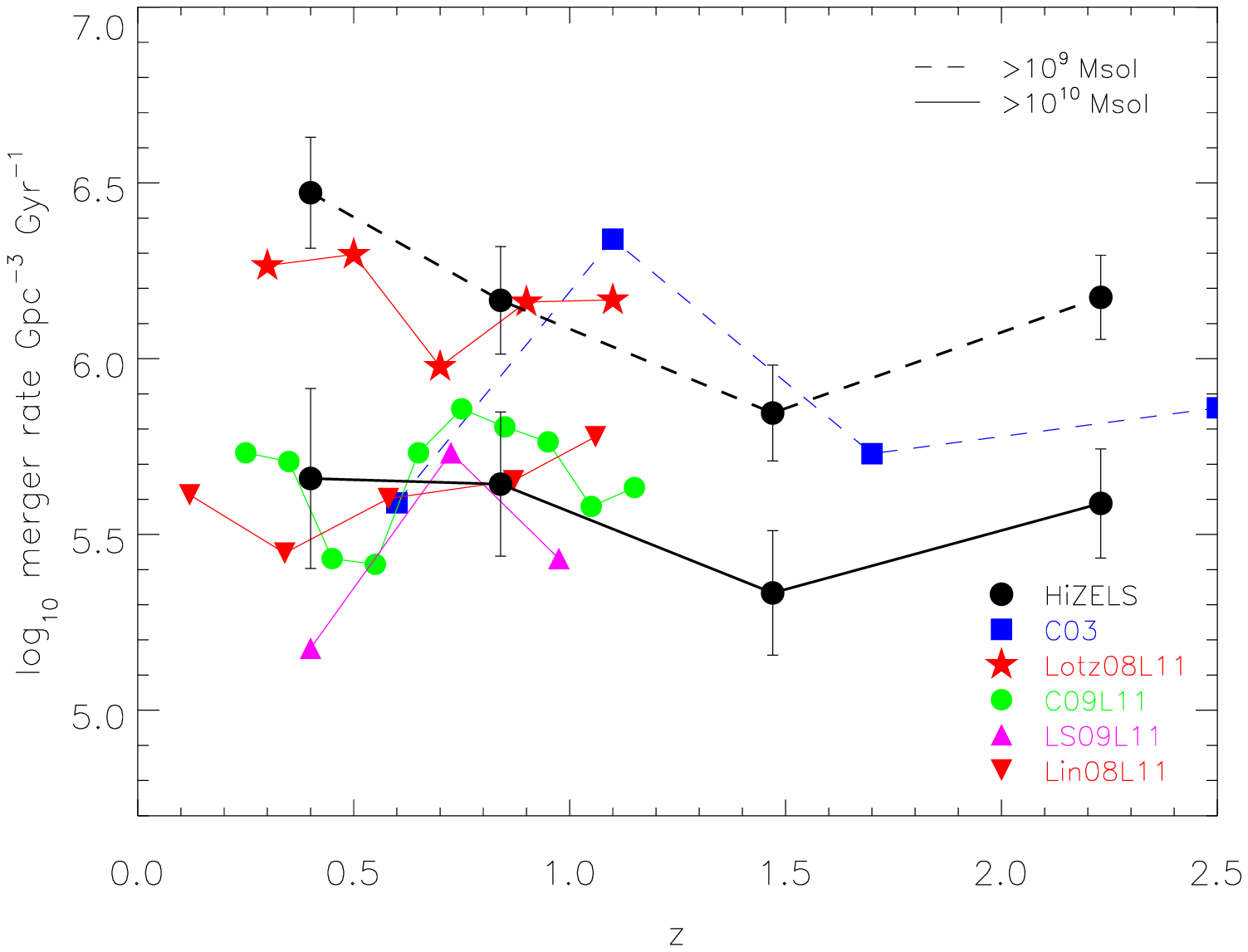} \includegraphics[scale=0.5]{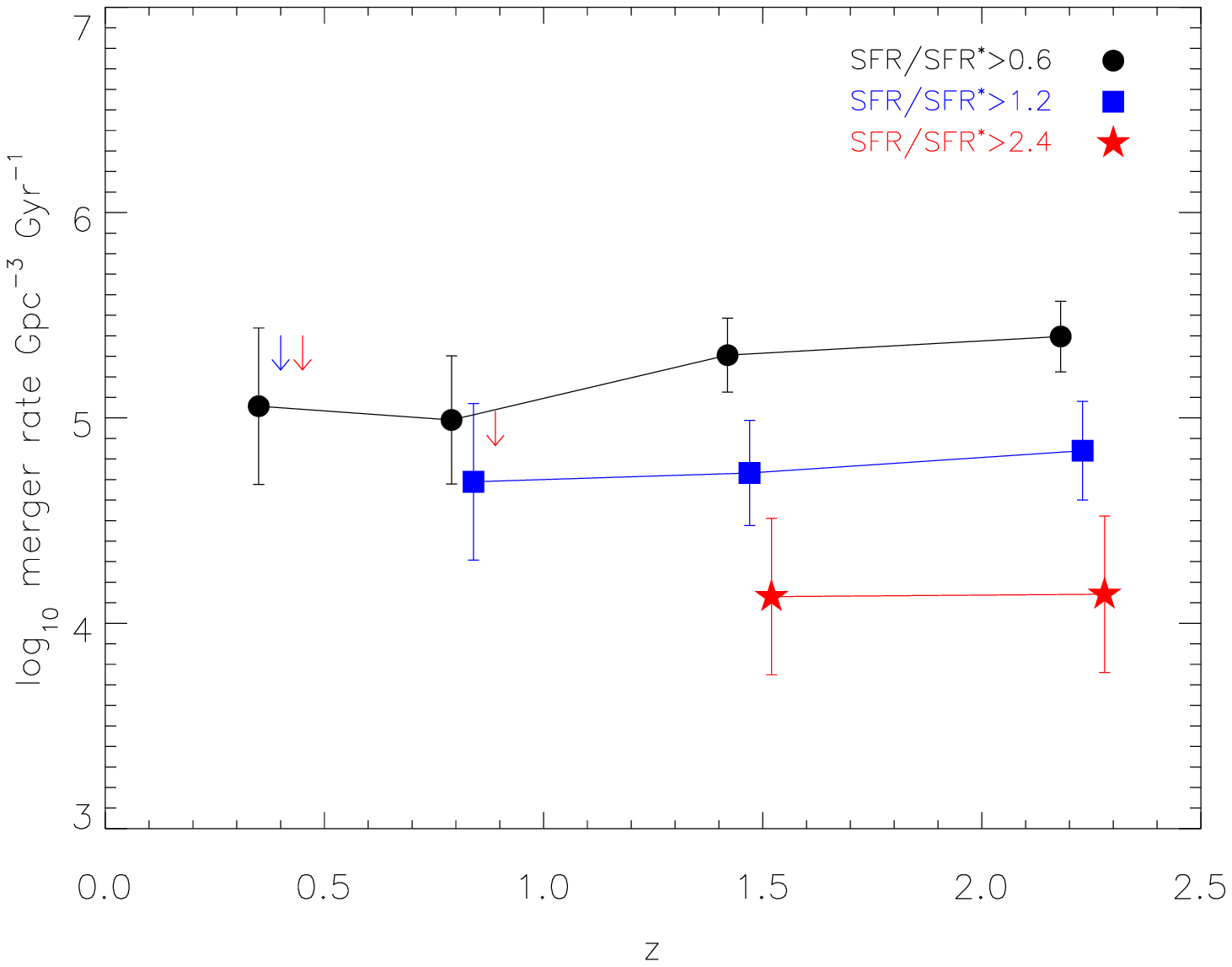}
\caption[]{{\it Left}: Merger rates for the HiZELS sample above a given mass against redshift. For comparison, we include merger rates derived from: close pairs (\citealt{lin2008}, Lin08L11); Gini/$\rm M_{20}$ (\citealt{lotz2008}, Lotz08L11); and galaxy asymmetry,  (\citealt{conselice2003,conselice2009,lopez2009}, labelled C03, C09L11 and LS09L11 respectively). The L11 denotes that these merger rates were originally sourced from their respective papers but have been corrected to the timescales calculated by \cite{lotz2011} using the galaxy evolution models of \cite{somerville2008}. The samples of Lin08L11, Lotz08L11, C09L11 and LS09L11 are all at $\rm M_{\star}>10^{10} \rm M_{\odot}$ while C03 is $M>10^{9} \rm M_{\odot}$. {\it Right}: The merger rates for HiZELS galaxies with $M_{\star}>10^{10} \rm M_{\odot}$ above a given epoch normalised star formation rate ($\rm ENSFR=SFR/SFR^{\star}({\it z})$). The points are offset by $\Delta z$ for clarity. From these plots one can see that there is no evidence for a significant evolution in merger rate when both the mass and the ENSFR of the galaxies are accounted for in the selection.}
   \label{fig:mrate}
\end{figure*}

For comparison with other surveys we initially cut our sample only on stellar mass. The merger rates for galaxies with $M>10^{9} \rm \,and \,10^{10} M_{\odot}$ are plotted in Figure \ref{fig:mrate} ({\it left}). Also plotted are values from \cite{conselice2003} ($M_{B}<-19$ which approximates $M>10^{9} \rm M_{\odot}$) and those with $M>10^{10} \rm M_{\odot}$ derived from Gini/$\rm M_{20}$ \citep{lotz2008}, close pairs  \citep{lin2008} and galaxy asymmetry from \citealt{conselice2009} and \cite{lopez2009}. These merger rates are corrected to the timescales calculated by \cite{lotz2011} using the galaxy evolution models of \cite{somerville2008}. 

Figure \ref{fig:mrate} ({\it left}) shows little evolution in merger rate with redshift and the results are generally in good agreement with those found in the studies of \cite{conselice2003,conselice2009,lin2008,lopez2009} where the redshift ranges overlap. The merger rates from \cite{lotz2008} are systematically higher, which may be because that sample is mass selected and therefore includes a significant contribution from merging red sequence galaxies which would not have been included in the HiZELS sample. There could also be secondary effects due to a mismatch in the stellar mass calculation between the studies, a different way of defining mergers through the $\rm M_{20}$ parameter or a differential in the timescales involved, so an offset is perhaps not unexpected. With the exception of the $z=0.4$ data point, which is significantly higher, the HiZELS merger rates for galaxies with $M>10^{9} \rm M_{\odot}$ are also in good agreement with those of the only study with this approximate mass limit \citep{conselice2003}. From this plot there is no strong evidence for an increase in the merger rate for mass-selected samples out to $z\sim2$. However, this comparison does not account for the SFR limits of the different surveys or the increase in typical sSFR with $z$ \citep{elbaz2011}.

The advantage of HiZELS over these earlier studies is that it is unbiased with respect to stellar mass and we derive the stellar mass and SFR from independent measurements, i.e. SED fitting and $\rm H{\alpha}$ flux. We can therefore consider both SFR and stellar mass independently to split the population into sub-samples based on these properties. As defined in \S\ref{sec:sample} we account for the increase in the typical sSFR with redshift by employing the ENSFR. Figure \ref{fig:mrate} ({\it right}) shows the population split into three ENSFR bins $>0.6$, 1.2 and 2.4, for which the HiZELS observations are complete at all redshifts. The first obvious thing to notice is that undulating shape of the plot with just a mass cut (Figure \ref{fig:mrate}, {\it left}) has disappeared. Instead the trends are flat, showing no evidence for an increase in the merger rate with increasing redshift for all masses and ENSFR cuts. This mass and ENSFR selected sample is a cleaner sample than those in Figure \ref{fig:mrate} ({\it left}) and so we suggest that the peak in the merger rate at $z\sim1$ seen for some comparison samples may be due to the mixing of a mass limit with an SFR selection function which strongly effects photometrically-selected galaxies.

From this merger analysis we can determine the total number of major mergers (with mass ratio $>1:10$) a galaxy of a given mass will undergo during the epoch covered by our study. Using Equation 11 from \cite{conselice2006} we find that one would expect $\sim3$ mergers per star-forming galaxy with $M\sim10^{10} \rm M_{\odot}$ between $z=2.23$ and $z=0.4$, or a merger every 2 Gyrs on average. We note that these numbers depend on the value of $\tau$ the timescale over which mergers can be observed using the $\rm M_{20}$ method (which we assume to be 0.2\,Gyr, \citealt{lotz2011} ) and therefore more generally there are $0.6\tau_{0}\tau^{-1}$ mergers between $z=2.23$ and $z=0.4$, corresponding to $0.1\tau_{0}\tau^{-1}$ mergers per Gyr, where $\tau_{0}=1\,\rm Gyr$. 

\section{Discussion \& Conclusions}
\label{sec:disc}

The HiZELS narrow-band H$\alpha$ survey selects star-forming galaxies within four well-defined volumes at $z\sim0.4-2.2$ and flux limits with an SFR indicator which is unbiased in terms of stellar mass and is independent of its determination. In this paper we have used these properties to understand the star-forming population and its merger rate to help illuminate the processes responsible for the up-turn in the SFRD with redshift.

By defining the epoch-normalised star-formation rate ($\rm ENSFR=SFR/SFR^{\star}(z)$) we account for the increase in the typical star-formation rate of galaxies with redshift. In \S\ref{sec:sample} we demonstrate that the number of galaxies above a given mass and ENSFR does not evolve significantly over the $\rm 6\,Gyr$ from $z=0.4$ to 2.23. We also note that the HiZELS sample has already been shown to accurately trace the increase of the SFRD with redshift and that there is no strong evolution in the normalisation of the H$\alpha$ luminosity function \citep{sobral2012b}. Taken, in combination this means the increase in the SFRD with redshift is not due to an increase in the number of star-forming galaxies of a given mass but instead must result from an increase in the amount of star formation in these galaxies. This can be described as an increase in the average sSFR for star-forming galaxies \citep{rodig2010,elbaz2011} without a significant increase in their number density. Also, we note that the $\rm SFR^{\star}$ (derived from $L\rm_{H\alpha}^{\star}$) evolves in the same way as the typical sSFR for star forming galaxies \citep{elbaz2011}, which implies that the luminosity of the knee in the H$\alpha$ luminosity function is evolving significantly more rapidly than the characteristic mass of the stellar mass function.

The size--mass relation for galaxies is assessed in \S\ref{sec:size}. In order to do this for a large sample we need to use wide-field ground-based imaging. Hence we confirm that we can reliably recover the galaxy size determined from the {\it HST} CANDELS imaging by deconvolving the affect of atmospheric seeing from the ground-based imaging. We find that the size--mass relation is surprisingly constant out to $z=2.23$, in agreement with the findings of \cite{barden2005, ichikawa2012} and at odds with the results of \cite{trujillo2007,mosleh2011}. The lack of strong size evolution at a given mass and the universal size--mass relation for star-forming galaxies in the range $0.4<z<2.23$ suggests that this population have not experienced significant size evolution, through mergers or star formation, during this period. Any evolution that does occur must thus act to move the galaxy along the locus of the relation. The slope of this relation is also shallow and thus low mass galaxies are not dramatically smaller than their higher mass counterparts. Even if there is no direct evolutionary connection between the galaxy populations at each epoch then this lack of change in typical size suggests a universal evolution scenario.

In order to study the merger rates of the HiZELS galaxies we test the Gini and $\rm M_{20}$ coefficients. By investigating these automated methods of determining merger classifications we find that {\sc sextractor} parameters that define the segmentation map employed in these analyses are the most important factor in how well the method performs (see Appendix \ref{sec:sim}). We find that, for the segmentation maps generated by our set of {\sc sextractor} parameters, the best delineation between mergers and non-mergers is $\rm M_{20}=-1.45$ while the Gini coefficient provides no useful information. We acknowledge that other authors have found this not to be the case with $\rm M_{20}$ and Gini being equally important in morphological classification \citep{lotz2004,wang2012} but we assume this is due to the differences in the construction of the segmentation maps (see Appendix \ref{sec:sim}) and potentially minor variations in the normalisation of the $\rm M_{20}$ and Gini values, depending upon the exact nature of the morphological code used. The $\rm M_{20}$ coefficient is found to be sensitive to mergers down to a mass ratio of $\sim1:10$  (in agreement with \cite{lotz2010}). We note here that not all mergers are star forming and as such we will miss `dry' mergers which do not induce activity, although obviously these will not be major contributors to the SFRD. As with the sizes we find that it is possible to use this morphological classification on ground-based data affected by atmospheric seeing, after applying a calibration derived from galaxies that are observed with both ground-based telescopes and {\it HST}. 

For the sample as a whole, without accounting for the $\rm H\alpha$ flux (SFR) limit or the increase in the sSFR of the star forming galaxies with redshift, we find that the merger fraction anti-correlates with both stellar mass and SFR. By combining these two results we find that the merger fraction correlates strongly with sSFR. This suggests that the more rapid the star formation is, the more likely it is to be driven by violent major mergers than secular processes. In fact we find that, $\sim50\%$, of starburst galaxies in our $z\sim2$ sample have major merger morphologies. Therefore to achieve such high sSFR, these galaxies are undergoing major merger driven and not `main-sequence' star formation. Interestingly we see no evolution in the merger fraction of starbursts with a constant $\sim40-50\%$ across all redshifts which suggests that merging is a universal process that can lead to a galaxy having enhanced sSFR for their epoch \citep{hopkins2006,kart2012}. 

Finally we consider the merger rates of star-forming galaxies initially only limiting our sample on stellar mass, where some previous studies have seen the characteristic merger rate increase to $z\sim1$. However these other studies use photometrically selected samples where the method of determining stellar mass is directly linked to the determination of the SFR. As these two parameters are independent in HiZELS we can also select on SFR for a fair comparison across the redshift range, while also accounting for the increase in sSFR for typical star-forming galaxies with redshift. By applying these selections we see little evidence for an increase in the merger rates of typical galaxies over the redshift range considered. Therefore even though there is an order of magnitude increase in typical SFR across the redshift range of our study this is not reflected in the merger rate. This is strong evidence that it is not major mergers that drive the increase in the SFRD with redshift, in contrast to the models of \cite{somerville2001} or \cite{hopkins2006} and as observed in part by \cite{conselice2003,conselice2008} and \cite{lin2008} who find some evidence for an increase in merging. Our result agrees with \cite{sobral2009} who find that the increase in SFRD between $z=0$ and $z=0.84$ was primarily due to regular (non-merging) galaxies.

Depending on the timescale $\tau$ for which it is possible to view a galaxy undergoing a major merger using the $\rm M_{20}$ parameter we find that star forming galaxies with mass $>10^{10} \rm M_{\odot}$ undergo $\sim0.6\tau_{0}\tau^{-1}$ (3 if $\tau=0.2$\,Gyr) mergers between $z=2.23$ and $z=0.4$, corresponding to $\sim0.1\tau_{0}\tau^{-1}$ (0.5 if $\tau=0.2$\,Gyr) mergers per galaxy per Gyr, where $\tau_{0}=1\,\rm Gyr$. From analysis of the mass function of galaxies in COSMOS at $z=0.35$--0.75, \cite{pozzetti2010} find merger rates of $\sim 0.1-0.4$ per galaxy per Gyr for galaxies with masses $\sim10^{10.5}-10^{11} \rm M_{\odot}$, in reasonable agreement with our findings (both are very sensitive to the choice of $\tau$). From a theoretical point of view \cite{hopkins2010} compile data from a number of simulations and models (see references therein). The predicted number of mergers per galaxy with mass $\sim10^{10}-10^{11} \rm M_{\odot}$ per Gyr is found to increase with redshift from a value of $\sim0.05$ at $z=0.4$ to $\sim0.25$ at $z=2.2$ apparently lower than the values we find. Again, this is dependent on $\tau$ so we are unable to provide solid constraints.

In summary we find that the increase in SFRD is due to an increase in the sSFR of typical star-forming galaxies. The process responsible for this increase is not major mergers as we find that the merger rate does not increase in step with the SFRD. We therefore conclude that secular processes such as disc instabilities and/or an increase in the effective fuel for star formation are the main driver of the increase in the SFRD with redshift as predicted or observed by others \citep{keres2005,dekel2009,bower2006,forster2011,Cacciato2012}. Although we note that it could also be driven by an increase in the minor merger rate (mass ratios $<1:10$) which this study is not sensitive to. 

We also find a constant merger fraction for starburst galaxies, in that around half are major mergers across all redshifts, demonstrating that extremely violent events are required for a galaxy to attain enhanced sSFR for their epoch and leave the `main-sequence'. Bringing these results together along with the lack of size evolution since at least $z=2.23$ we can say that many of the properties of star forming galaxies are surprisingly constant over the $\sim6$ Gyr covered in this study.

\vspace{1in}
\noindent{\bf ACKNOWLEDGEMENTS}

We first thank the anonymous referee for improving the clarity of this paper. J.P.S. acknowledges STFC for financial support. D.S. acknowledges the award of a NOVA fellowship. I.R.S. acknowledges support from STFC and the Leverhulme Trust. We also thank James Mullaney and Mark Swinbank for useful discussions.

The United Kingdom Infrared Telescope is operated by the Joint Astronomy Centre on behalf of the Science and Technology Facilities Council of the U.K.

Based on observations obtained with WIRCam, a joint project of CFHT, Taiwan, Korea, Canada, France, at the Canada-France-Hawaii Telescope (CFHT) which is operated by the National Research Council (NRC) of Canada, the Institute National des Sciences de l'Univers of the Centre National de la Recherche Scientifique of France, and the University of Hawaii. This work is based in part on data products produced at TERAPIX, the WIRDS (WIRcam Deep Survey) consortium, and the Canadian Astronomy Data Centre. This research was supported by a grant from the Agence Nationale de la Recherche ANR-07-BLAN-0228.

This work is based on observations taken by the CANDELS Multi-Cycle Treasury Program with the NASA/ESA HST, which is operated by the Association of Universities for Research in Astronomy, Inc., under NASA contract NAS5-26555.

\bibliographystyle{mn2e}
\bibliography{hizels}

\appendix
\section{$\rm M_{20}$ simulations}
\label{sec:sim}
We discuss here the effect of the {\sc sextractor} property {\sc deblend\_mincont} on the segmentation map and derived $\rm M_{20}$ value.  In \S\ref{sec:morphcal} we find that by analysing the $\rm M_{20}$ values and visual classification of a sample of $z=1.4-2.5$ star forming galaxies, a segmentation map generated with a value of {\sc deblend\_mincont}=0.1 provides a demarcation between major mergers and non-mergers, with a boundary found at $\rm M_{20}\sim-1.5$ (we later fix this value to $-1.45$). 

\begin{figure}
   \centering
\includegraphics[scale=0.5]{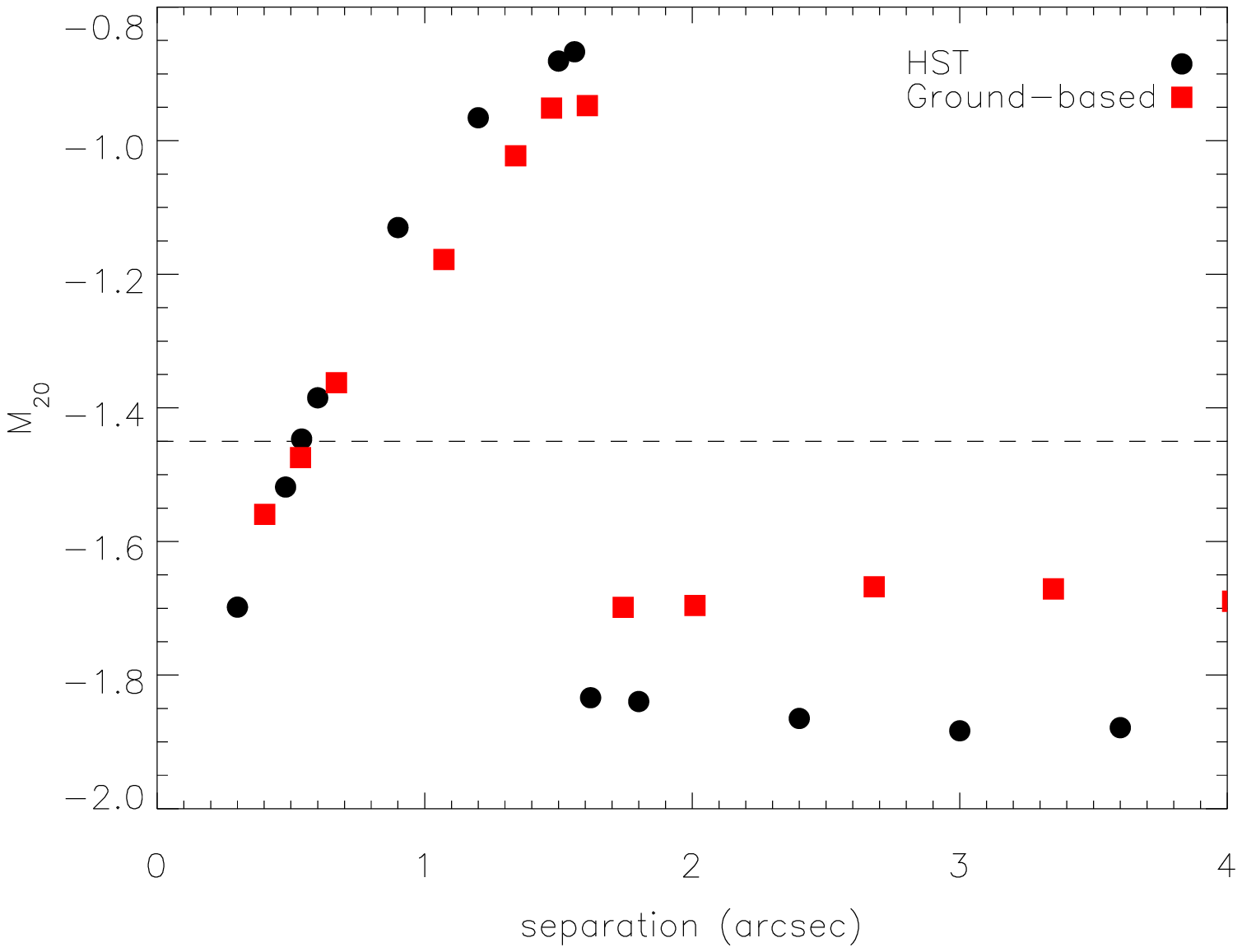} 
\caption[]{The $\rm M_{20}$ value plotted against separation derived from a simple simulation of two identical face-on, disc galaxies approaching each other, as described in the text. The simulated ground based data are represented by red squares and the simulated $HST$ data are black circles. To make the separations at which the $\rm M_{20}$ value jumps to be the same, we adopt a {\sc deblend\_mincont}=0.10 and 0.03 for the creation of the space- and ground-based {\sc sextractor} segmentation maps respectively. The horizontal line at $\rm M_{20}=-1.45$ represents the boundary between mergers above and non-merger below which we adopt throughout the paper. }
   \label{fig:db01}
\end{figure}

To quantify how the $\rm M_{20}$ value relates to a major merger we run some very basic simulations. The simulations comprise of moving one artificial galaxy towards another and plotting the variation of $\rm M_{20}$ with distance. The artificial observations are created using the {\sc galfit} software with both galaxies being face on discs (i.e. S\'{e}rsic index, $n=1$) of the same magnitude and half-light radius. We first perform an analysis appropriate to the high redshift star forming galaxies for which {\sc deblend\_mincont}=0.1 is found to efficiently select mergers. For the simulations we use the appropriate values of the sky noise, magnitude zero point and PSF of the observation we are simulating. The results of this simulation are presented in the black points on Figure \ref{fig:db01}. This demonstrates that for a {\sc deblend\_mincont}=0.1 and CANDELS {\it HST} data the $\rm M_{20}$ value remains low and consistent with being a non-merger for galaxy separations down to $\sim 1.6$ arcsec ($\sim$13\,\rm kpc at $z=0.84-2.23$), as the galaxies have individual segmentation maps. Once this separation drops below this value the two galaxies share the same segmentation map and thus the $\rm M_{20}$ value jumps dramatically as the top 20\% of the light is now spread over two locations rather than one. As the separation decreases further this value lowers until a point is reached where the top 20\% of the light is essentially co-located at the centre of a single bright galaxy and as such the $\rm M_{20}$ curve resembles a `shark fin'. The distance over which this system would be classed as a merger is then $\sim1$ arcsec which at $z=0.84-2.23$ corresponds to a distance of $\sim8\rm\,kpc$. 

Potentially the most important factor influencing the $\rm M_{20}$ value for a given galaxy is whether it is derived from the space-based {\it HST} data as discussed above or from the ground-based imaging with a significantly larger PSF and different background characteristics. We first investigate this by using the original space-based value of {\sc deblend\_mincont}=0.1 on the ground-based data and find that this gives a factor of $\sim2$ larger range in separation over which the galaxies would be classified as a merger and would thus result in an increase in merger numbers relative to the {\it HST} imaging. A value of {\sc deblend\_mincont}=0.03 accounts for this difference, equalling the separation over which a `merger' occurs with the results plotted as red squares in Figure \ref{fig:db01}. This plot confirms the slope in the relation seen between ground and space-based derived $\rm M_{20}$ seen in Figure \ref{fig:ginimatch} and used to calibrate the ground-based values.

For the lower redshift $z=0.4$ sample, this angular distance range corresponds to $5.3\rm\,kpc$ and as such may miss some galaxies that would have been classed as mergers in the higher $z$ samples. We therefore alter the value of {\sc deblend\_mincont} to 0.11 for the CANDELS and 0.04 for ground-based imaging to account for this so that the same separation in kpc is used at each redshift. 

By varying the relative magnitudes of the galaxies and assuming the flux is linearly proportional to the mass and the size is proportional to the square root of the mass we test what mass ratio of mergers can be seen with this method. The result is that the $\rm M_{20}$ coefficient is sensitive to mergers with a luminosity (mass) ratio down to $\sim1:10$ (in agreement with the simulations of  \citealt{lotz2010}). For mass ratios less than this the $\rm M_{20}$ coefficient does not increase significantly when the two galaxies share the same segmentation map.

\end{document}